\documentclass[draftclsnofoot,onecolumn]{IEEEtran}
\usepackage[latin9]{inputenc}
\usepackage{color}
\usepackage{units}
\usepackage{amsmath}
\usepackage{amsthm}
\usepackage{amssymb}
\usepackage{graphicx}
\usepackage{esint}
\usepackage[unicode=true,
 bookmarks=true,bookmarksnumbered=false,bookmarksopen=false,
 breaklinks=false,pdfborder={0 0 1},backref=false,colorlinks=true]
 {hyperref}
\hypersetup{
 linkcolor=blue,urlcolor=blue,citecolor=red,anchorcolor=blue}

\makeatletter
\theoremstyle{plain}
\newtheorem{thm}{\protect\theoremname}
\theoremstyle{definition}
\newtheorem{defn}[]{\protect\definitionname}
\theoremstyle{plain}
\newtheorem{prop}[]{\protect\propositionname}
\theoremstyle{plain}
\newtheorem{lem}[]{\protect\lemmaname}
\theoremstyle{plain}
\newtheorem{cor}[]{\protect\corollaryname}
\theoremstyle{remark}
\newtheorem{rem}[]{\protect\remarkname}

\usepackage{bm}
\usepackage{mathabx}

\usepackage{subfigure}
\usepackage{epstopdf}
\usepackage{float}
\usepackage{subfloat}
\usepackage{array}
\usepackage{tabularx}
\usepackage{multirow}
\usepackage{tikz}
\usepackage{algorithmic}

\newcommand{\Ebb}{\mathbb{E}}

\newcommand{\Pbb}{\mathbb{P}}

\newcommand{\Ccal}{\mathcal{C}}

\newcommand{\Lcal}{\mathcal{L}}

\newcommand{\Scal}{\mathcal{S}}
\newcommand{\Tcal}{\mathcal{T}}
\newcommand{\Ucal}{\mathcal{U}}

\newcommand{\Wcal}{\mathcal{W}}
\newcommand{\Xcal}{\mathcal{X}}

\tikzstyle{arw}=[->,>=latex]
\tikzstyle{node}=[draw,rectangle,rounded corners, minimum width=1cm,minimum height =.75 cm]

\usepackage{color}%May be necessary if you want to color links

\usepackage{adjustbox}

\usepackage{algorithmic}

\makeatother

\providecommand{\corollaryname}{Corollary}
\providecommand{\definitionname}{Definition}
\providecommand{\lemmaname}{Lemma}
\providecommand{\propositionname}{Proposition}
\providecommand{\remarkname}{Remark}
\providecommand{\theoremname}{Theorem}

\begin{document}

\title{Joint Source-Channel Secrecy Using Uncoded Schemes: Towards Secure
Source Broadcast}

\author{Lei Yu, Houqiang Li, \textit{Senior} \textit{Member, IEEE, }and Weiping
Li, \textit{Fellow, IEEE}\thanks{L. Yu is with the Department of Electrical and Computer Engineering,
National University of Singapore, Singapore (e-mail: leiyu@nus.edu.sg).
This work was done when he was at University of Science and Technology
of China. H. Li and W. Li are with the Department of Electronic Engineering
and Information Science, University of Science and Technology of China,
Hefei, China (e-mail: \{lihq,wpli\}@ustc.edu.cn).}}
\maketitle
\begin{abstract}
This paper investigates a  joint source-channel secrecy problem
for the Shannon cipher broadcast system. We suppose list secrecy is
applied, i.e., a wiretapper is allowed to produce a list of reconstruction
sequences and the secrecy is measured by the minimum distortion over
the entire list. For discrete communication cases, we propose a permutation-based
uncoded scheme, which cascades a random permutation with a symbol-by-symbol
mapping. Using this scheme, we derive an inner bound for the admissible
region of secret key rate, list rate, wiretapper distortion, and distortions
of legitimate users. For the converse part, we easily obtain an outer
bound for the admissible region from an existing result. Comparing
the outer bound with the inner bound shows that the proposed scheme
is optimal under certain conditions. Besides, we extend the proposed
scheme to the scalar and vector Gaussian communication scenarios,
and characterize the corresponding performance as well. For these
two cases, we also propose another uncoded scheme, orthogonal-transform-based
scheme, which achieves the same performance as the permutation-based
scheme. Interestingly, by introducing the random permutation or the
random orthogonal transform into the traditional uncoded scheme, the
proposed uncoded schemes, on one hand, provide a certain level of
secrecy, and on the other hand, do not lose any performance in terms
of the distortions for legitimate users.
\end{abstract}


\section{Introduction}

\label{sec:introduction}

Investigations on joint source-channel coding (JSCC) could trace back
to Shannon's pioneering work \cite{Shannon49-2}, where a geometric
method was developed to design a communication system. For the JSCC
of transmitting a Gaussian source over a Gaussian broadcast channel,
Goblick observed \cite{Goblick} that when the source and channel
bandwidths are matched (i.e., one channel use per source sample),
directly sending a scaled version of the source samples on the channel
(i.e., linear scheme) is in fact optimal; while for this case the
separation scheme that cascades source coding with channel coding
indeed suffers a performance loss \cite{Gastpar}. For vector Gaussian
communication cases, the optimal linear coding was studied in \cite{Lee76}.
In general, the schemes that consist of symbol-by-symbol mappings
(not limited to the linear one) are named \emph{uncoded schemes. }The
optimality of uncoded schemes for the general source-channel pair
has been investigated in \cite{Gastpar}, which showed that the Shannon
limit can be achieved by uncoded schemes only when the source and
channel satisfy a certain probabilistic matching condition. To further
improve the performance for mismatched source-channel pairs, the hybrid
coding (or hybrid digital-analog coding) has been studied in \cite{Shamai}-\cite{Yu2016-1},
which combines the traditional digital coding and symbol-by-symbol
mapping together. As for the converse part of JSCC problem, Reznic
\emph{et al.} \cite{Reznic} and Tian \emph{et al.} \cite{Tian} derived
some nontrivial converse results for Gaussian source broadcast problem.
Besides, Yu \emph{et al.} \cite{Yu2016,Yu2016-1} generalized the
achievability and converse results for the Gaussian communication
to the general source-channel case.

On information-theoretic security, the Shannon cipher system (the
noisy broadcast version depicted in Fig. \ref{fig:Shannon}) was first
investigated in Shannon's pioneering work \cite{Shannon49}, where
a sender A communicates with a legitimate receiver B secretly by exploiting
a shared secret key. For lossy source communication, wiretapper might
only want to decrypt a lossy version of the source. Schieler \emph{et
al.} \cite{Schieler} studied a distortion-based secrecy measure in
the Shannon cipher system around the assumption that the wiretapper
has ability to conduct list decoding with fixed list size, and the
secrecy is measured by the minimum distortion over the entire list.
Yu \emph{et al.} \cite{Yu-1} showed that the systems with this secrecy
measure are equivalent to those with secrecy measured by a new quantity
\emph{lossy-equivocation}, which could be considered as a lossy extension
of the traditional equivocation. Hence the list secrecy is closely
related to the traditional equivocation as well. Furthermore, Yu \emph{et
al.} used this secrecy measure to study the problem of \emph{source-channel
secrecy} for the Shannon cipher system, and showed that for the source-channel
pair satisfying certain conditions, an uncoded scheme could outperform
the separate one.

JSCC improves the robustness of communication or the performance of
broadcast, while secrecy coding improves the security of communication
by exploiting the secret key and/or the wiretap channel. Therefore,
intuitively the robustness and the security could be obtained simultaneously
if we combine JSCC and secrecy coding together. This \emph{joint source-channel
secrecy} (JSCS) problem has been considered in several works already.
Yamamoto in \cite{Yamamoto} studied the secure lossy transmission
over the noisy wiretap channel with secrecy measured by the wiretapper's
best reconstruction distortion. However, it is shown in \cite{Schieler}
this secrecy measure is cheap and fragile, since only one bit of secret
key suffices to achieve the optimality of secrecy, and meanwhile,
only one bit of additional information for the wiretapper suffices
to decrypt this optimal encryption scheme. A different formulation
of the problem was considered in \cite{Wilson}, where the authors
assumed there is a fixed information leakage to the wiretapper and
wish to minimize the distortion at the legitimate receiver, while
at the same time providing a graceful distortion degradation when
there is an SNR (Signal Noise Ratio) mismatch. They showed that, for
a positive leakage, this can be achieved by a hybrid digital-analog
coding. This scenario was extended to consider side information at
the receiver in \cite{Bagherikaram} or side information at the sender
in \cite{Bagherikaram13}.

Analog encryption (or analog scrambling) technologies, e.g., sign-change
based scheme \cite{Kak}, permutation based scheme \cite{Kak} and
bandwidth-keeping scheme \cite{Wyner79}, can be seen as uncoded JSCS
schemes as well, although they are not designed for a specified source-channel
pair. Sign-change based scheme improves secrecy by changing the sign
of each sample according to the secret key. But owing to at most one
bit secret key used per sample, this scheme could not provide higher
secrecy even with a higher key rate available. The permutation based
scheme improves secrecy by shuffling the positions of samples. Unlike
the sign-change based scheme, it supports any arbitrarily high key
rate. Furthermore, Kang and Liu \cite{Kang} recently applied the
permutation operation in a digital encryption scheme, and showed that
the permutation is another powerful encryption technique (besides
the one-time pad) to achieve the optimality of secrecy.

\subsection{Contributions}

In this paper, we consider the joint source-channel secrecy problem
of secure source broadcast in the bandwidth-matched Shannon cipher
system (see Fig. \ref{fig:Shannon}). The list secrecy \cite{Schieler}
is used to measure secrecy, that is, the wiretapper is allowed to
conduct list decoding with fixed list size, and the secrecy is measured
by the minimum distortion over the entire list. We study an achievable
 region of secret key rate, list rate, wiretapper distortion, and
distortions of all legitimate users and show optimality under certain
conditions. Our contributions are as follows:
\begin{enumerate}
\item For the discrete source case, we propose a permutation-based uncoded
scheme, which cascades a random permutation with a symbol-by-symbol
mapping. Our scheme differs from the permutation based scheme proposed
in \cite{Kang} in two main aspects: 1) our scheme, coupling a permutation
operation with a traditional \emph{uncoded scheme,} is designed for
the \emph{source-channel secrecy problem},\emph{ }however, the scheme
in \cite{Kang} couples a permutation operation with a  \emph{digital
scheme, }and is designed for the\emph{ source-secrecy coding problem};
2) in addition to the finite alphabet case, we also extend the scheme
to source-channel pairs with countably infinite alphabets and Gaussian
source-channel pairs, which require us to use some more powerful techniques,
including unified typicality, $\mathsf{d}-$tilted information, and
geometric analysis.  By analyzing the proposed scheme, we provide
an inner bound for the admissible region. For the converse part, we
give an outer bound by using our recent result \cite{Yu-1}. Comparing
the outer bound with the inner bound shows that the proposed scheme
is optimal under certain conditions.
\item We extend the proposed scheme to scalar and vector bandwidth-matched
Gaussian communication scenarios. For these two cases, we also propose
another uncoded scheme, orthogonal-transform-based scheme, which achieves
the same inner bounds as the one achieved by the permutation-based
scheme. Interestingly, by introducing the random permutation or the
random orthogonal transform into the traditional uncoded scheme, the
proposed uncoded schemes, no matter for the discrete source case or
the Gaussian source-channel case, on one hand, provide a certain level
of secrecy, and on the other hand, do not lose any performance in
terms of the distortions for legitimate users.
\end{enumerate}
Schieler and Cuff \cite{Schieler} studied the list secrecy problem
for the \emph{noiseless} \emph{point-to-point}\footnote{Here the word \emph{noiseless} means the wiretap channel is noiseless,
and the word \emph{point-to-point} means there is only one legitimate
user in the system.} version of Shannon cipher system, and showed a digital scheme, in
which the secret key is used to choose a source codebook to code the
source sequence, is optimal. For this problem, a separate coding,
cascading source coding and one-time pad, has been proven optimal
as well \cite{Yu-1}.  Yu \emph{et al.} \cite{Yu-1} extended this
problem to the \emph{noisy} channel case, and showed that the separate
strategy (cascading source coding, one-time pad, and channel coding)
is suboptimal in general and a single-letter uncoded scheme could
outperform the separate scheme. In this paper we extend the problem
to \emph{noisy broadcast} scenarios, and propose two kind of uncoded
schemes that adopt two different encryption strategies\textemdash random
permutation and random orthogonal-transform (instead of the traditional
one-time pad encryption). We show the proposed uncoded schemes could
achieve the optimality under certain cases.

\emph{}

The rest of the paper is organized as follows. Section II formulates
the joint source-channel secrecy problem. Section III proposes a permutation-based
uncoded scheme for the discrete communication, and analyzed the corresponding
performance. Sections IV and V extend the proposed scheme to the scalar
and vector Gaussian communications respectively, and another scheme,
orthogonal-transform based scheme, is also proposed in these two sections.
Finally, Section VI concludes the paper.

\section{\label{sec:Problem-Formulation}Problem Formulation}

\subsection{Problem setup}

\begin{figure}
\centering\includegraphics[width=0.5\textwidth]{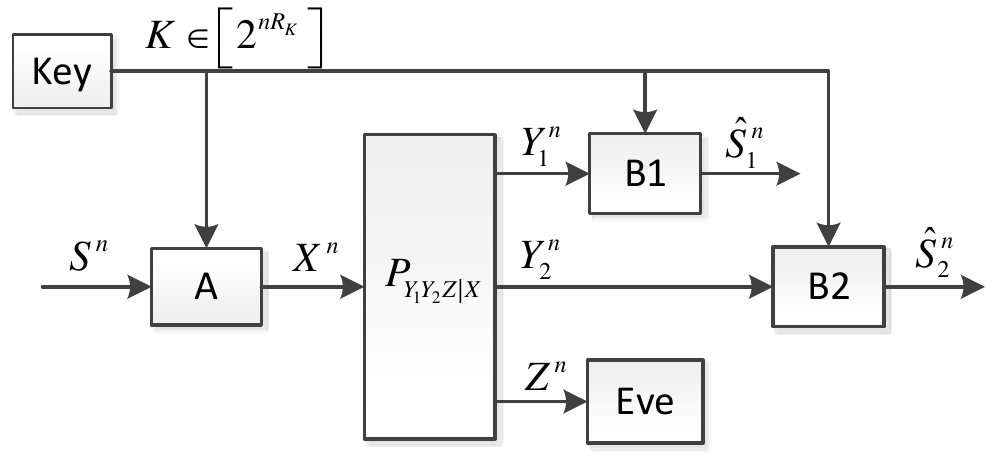}
\protect\caption{\label{fig:Shannon}The Shannon cipher broadcast system{\small{}. }}
\end{figure}

Consider a bandwidth-matched\footnote{\label{fn:Although-here-we}Although here we consider a bandwidth-matched
communication system, our results in this section are easy to be extended
to any bandwidth-mismatched system since any system with source-channel
bandwidth ratio $\frac{n_{s}}{n_{c}}$ can be converted into a bandwidth-matched
system, by considering $n_{s}$ source symbols and $n_{c}$ channel
symbols as a source supersymbol and a channel supersymbol, respectively.
} Shannon cipher broadcast system with two legitimate users\footnote{Although we only consider the system with two legitimate users, our
results derived in this paper can be easily extended to the cases
with more legitimate users.} shown in Fig. \ref{fig:Shannon}, where a sender A and two legitimate
receivers B1 and B2 share a secret key $K$ that is uniformly distributed
over $\left[2^{nR_{\mathsf{K}}}\right]$\footnote{In this paper, the set $\left\{ 1,...,m\right\} $ is sometimes denoted
by $[m]$.} and independent of a source $S^{n}$. The sender A observes the discrete
memoryless (DM) source sequence $S^{n}$ with each element i.i.d.
(independent and identically distributed) according to $P_{S}$, and
then transmits it to the legitimate users B1 and B2 over a DM wiretap
broadcast channel $P_{Y_{1}Y_{2}Z|X}$ confidentially by utilizing
the secret key and the wiretap channel. Finally, the legitimate users
B1 and B2 produce source reconstructions $\widehat{S}_{1}^{n}$ and $\widehat{S}_{2}^{n}$,
respectively.

\begin{defn}
An $\left(n,R_{\mathsf{K}}\right)$ block code consists of\\
 1) Encoder: $\varphi:\mathcal{S}^{n}\times\left[2^{nR_{\mathsf{K}}}\right]\mapsto\mathcal{X}^{n}$;\\
 2) Decoders: $\psi_{i}:\mathcal{Y}_{i}^{n}\times\left[2^{nR_{\mathsf{K}}}\right]\mapsto\mathcal{\widehat{S}}_{i}^{n}$,
$i=1,2$.\\
 The encoder and decoders can be stochastic.
\end{defn}
Another output $Z^{n}$ of the channel is accessed by a wiretapper
Eve. Based on $Z^{n}$, the wiretapper produces a list $\Lcal(Z^{n})\subseteq\widecheck{\mathcal{S}}^{n}$
and the induced distortion is set to the minimum one over the entire
list, i.e., $\min_{\widecheck{s}^{n}\in\Lcal(Z^{n})}d_{\mathsf{E}}(S^{n},\widecheck{s}^{n}),$
where $d_{\mathsf{E}}\left(s^{n},\widecheck{s}^{n}\right)\triangleq\frac{1}{n}\sum_{t=1}^{n}d_{\mathsf{E}}\left(s_{t},\widecheck{s}_{t}\right)$
is a distortion measure for the wiretapper. For given distortion levels
$D_{0},D_{1},D_{2}$, Nodes A and B1, B2 want to communicate the source
within distortions $D_{1},D_{2}$ (for B1 and B2 respectively) by
exploiting the secret key and the wiretap channel, while ensuring
that the wiretapper's strategy always suffers distortion above $D_{0}$
with high probability.
\begin{defn}
\label{def:list}The tuple $\left(R_{\mathsf{K}},R_{\mathsf{L}},D_{0},D_{1},D_{2}\right)$
is achievable if there exists a sequence of $(n,R_{\mathsf{K}})$ codes such
that $\forall\epsilon>0$, \\
 1) Distortion constraint:
\begin{align}
\Pbb\Big[d_{\mathsf{B}}(S^{n},\widehat{S}_{i}^{n})\le D_{i}+\epsilon\Big] & \xrightarrow{n\to\infty}1,\:i=1,2;\label{eq:-42}
\end{align}
where $d_{\mathsf{B}}\left(s^{n},\widehat{s}^{n}\right)\triangleq\frac{1}{n}\sum_{t=1}^{n}d_{\mathsf{B}}\left(s_{t},\widehat{s}_{t}\right)$\footnote{For simplicity, we only consider the legitimate users have the same
distortion measure. Note that our results derived in this paper still
hold for the case with different distortion measures.} is a distortion measure for the legitimate users;\\
 2) Secrecy constraint:
\begin{equation}
\min_{\substack{\Lcal_{n}(z^{n}):\\
\limsup_{n\rightarrow\infty}\frac{1}{n}\log|\Lcal_{n}|\leq R_{\mathsf{L}}-\epsilon
}
}\Pbb\Big[\min_{\widecheck{s}^{n}\in\Lcal(Z^{n})}d_{\mathsf{E}}(S^{n},\widecheck{s}^{n})\geq D_{0}-\epsilon\Big]\xrightarrow{n\to\infty}1.\label{listsecrecylossy}
\end{equation}
\end{defn}
\begin{defn}
\label{def:The-admissible-region}The admissible region $\mathcal{R}\triangleq\left\{ \textrm{Achievable }\left(R_{\mathsf{K}},R_{\mathsf{L}},D_{0},D_{1},D_{2}\right)\right\} $.
\end{defn}
We assume that the wiretapper knows the $(n,R_{\mathsf{K}})$ code and the
distributions $P_{S}$ and $P_{Y_{1}Y_{2}Z|X}$.

\subsection{Henchman problem}

\begin{figure}
\centering\includegraphics[width=0.5\textwidth]{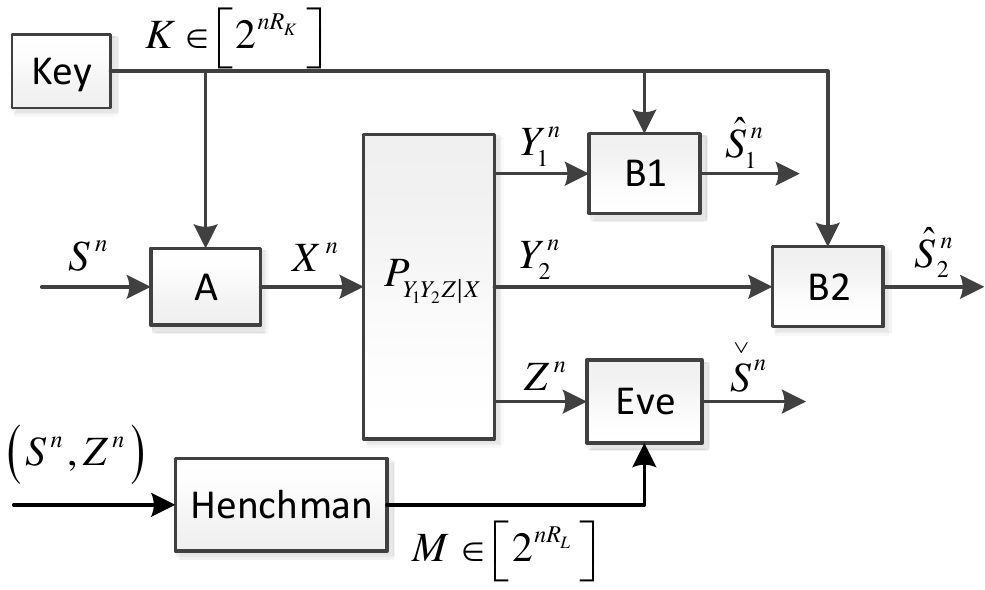}
\protect\caption{\label{fig:henchman}Henchman problem. }
\end{figure}

The problem above is equivalent to the henchman problem \cite{Schieler},
in which wiretapper reconstructs a single sequence with the help of
a rate-limited henchman who can access to the source $S^{n}$ and
the wiretapper\textquoteright s observation $Z^{n}$. As depicted
in Fig. \ref{fig:henchman}, the wiretapper receives the best possible
$nR_{n}$ bits from the henchman to assist in producing a reconstruction
sequence $\widecheck{S}^{n}$.

\begin{defn}
The $R_{n}$ henchman code (Hcode) of a $\left(n,R_{\mathsf{K}}\right)$ block
code consists of\\
 1) Encoder: $\varphi_{\mathsf{H}}:\mathcal{S}^{n}\times\mathcal{Z}^{n}\mapsto\left[2^{nR_{n}}\right]$;\\
 2) Decoder: $\psi_{\mathsf{H}}:\left[2^{nR_{n}}\right]\times\mathcal{Z}^{n}\mapsto\mathcal{\widecheck{S}}^{n}$.
\end{defn}
We assume that the wiretapper and henchman are aware of the $\left(n,R_{\mathsf{K}}\right)$
block code adopted by Nodes A and B, and they cooperate to design
a henchman code based on the $\left(n,R_{\mathsf{K}}\right)$ block code.
\begin{defn}
\label{def:henchman}The tuple $\left(R_{\mathsf{K}},R_{\mathsf{L}},D_{0},D_{1},D_{2}\right)$
is achievable in the henchman problem if there exists a sequence of
$(n,R_{\mathsf{K}})$ codes such that $\forall\epsilon>0$, \\
 1) Distortion constraint: \eqref{eq:-42}; \\
 2) Secrecy constraint:
\begin{equation}
\min_{\substack{R_{n}\mathsf{Hcodes}:\\
\limsup_{n\rightarrow\infty}R_{n}\leq R_{\mathsf{L}}-\epsilon
}
}\Pbb\Big[d_{\mathsf{E}}(S^{n},\widecheck{S}^{n})\geq D_{0}-\epsilon\Big]\xrightarrow{n\to\infty}1.\label{mainobj}
\end{equation}
\end{defn}
The equivalence between the list secrecy problem and the henchman
problem, shown in the following proposition, has been proven by Schieler
and Cuff \cite[Prop. 1]{Schieler}.
\begin{prop}
\label{prop:-The-tuple}\cite{Schieler} The tuple $\left(R_{\mathsf{K}},R_{\mathsf{L}},D_{0},D_{1},D_{2}\right)$
is achievable in the list reconstruction problem if and only if it
is achievable in the henchman problem.
\end{prop}
Furthermore, the list secrecy problem and henchman problem are also
equivalent to a lossy-equivocation secrecy problem; see \cite{Yu-1}.

In addition to the DM system, we also consider the Shannon cipher
system with a Gaussian source $S\sim\mathcal{N}\left(0,\lambda\right)$
transmitted over a power-constrained Gaussian wiretap broadcast channel
\begin{align}
Y_{i} & =X+V_{i},i=1,2,\\
Z & =X+V_{0},
\end{align}
where $V_{i},i=0,1,2$ are zero-mean additive Gaussian noises with
variances $N_{i},i=0,1,2$, independent of $X$. For this case, the
constraint on channel input power
\begin{equation}
\mathbb{P}\bigl[\rho\left(X^{n}\right)\le P+\epsilon\bigr]\xrightarrow{n\to\infty}1,\forall\epsilon>0,\label{eq:powerconstraint}
\end{equation}
should be added to Definitions \ref{def:list} and \ref{def:henchman},
where $\rho\left(x^{n}\right)=\frac{1}{n}\sum_{i=1}^{n}x_{i}^{2}$.
For the system involving the channel power constraint, Proposition
\ref{prop:-The-tuple} still holds.

\section{Discrete Communication}

\label{sec:joint-1}

\subsection{Permutation based Scheme (Finite Alphabets)}

\label{subsec:encryption-1} In this section, we propose a secure
uncoded scheme by coupling the permutation operation with the traditional
uncoded JSCC scheme. The uncoded scheme for JSCC system (with two
receivers) consists of three symbol-by-symbol mappings: $x\left(s\right),\widehat{s}_{1}(y_{1}),\widehat{s}_{2}(y_{2})$.
The induced distortions are $D_{i}=\mathbb{E}d_{\mathsf{B}}(S,\widehat{s}_{i}(Y_{i})),i=1,2$.
It is easy to show that we can benefit from replacing the encoder
$x\left(s\right)$ with a stochastic one $P_{X|S}$ when secrecy is
considered for the system. On the other hand, observe that $d_{\mathsf{B}}(s^{n},\widehat{s}_{i}^{n})=\frac{1}{n}\sum_{t=1}^{n}d_{\mathsf{B}}(s_{t},\widehat{s}_{i,t})=\mathbb{E}_{T_{s^{n},\widehat{s}_{i}^{n}}}d_{\mathsf{B}}(s,\widehat{s}_{i})$,
where $T_{s^{n},\widehat{s}_{i}^{n}}$ denotes the joint type (empirical
distribution) of $(s^{n},\widehat{s}_{i}^{n})$. That is, the induced
distortions only depend on the joint type of source and reconstruction
sequences. Therefore, if we want to improve the secrecy of a scheme
and at the same time retain the induced distortions unchanged, we
only need to require the encryption and decryption operations does
not change the joint type of source and reconstruction sequences.
That is, for the encryption $s^{\prime n}\left(s^{n},k\right)$ and
the decryption $\widehat{s}_{i}^{n}\left(\widehat{s}_{i}^{\prime n},k\right)$,
we require $T_{s^{\prime n},\widehat{s}_{i}^{\prime n}}=T_{s^{n},\widehat{s}_{i}^{n}}$.
To that end, here we consider a random permutation as the encryption
operation, and the inverse permutation as the decryption operation.
Obviously, the permutation and its inverse operation do not change
the joint type of the source sequence and its reconstructions.

\emph{Codebook (Public Key) Generation:} Generate a permutation set
$\mathcal{C}=\left\{ \Psi_{k},k\in\left[2^{nR_{\mathsf{K}}}\right]\right\} $
with each element uniformly at random and independently selected from
the set of permutations of $\left[n\right]$ (which is denoted as
$\mathcal{\mathfrak{S}}_{n}$). As a public key, the codebook $\mathcal{C}$
is revealed to the sender and all the receivers (including the wiretapper).

\emph{Encoding:} Upon observing a source sequence $s^{n}$ and a key
$k$ , the encoder first generates
\begin{equation}
s^{\prime n}=\Psi_{k}\left(s^{n}\right),
\end{equation}
and then generates $x^{n}$ according to $\prod_{t=1}^n P_{X|S}\left(x_t|s^{\prime}_t\right)$.
%$\left[\prod P_{X|S}\right]\left(x^{n}|s^{\prime n}\right)$.
Here for a permutation sequence $\Psi_{k}=\sigma^{n}$, $\Psi_{k}\left(s^{n}\right)\triangleq(s_{\sigma_{1}},s_{\sigma_{2}},...,s_{\sigma_{n}})$
denotes the permutation operation on $s^{n}$ (more precisely, on
the indices of $s^{n}$) respect to the permutation sequence $\Psi_{k}$.\footnote{In this paper, the permutation sequence is termed as \emph{permutation
sequence}, and to distinguish from it, the permutation mapping from
one sequence to another sequence is termed as \emph{permutation operation}.
When there is no disambiguation, we call both of them \emph{permutation.}
  }

\emph{Decoding (for Legitimate Users):} For the legitimate user $\textrm{B}i$,
$i=1,2,$ upon the received sequence $y_{i}^{n}$ and the key $k$,
the decoder first reconstructs $s^{\prime n}$ as
\begin{equation}
\widehat{s}_{i,t}^{\prime}=\widehat{s}_{i}(y_{i,t}),t\in[n],
\end{equation}
by using the symbol-by-symbol mapping $\widehat{s}_{i}\left(\cdot\right)$,
and then produces
\begin{equation}
\widehat{s}_{i}^{n}=\Psi_{k}^{-1}(\widehat{s}_{i}^{\prime n}),
\end{equation}
by using the inverse permutation operation $\Psi_{k}^{-1}\left(\cdot\right)$
of $\Psi_{k}\left(\cdot\right)$.

The proposed scheme above cascades a random permutation operation
with the traditional uncoded JSCC scheme. The uncoded JSCC part provide
a graceful degradation of the source for legitimate users with different
channel qualities. The random permutation operation part that shifts
the sequence in the same type provides a certain level of secrecy.
Next we will analyze the asymptotic performance of the proposed scheme
as blocklength $n$ goes to infinity. At first, we need introduce
some basic properties of the random codebook $\mathcal{C}$.

Observe that for any permutation sequence $\Psi$, the mapping between $\Psi\left(\cdot\right)$
and $\Psi^{-1}\left(\cdot\right)$ is bijective, hence we have the
following lemma.
\begin{lem}
\label{lem:permutation} Suppose $\Psi$ is a permutation sequence uniformly
at random selected from $\mathcal{\mathfrak{S}}_{n}$, the set of
permutations of $\left[n\right]$. Then $\Psi^{-1}$ is also uniformly
distributed on $\mathcal{\mathfrak{S}}_{n}$, and moreover for any
permutation sequence $\psi\in\mathcal{\mathfrak{S}}_{n}$, both $\Psi\left(\psi\right)$
and $\Psi^{-1}\left(\psi\right)$ also have the uniform distribution
on $\mathcal{\mathfrak{S}}_{n}$.
\end{lem}
Utilizing Lemma \ref{lem:permutation}, we can establish the following
lemma.
\begin{lem}
\label{lem:Random-permutation} Suppose $\Psi$ is a permutation sequence uniformly
at random selected from $\mathcal{\mathfrak{S}}_{n}$. Then $\Psi\left(s^{n}\right)$
transforms an arbitrary sequence $s^{n}\in\mathcal{S}^{n}$ into a
random sequence that is uniformly distributed on the set of sequences
of type $T_{s^{n}}$ (the type class of $T_{s^{n}}$). Moreover, for
finite $\mathcal{S}$, the set of sequences of type $T_{s^{n}}$ has
cardinality $2^{n\left(H(T_{s^{n}})-o(1)\right)}$, and hence
\begin{equation}
\mathbb{P}\left[\Psi\left(s^{n}\right)=s^{\prime n}\right]=2^{-n\left(H(T_{s^{n}})-o(1)\right)}1\left\{ T_{s^{\prime n}}=T_{s^{n}}\right\} ,\;s^{\prime n}\in\mathcal{S}^{n},
\end{equation}
where $o(1)$ denotes a term tending zero as $n\to\infty$.
\end{lem}
\begin{IEEEproof}
%The former part of this lemma is obvious, so we only prove the
\begin{align}
\mathbb{P}\left[\Psi\left(s^{n}\right)=s^{\prime n}\right] & =\sum_{\psi\in\mathcal{\mathfrak{S}}_{n}:\psi\left(s^{n}\right)=s^{\prime n}}\mathbb{P}\left(\Psi=\psi\right)\\
 & =\sum_{\psi\in\mathcal{\mathfrak{S}}_{n}:\psi\left(s^{n}\right)=s^{\prime n}}\frac{1}{n!}\\
 & =\frac{\prod_{s\in\mathcal{S}}\left(nT_{s^{n}}\left(s\right)\right)!}{n!}1\left\{ T_{s^{\prime n}}=T_{s^{n}}\right\} \\
 & =\frac{1\left\{ T_{s^{n}}=T_{s^{\prime n}}\right\} }{\left|\left\{ s^{\prime n}\in\mathcal{S}^{n}:T_{s^{\prime n}}=T_{s^{n}}\right\} \right|},\label{eq:-48}
\end{align}
where \eqref{eq:-48} follows from $\left|\left\{ s^{\prime n}\in\mathcal{S}^{n}:T_{s^{\prime n}}=T_{s^{n}}\right\} \right|=\frac{n!}{\prod_{s\in\mathcal{S}}\left(nT_{s^{n}}\left(s\right)\right)!}$.
This implies $\Psi\left(s^{n}\right)$ transforms an arbitrary sequence
$s^{n}\in\mathcal{S}^{n}$ into a random sequence uniformly distributed
on the set of sequences of type $T_{s^{n}}$.

From the type counting lemma \cite[Lem. 2.3]{Csiszar}, we have that
for finite $\mathcal{S}$,
\begin{equation}
\left(n+1\right)^{-\left|\mathcal{S}\right|}2^{nH(T_{s^{n}})}\leq\left|\left\{ s^{\prime n}\in\mathcal{S}^{n}:T_{s^{\prime n}}=T_{s^{n}}\right\} \right|\leq2^{nH(T_{s^{n}})}.\label{eq:-26}
\end{equation}
Hence $\left|\left\{ s^{\prime n}\in\mathcal{S}^{n}:T_{s^{\prime n}}=T_{s^{n}}\right\} \right|=2^{n\left(H(T_{s^{n}})-o(1)\right)}$.
Combining it with \eqref{eq:-48} gives us
\begin{equation}
\mathbb{P}\left[\Psi\left(s^{n}\right)=s^{\prime n}\right]=2^{-n\left(H(T_{s^{n}})-o(1)\right)}1\left\{ T_{s^{\prime n}}=T_{s^{n}}\right\} .
\end{equation}
\end{IEEEproof}
Lemma \ref{lem:Random-permutation} shows a nice property of the random
permutation operation: The resulting sequence will be uniformly distributed
on the set of sequences of type $T_{s^{n}}$ for the input sequence
$s^{n}$, if the permutation is randomly and uniformly chosen from
the set of permutations of $\left[n\right]$. Utilizing this property,
we can characterize the performance of the proposed scheme, as shown
in the following theorem. The proof of this theorem is given in Appendix
\ref{sec:Proof-of-TheoremDM}.
\begin{thm}[Permutation based Scheme for Finite Alphabets]
\label{thm:proposedDM} For DM communication with finite alphabets
($\mathcal{S},\widecheck{\mathcal{S}},\mathcal{X},\mathcal{Z},\mathcal{Y}_{i},\mathcal{\widehat{S}}_{i},i=1,2$
are all finite), the permutation based scheme above achieves the region
$\mathcal{R}^{\mathsf{(i)}}\subseteq\mathcal{R}$, where
\begin{equation}
\mathcal{R}^{\mathsf{(i)}}\triangleq\underset{P_{X|S}}{\bigcup}\left\{ \begin{array}{l}
\left(R_{\mathsf{K}},R_{\mathsf{L}},D_{0},D_{1},D_{2}\right):\\
D_{i}\geq\min_{\widehat{s}_{i}}\mathbb{E}d_{\mathsf{B}}(S,\widehat{s}_{i}(Y_{i})),i=1,2,\\
R_{\mathsf{L}}\leq\min\left\{ R_{\mathsf{K}}+R_{S|Z}(D_{0}),R_{S}(D_{0})\right\}
\end{array}\right\} ,\label{eq:-33}
\end{equation}
with
\begin{equation}
\left(S,Y_{1},Y_{2},Z\right)\sim\sum_{x}P_{S}P_{X|S}P_{Y_{1}Y_{2}Z|X},\label{eq:-30}
\end{equation}
\begin{align}
R_{S}\left(D\right) & =\mathop{\min}\limits _{P_{\widecheck{S}|S}:\mathbb{E}d_{\mathsf{E}}(S,\widecheck{S})\le D}I(S;\widecheck{S})\label{eq:rd}
\end{align}
denoting the rate-distortion function of $S$, and
\begin{align}
R_{S|Z}\left(D\right) & =\mathop{\min}\limits _{P_{\widecheck{S}|SZ}:\mathbb{E}d_{\mathsf{E}}(S,\widecheck{S})\le D}I(S;\widecheck{S}|Z)\label{eq:rd-si-1}
\end{align}
denoting the conditional rate-distortion function of $S$ given two-sided
information $Z$.
\end{thm}
Note that for the $\mathcal{R}^{\mathsf{(i)}}$ above, the components $\left(D_{1},D_{2}\right)$
and the components $\left(R_{\mathsf{K}},R_{\mathsf{L}},D_{0}\right)$ depend on each
other through $P_{X|S}$. Observe that for a given $P_{X|S}$, $\min_{\widehat{s}_{i}}\mathbb{E}d_{\mathsf{B}}(S,\widehat{s}_{i}(Y_{i})),i=1,2$
are the minimal distortions that the legitimate users can achieve
even for the non-secrecy communication case. On the other hand, $\min\left\{ R_{\mathsf{K}}+R_{S|Z}(D_{0}),R_{S}(D_{0})\right\} $
is larger than $R_{S|Z}(D_{0})$, the optimal $R_{\mathsf{L}}$ can be achieved
by uncoded schemes when there is no key. Hence  compared with traditional
uncoded schemes, the proposed scheme, on one hand, improves improve
the performance of secrecy to a certain extent, and on the other hand,
does not lose any performance in terms of the distortions of legitimate
users.

The first constraint of $\mathcal{R}^{\mathsf{(i)}}$ is consistent with the
performance of traditional uncoded schemes. The second constraint
of $\mathcal{R}^{\mathsf{(i)}}$ , roughly speaking, follows from the following
argument. On one hand, the henchman and the wiretapper can ignore
the signal $Z^{n}$ altogether and use a $R_{S}(D_{0})$-rate point-to-point
source code to describe $S^{n}$ within distortion $D_{0}$. On the
other hand, the proposed scheme forces the wiretapper's optimal strategy
to be an indirect guessing strategy: First, the wiretapper decrypts
the secret key by using $R_{\mathsf{K}}$ rate; then upon the observation $Z^{n}$,
the wiretapper reconstructs the sequence $S^{\prime n}$ within distortion
$D_{0}$ by using rate $R_{S|Z}(D_{0})$ (denote the reconstruction
as $\widecheck{S}^{\prime n}$); finally, upon the secret key and $\widecheck{S}^{\prime n}$,
the wiretapper reconstructs the source $S^{n}$ as $\widecheck{S}^{n}=\Psi_{k}^{-1}(\widecheck{S}^{\prime n})$.
Obviously the distortion between $S^{n}$ and $\widecheck{S}^{n}$ is
the same as that between $S^{\prime n}$ and $\widecheck{S}^{\prime n}$,
since the average distortion only depends the joint type of the sequences.
Hence the wiretapper needs rate $R_{\mathsf{K}}+R_{S|Z}(D_{0})$ to achieve
the distortion $D_{0}$.

Now we consider a special case: sending a binary source over a binary
wiretap broadcast channel. For the binary communication, the source
is a Bernoulli source $S\sim\textrm{Bern}\left(\frac{1}{2}\right)$
with the Hamming distortion measure $d_{\mathsf{B}}(s,\widehat{s})=d_{\mathsf{E}}(s,\widehat{s})\triangleq0,\textrm{ if }s=\widehat{s};1,\textrm{ otherwise}$.
The binary wiretap broadcast channel is $Y_{i}=X\oplus V_{i},i=1,2,\:Z=X\oplus V_{0}$
with $V_{i}\sim\textrm{Bern}\left(p_{i}\right),V_{0}\sim\textrm{Bern}\left(p_{0}\right),0\leq p_{0},p_{1},p_{2}\leq\frac{1}{2}$.
Set $X=S\oplus E$ with $E\sim\textrm{Bern}(p^{\prime})$. Then from
Theorem \ref{thm:proposedDM}, we get the following corollary.
\begin{cor}[Binary Communication]
\label{cor:proposedbinary} For the binary communication above, we
have $\mathcal{R}^{\mathsf{(i)}}\subseteq\mathcal{R}$, where
\[
\mathcal{R}^{\mathsf{(i)}}\triangleq\underset{0\leq p^{\prime}\leq\frac{1}{2}}{\bigcup}\left\{ \begin{array}{l}
\left(R_{\mathsf{K}},R_{\mathsf{L}},D_{0},D_{1},D_{2}\right):\\
D_{i}\geq p^{\prime}\star p_{i},i=1,2,\\
R_{\mathsf{L}}\leq\min\left\{ R_{\mathsf{K}}+\left[H_{2}\left(p^{\prime}\star p_{0}\right)-H_{2}\left(D_{0}\right)\right]^{+},\left[1-H_{2}\left(D_{0}\right)\right]^{+}\right\}
\end{array}\right\} ,
\]
with $\left[x\right]^{+}\triangleq\max\left\{ 0,x\right\} $, $\star$
denoting the binary convolution, i.e.,
\begin{equation}
x\star y=(1-x)y+x(1-y),\label{eq:star}
\end{equation}
and $H_{2}$ denoting the binary entropy function, i.e.,
\begin{equation}
H_{2}(p)=-p\log p-(1-p)\log(1-p).\label{eq:binaryentropy}
\end{equation}
\end{cor}

\subsection{Permutation based Scheme (More General Alphabets)}

Theorem \ref{thm:proposedDM} can be extended to more general alphabets
cases, as shown in the following theorem. The proof of this theorem
is given in Appendix \ref{sec:Proof-of-TheoremGDM}.
\begin{thm}[Permutation based Scheme for More General Alphabets]
\label{thm:proposedGDM} Assume $\mathcal{S}$ is countable, $\widecheck{\mathcal{S}}$
is finite, and $\mathcal{X},\mathcal{Z},\mathcal{Y}_{i},\mathcal{\widehat{S}}_{i},i=1,2$
are general\footnote{An alphabet is countable means that it is either finite or countably
infinite. An alphabet is general means that it is either countable
or uncountable (e.g., continuous).}. Assume $H\left(S\right)$ is finite, and $P_{S}$ satisfies
\begin{align}
N_{P_{S}}\left(\frac{1}{n}\right) & =o\left(\frac{n}{\log n}\right),\label{eq:-46}\\
\Phi_{P_{S}}\left(\frac{1}{n}\right) & =o\left(\frac{1}{\log n}\right),\label{eq:-49}\\
\widetilde{N}_{P_{S}}\left(\frac{\delta}{\log n}\right) & =o\left(\frac{n}{\log^{2}n}\right),\forall0<\delta\leq1,\label{eq:-47}
\end{align}
where $N_{P_{S}}\left(\alpha\right)\triangleq\left|\left\{ s:P_{S}\left(s\right)\geq\alpha\right\} \right|$
denotes the number of probability values that are not smaller than
$\alpha$,  $\Phi_{P_{S}}\left(\alpha\right)\triangleq\sum_{s:P_{S}\left(s\right)<\alpha}P_{S}\left(s\right)$
denotes the sum of probability values that are smaller than $\alpha$,
$\widetilde{N}_{P_{S}}\left(\beta\right)\triangleq\min_{\alpha:\Phi_{P_{S}}\left(\alpha\right)\leq\beta}N_{P_{S}}\left(\alpha\right)$
 denotes the minimum number $N$ such that the sum of the probability
values except $N$ largest ones is not larger than $\beta$. Then
Theorem \ref{thm:proposedDM}  still holds.
\end{thm}
\begin{rem}
The conditions \eqref{eq:-46}-\eqref{eq:-47} is equivalent to as
$x\downarrow0$,\footnote{This claim holds when we ignore $n$ is an integer in \eqref{eq:-46}-\eqref{eq:-47}.}
\begin{align}
N_{P_{S}}\left(x\right) & =o\left(\frac{1}{x\log\frac{1}{x}}\right),\label{eq:-46-3}\\
\Phi_{P_{S}}\left(x\right) & =o\left(\frac{1}{\log\frac{1}{x}}\right),\label{eq:-49-3}\\
\widetilde{N}_{P_{S}}\left(x\right) & =o\left(x^{2}e^{\frac{\delta}{x}}\right),\forall0<\delta\leq1.\label{eq:-47-3}
\end{align}
\end{rem}
\begin{rem}
\label{rem:The-conditions--}The conditions \eqref{eq:-46}-\eqref{eq:-47}
require that the sequence $P_{S}\left(s\right),s\in\mathcal{S}$ should
vanish as fast as possible. Obviously, \eqref{eq:-46}-\eqref{eq:-47}
hold for any finite $\mathcal{S}$. Besides, for any countably infinite
$\mathcal{S}$, it is easy to verify that any distribution $P_{S}$
such that $P_{S}\left(s\right)=o\left(s^{-\alpha}\right),s=1,2,...$\footnote{Without loss of generality, any countably infinite $\mathcal{S}$
can be converted into $\left\{ 1,2,3,...\right\} $ by some bijective
mapping. } for some $\alpha>1$ satisfies \eqref{eq:-46}-\eqref{eq:-47} as
well. However, if $P_{S}\left(s\right)$ converges slower than or
as slow as $\frac{1}{s}$, then $\sum_{s\geq1}P_{S}\left(s\right)$
does not converge, and hence $P_{S}$ cannot be a probability distribution.
This implies Theorem \ref{thm:proposedGDM} holds for almost all probability
distributions.
\end{rem}
Note that for a countably infinite alphabet $\mathcal{S}$, we need
the conditions \eqref{eq:-46}-\eqref{eq:-47} to guarantee the existence
of a high-probability set (unified typicality set), for each sequence
of which, Lemma \ref{lem:Random-permutation} still holds.  This
further makes Theorem \ref{thm:proposedGDM} hold, just as done for
the finite alphabets case.

\subsection{Outer Bound}

For the system with a single legitimate user (remove the legitimate
user B2 from the system in Fig. \ref{fig:Shannon}), the following
outer bound for the admissible region of $\left(R_{\mathsf{K}},R_{\mathsf{L}},D_{0},D_{1}\right)$
has been proven by us recently \cite{Yu-1}.
\begin{lem}
\cite{Yu-1}\label{lem:singleuser-1} For the DM communication with
only one legitimate user,
\[
\mathcal{R}\subseteq\mathcal{R}^{\mathsf{(o)}}\triangleq\underset{P_{\widehat{S}_{1}|S}}{\bigcup}\left\{ \begin{array}{l}
\left(R_{\mathsf{K}},R_{\mathsf{L}},D_{0},D_{1}\right):C_{1}\geq I(S;\widehat{S}_{1}),\\
D_{1}\geq\mathbb{E}d_{\mathsf{B}}(S,\widehat{S}_{1}),\\
R_{\mathsf{L}}\leq\min\Bigl\{ R_{\mathsf{K}}+\Gamma\left(I(S;\widehat{S}_{1}),P_{Y_{1}|X},P_{Z|X}\right)+R_{S|\widehat{S}_{1}}(D_{0}),\\
\qquad R_{S}(D_{0})\Bigr\}
\end{array}\right\} ,
\]
where $C_{1}$ denotes the channel capacity for the legitimate user,
and
\begin{align}
\Gamma\left(R,P_{Y|X},P_{Z|X}\right) & \triangleq\min_{\substack{Q_{YZ|X}:Q_{Y|X}=P_{Y|X},\\
Q_{Z|X}=P_{Z|X}
}
}\max_{Q_{X}:I_{Q}\left(X;Y\right)\geq R}I_{Q}\left(X;Y|Z\right)
\end{align}
with $I_{Q}\left(\cdot\right)$ denoting the mutual information under
distribution $Q_{X}Q_{YZ|X}$, is a function specified by the wiretap
channel.
\end{lem}
The first two constraints of $\mathcal{R}^{\mathsf{(o)}}$ follow from the
source-channel coding theorem \cite{Gamal}, and the last constraint
follows from  an indirect decryption strategy for the wiretapper:
Roughly speaking, the wiretapper first reconstructs $\widehat{S}_{1}^{n}$
using rate $\Gamma(I(S;\widehat{S}_{1}),P_{Y_{1}|X},P_{Z|X})$, next decrypts
the secret key using rate $R_{\mathsf{K}}$, then upon $Y_{1}^{n}$ and secret
key, produces the legitimate user's reconstruction $\widehat{S}_{1}^{n}$,
and finally upon $\widehat{S}_{1}^{n}$ produces a final reconstruction
$\widecheck{S}^{n}$ using rate $R_{S|\widehat{S}_{1}}(D_{0})$. The details
can be seen in \cite{Yu-1}.

By applying this lemma to the system with two legitimate users (the
system considered in this paper), the following outer bound is immediate.
\begin{thm}[Outer Bound]
\label{thm:broadcast-1} For the DM communication (with two legitimate
users),
\[
\mathcal{R}\subseteq\mathcal{R}^{\mathsf{(o)}}\triangleq\underset{P_{\widehat{S}_{1}\widehat{S}_{2}|S}}{\bigcup}\left\{ \begin{array}{l}
\left(R_{\mathsf{K}},R_{\mathsf{L}},D_{0},D_{1},D_{2}\right):C_{i}\geq I(S;\widehat{S}_{i}),\\
D_{i}\geq\mathbb{E}d_{\mathsf{B}}(S,\widehat{S}_{i}),i=1,2,\\
R_{\mathsf{L}}\leq\min\left\{ R_{1},R_{2},R_{S}(D_{0})\right\}
\end{array}\right\} ,
\]
where $C_{i}$ denotes the channel capacity for the legitimate user
$i$, and
\begin{equation}
R_{i}=R_{\mathsf{K}}+\Gamma\left(I(S;\widehat{S}_{i}),P_{Y_{i}|X},P_{Z|X}\right)+R_{S|\widehat{S}_{i}}(D_{0}),i=1,2.
\end{equation}
\end{thm}
When specialized to the binary communication, we have the following
corollary.
\begin{cor}[Binary Communication]
\label{cor:proposed} For binary communication,
\[
\mathcal{R}\subseteq\mathcal{R}^{\mathsf{(o)}}\triangleq\left\{ \begin{array}{l}
\left(R_{\mathsf{K}},R_{\mathsf{L}},D_{0},D_{1},D_{2}\right):\\
D_{i}\geq p_{i},i=1,2,\\
R_{\mathsf{L}}\leq\min\left\{ R_{1},R_{2},\left[1-H_{2}\left(D_{0}\right)\right]^{+}\right\}
\end{array}\right\} .
\]
where
\begin{equation}
R_{i}=R_{\mathsf{K}}+\left[H_{2}\left(p_{0}\right)-H_{2}\left(p_{i}\right)\right]^{+}+\left[H_{2}\left(D_{i}\right)-H_{2}\left(D_{0}\right)\right]^{+},i=1,2.
\end{equation}
\end{cor}
Comparing Theorem \ref{cor:proposedbinary} and Corollary \ref{cor:proposed},
we can identify the optimality of the proposed scheme for binary communication.
\begin{thm}[Optimality of the Proposed Scheme]
\label{thm:optimalitybinary} For the binary communication (with
2 legitimate users), the proposed uncoded scheme is optimal if $p_{0}\leq p_{i}\leq D_{i}\leq D_{0}$
or $p_{0}\geq p_{i}=D_{i}\geq D_{0}$ holds for $i=1$ or $2$.
\end{thm}
\begin{rem}
\label{rem:Theorem--implies}Theorem \ref{thm:optimalitybinary} implies
under conditions that compared with one of legitimate users, the wiretapper
has a better channel and wants to produce a worse reconstruction,
or the legitimate user's distortion is restricted to be the Shannon
limit and meanwhile the wiretapper has a worse channel and wants to
produce a better reconstruction, the proposed uncoded scheme is optimal.
It is worth noting that these optimality conditions do not include
the practical scenario in which the wiretapper has a worse channel
than the legitimate users and a higher distortion requirement. But it does not mean our scheme is not optimal for the practical scenario. We believe that for
the binary broadcast communication without secrecy requirement, the
proposed uncoded scheme with $p^{\prime}=0$ and with no permutation
operation is the \emph{unique} scheme to achieve the Shannon limits for both
the legitimate users. If so, when the secrecy requirement is involved,
the proposed scheme is optimal as well, no matter what the wiretapper's channel condition is and what his desired distortion level is. This is because $R_{\mathsf{K}}$ rate
of secret key could   increase $R_{\mathsf{L}}$ at most by  $R_{\mathsf{K}}$, and our scheme satisfies this point. Of course, we need a rigorous proof about this claim, but unfortunately, now we have no idea how to  prove it.
%cannot prove it now since it is not easy.

\end{rem}
We know that when there is no secrecy constraint, the traditional
uncoded scheme could outperform the separate scheme for broadcast
communication scenarios. It is not surprising that when secrecy constraint
is involved, the proposed uncoded scheme still could outperform the
separate scheme. However, surprisingly, the example given in \cite{Yu-1}
shows the proposed uncoded scheme may strictly outperform the separate
coding even for the secure \emph{point-to-point} communication (with
only one legitimate user).

\section{Scalar Gaussian Communication}

\label{sec:joint}In this section, we consider a Gaussian source $S\sim\mathcal{N}\left(0,\lambda\right)$
transmitted over a bandwidth-matched\footnote{Although we can also convert a bandwidth-mismatched Gaussian system
into a bandwidth-matched system, just as done in Remark \ref{rem:The-conditions--},
our results in this section cannot be easily extended to the bandwidth-mismatched
system since the linear coding used in our schemes is specified for
the bandwidth-matched one.} and power-constrained Gaussian wiretap broadcast channel (the average
input power is constrained by $P$). The distortion measures are set
to $d_{\mathsf{B}}\left(s,\widehat{s}\right)=d_{\mathsf{E}}\left(s,\widehat{s}\right)=d\left(s,\widehat{s}\right)\triangleq\left(s-\widehat{s}\right)^{2}$.

For this communication system, we provide two uncoded schemes. The
first one is just the scheme proposed in previous section. Next we
will show that the permutation based scheme also works in the Gaussian
communication case. The other one is an orthogonal-transform based
scheme, which cascades a random orthogonal transform (instead of random
permutation operation) with a symbol-by-symbol mapping.

\subsection{Permutation based Scheme}

\begin{figure}
\centering \includegraphics[width=0.55\textwidth]{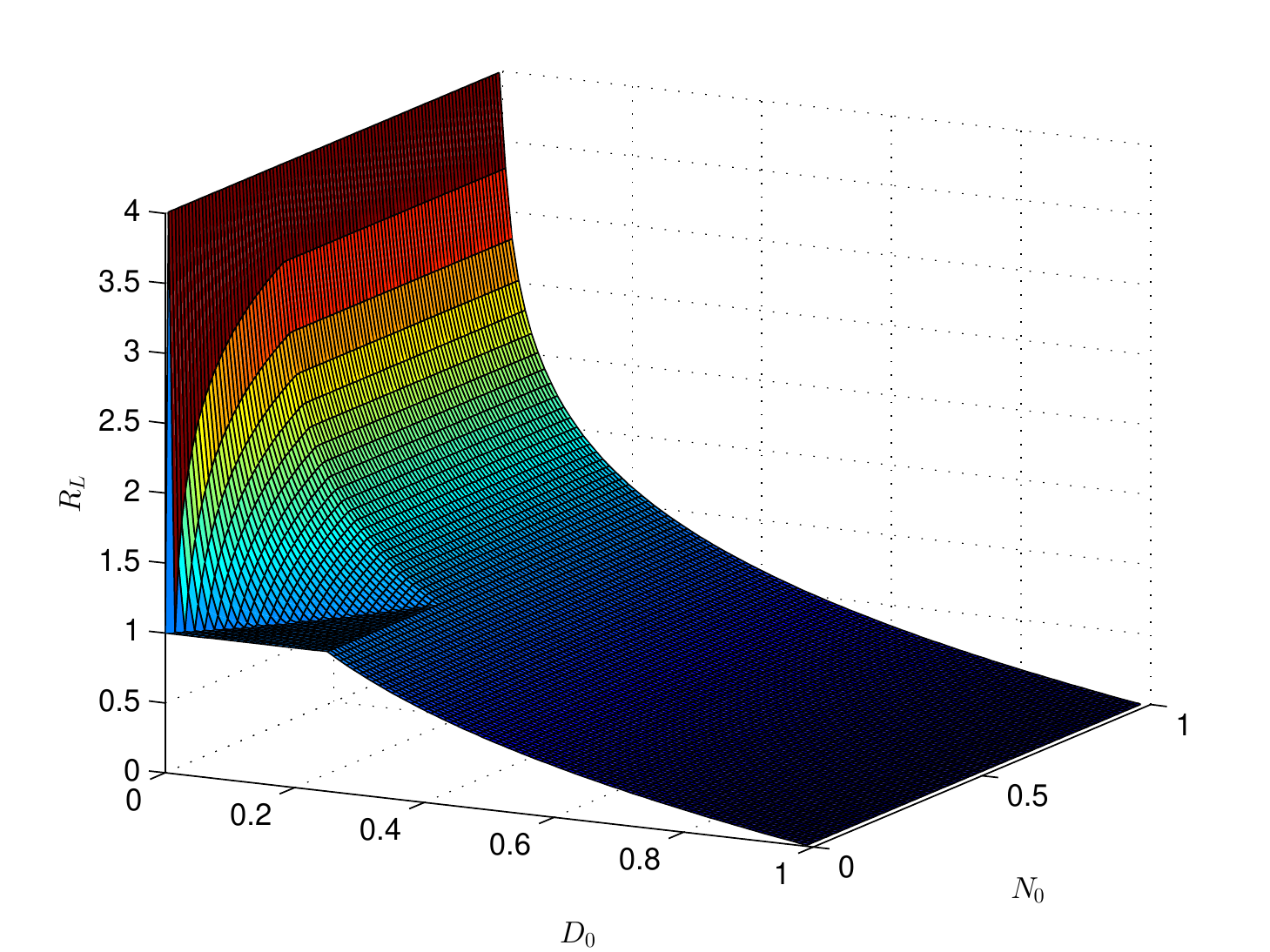}

\protect\caption{\label{fig:Gaussian}The region in Theorem \ref{thm:proposedGaussian}
with $\lambda=1$ and $P^{\prime}=1$\textit{.}}
\end{figure}

It has been shown that linear coding is optimal for the bandwidth-matched
Gaussian broadcast communication when there is no secrecy requirement
\cite{Goblick}. Hence we set $P_{X|S}$ and $\widehat{s}_{i}(y_{i}),i=1,2$
to the linear functions $x=\alpha s,\,\widehat{s}_{i}=\beta_{i}y_{i},\,i=1,2$
in the proposed scheme for DM communications, where $\alpha=\sqrt{\frac{P^{\prime}}{\lambda}}$
with $0\leq P^{\prime}\leq P$ and $\beta_{i}=\frac{\sqrt{\lambda P^{\prime}}}{P^{\prime}+N_{i}}$.
Then we apply this permutation based scheme to the Gaussian communication.
The performance of this scheme is provided in the following theorem,
the proof of which is given in Appendix \ref{sec:Proof-of-TheoremGauss}.
\begin{thm}[Permutation based Scheme]
\label{thm:proposedGaussian} For the Gaussian communication, the
proposed permutation based scheme achieves the region $\mathcal{R}^{\mathsf{(i)}}\subseteq\mathcal{R}$,
where
\begin{equation}
\mathcal{R}^{\mathsf{(i)}}\triangleq\underset{0\leq P^{\prime}\leq P}{\bigcup}\left\{ \begin{array}{l}
\left(R_{\mathsf{K}},R_{\mathsf{L}},P,D_{0},D_{1},D_{2}\right):\\
D_{i}\geq\frac{\lambda N_{i}}{P^{\prime}+N_{i}},i=1,2,\\
R_{\mathsf{L}}\leq\min\left\{ R_{\mathsf{K}}+\frac{1}{2}\log^{+}\left(\frac{\lambda N_{0}}{D_{0}\left(P^{\prime}+N_{0}\right)}\right),\frac{1}{2}\log^{+}\left(\frac{\lambda}{D_{0}}\right)\right\}
\end{array}\right\} ,\label{eq:-35}
\end{equation}
with $\log^{+}x\triangleq\max\left\{ 0,\log x\right\} $.
\end{thm}
\begin{rem}
The $\mathcal{R}^{\mathsf{(i)}}$ here is just the one given in Theorem \ref{thm:proposedDM}
with $P_{X|S}$ and $\widehat{s}_{i}(y_{i}),i=1,2$ set to $x=\alpha s$
and $\widehat{s}_{i}=\beta_{i}y_{i},\,i=1,2$, respectively. This is because
they are achieved by the same scheme.
\end{rem}
\begin{rem}
The first constraint of $\mathcal{R}^{\mathsf{(i)}}$ is consistent with the
performance of linear coding \cite{Goblick}. The second constraint
of $\mathcal{R}^{\mathsf{(i)}}$ follows from the similar argument to the DM
case.
\end{rem}
Note that for $\mathcal{R}^{\mathsf{(i)}}$, $P^{\prime}$ is a variable. Moreover,
the region of $\left(D_{1},D_{2}\right)$ and the region of $\left(R_{\mathsf{K}},R_{\mathsf{L}},D_{0}\right)$
depend on each other through $P^{\prime}$ which satisfies $0\leq P^{\prime}\leq P$.
This finding is similar to the discrete communication case. Given
$\left(R_{\mathsf{K}},D_{0}\right)$, the minimum of achievable $D_{1}$ (or
$D_{2}$) and the maximum of achievable $R_{\mathsf{L}}$ are both decreasing
in $P^{\prime}$, which implies for the proposed scheme, transmitting
the source using a larger power results in smaller distortions for
legitimate users, but also leads to decrypting the source more easily
for the wiretapper. The proposed scheme with $P^{\prime}=P$, on one
hand, provides a certain level of secrecy, and on the other hand,
it achieves the Shannon's distortion limits for both legitimate users.
The region in Theorem \ref{thm:proposedGaussian} with $\lambda=1$
and $P^{\prime}=1$ is illustrated in Fig. \ref{fig:Gaussian}. Given
$P^{\prime}$, $\left(D_{1},D_{2}\right)$ has no effect on the $\left(R_{\mathsf{K}},R_{\mathsf{L}},D_{0}\right)$
tradeoff.

\subsection{Orthogonal-Transform based Scheme}

\label{subsec:encryption} The proposed scheme above uses a random
permutation operation (which shuffles the sequence within the same
type class) to improve the level of secrecy. It works not only for
the discrete communication but also for the continuous communication,
such as the Gaussian communication. In this subsection we propose
another secure uncoded scheme for the Gaussian communication case
which is designed from a geometric point of view.

To give an interpretation for the motivation of our proposed scheme,
we consider a special case where the wiretapper has a noiseless channel.
Apply linear coding to the Gaussian communication, then  we know
that given the Euclidean norm, the sequence of a Gaussian source uniformly
distributed on some sphere, and so are the sequences of channel input,
outputs, and source reconstructions. Assume we generate a set of
bijective transforms (as a codebook), and randomly choose one of them
(according to the key) to transform the  source sequence before
applying linear coding on it. To keep the power unchanged, these transforms
are required to map a sphere into itself. On the other hand, by using
the secret key the legitimate users could transform it back.  Hence
the induced distortions of legitimate users do not change as well.
Furthermore, without knowing the secret key but with knowing the norm
of the source sequence and the codebook, in the view of the wiretapper,
the source sequence is uniformly distributed over the vectors that
are possible to generate the channel output (wiretapper's observation)
through some key values. To make the wiretapper guess the source
as difficultly as possible, these vectors should be ``uniformly''
(at equal distance) located on the sphere.  This is because if so,
the wiretapper has to cover either all these vectors or the whole
sphere to meet the decryption requirement.  It can be shown the orthogonal
transform is one of such transforms. Hence  it is adopted in our
second scheme.

\emph{}

\emph{Codebook (Public Key) Generation:} Generate random $n\times n$
matrices $Q_{k},k\in\left[\mbox{2}^{nR_{\mathsf{K}}}\right]$ independently
whose elements are generated i.i.d. according to $\mathcal{N}\left(0,1\right)$.
Then apply Gram-Schmidt orthonormalization process to the columns
of each matrix, hence all the resulting matrices are orthogonal and
they constitute a subset of orthogonal matrices $\mathcal{C}=\left\{ \Psi_{k},k\in\left[\mbox{2}^{nR_{\mathsf{K}}}\right]\right\} $.
As a public key, the codebook $\mathcal{C}$ is revealed to the sender
and all the receivers (including the wiretapper).

\emph{Encoding:} Upon observing a source sequence $s^{n}$ and a key
$k$, the encoder generates $x^{n}$ as follows.
\begin{equation}
x^{n}=\alpha\Psi_{k}s^{n},
\end{equation}
where $\alpha=\sqrt{\frac{P^{\prime}}{\lambda}}$ with $0\leq P^{\prime}\leq P$.

\emph{Decoding (for Legitimate Users):} For legitimate user $\textrm{B}i$,
$i=1,2$, upon the received sequence $y_{i}^{n}$ and the key $k$,
the decoder reconstructs the source as follows.
\begin{equation}
\widehat{s}_{i}^{n}=\beta_{i}\Psi_{k}^{T}y_{i}^{n},
\end{equation}
where $\beta_{i}=\frac{\sqrt{\lambda P^{\prime}}}{P^{\prime}+N_{i}}$,
and $\Psi_{k}^{T}$ denotes the transpose of the matrix $\Psi_{k}$.

Next we will analyze the asymptotic performance of this scheme. Similar
to the case of permutation based scheme, we need first introduce some
basic properties of the random codebook $\mathcal{C}$.
\begin{lem}
\cite{Eaton83}\label{lem:randmat} Suppose $Q$ is a random $n\times n$
matrix with each element independently distributed according to Gaussian
distribution $\mathcal{N}(0,1)$. Let $Q_{1},Q_{2},\cdots,Q_{n}$
be the columns of $Q$ and let $\Psi$ be the random matrix whose
columns are obtained by applying the Gram-Schmidt orthonormalization
procedure to $Q_{1},Q_{2},\cdots,Q_{n}$. Then both $\Psi$ and $\Psi^{T}$
have the uniform distribution (Haar measure under orthogonal transform)
on the set of $n\times n$ orthogonal matrices $\mathcal{F}\left(n\right)$,
and moreover for any orthogonal matrix $A$, both $A\Psi$ and $\Psi A$
also have the uniform distribution on $\mathcal{F}\left(n\right)$.
\end{lem}
Utilizing Lemma \ref{lem:randmat}, we can establish the following
lemma.
\begin{lem}
\label{lem:Random-matrix} Random orthogonal transform $x^{n}=\Psi s^{n}$
with $\Psi$ uniformly distributed on orthogonal matrices set $\mathcal{F}\left(n\right)$,
transforms an arbitrary vector $s^{n}\in\mathbb{R}^{n}$ into a random
vector that is uniformly distributed on the $\left(n-1\right)$-sphere
with radius $\left\Vert s^{n}\right\Vert $.
\end{lem}
\begin{IEEEproof}
From Lemma \ref{lem:randmat}, without loss of generality we can assume
$\Psi$ is obtained in the manner described in Lemma \ref{lem:randmat}.
Let $\Psi_{1},\Psi_{2},\cdots,\Psi_{n}$ be the columns of $\Psi$.
From Gram-Schmidt orthonormalization, we know that $\Psi_{1}=\frac{Q_{1}}{\left\Vert Q_{1}\right\Vert }$,
and for any rotation matrix (or more generally, orthogonal matrix)
$A$, $A\Psi_{1}=\frac{AQ_{1}}{\left\Vert Q_{1}\right\Vert }=\frac{AQ_{1}}{\left\Vert AQ_{1}\right\Vert }$.
On the other hand, $Q_{1}$ is a random vector with each element i.i.d.
$\sim\mathcal{N}(0,1)$, and it is easy to verify that for any rotation
matrix $A$, $AQ_{1}$ has the same distribution as $Q_{1}$, i.e.,
a normally distributed random vector is invariant to rotation. Therefore,
$A\Psi_{1}$ has the same distribution as $\Psi_{1}$, i.e., $\Psi_{1}$
is also invariant to rotation. This implies $\Psi_{1}$ is uniformly
distributed on the unit $\left(n-1\right)$-sphere. In addition, observe
$\Psi\left(1,0,\cdots,0\right)^{T}=\Psi_{1}$. Hence the random matrix
$\Psi$ transforms vector $\left(1,0,\cdots,0\right)^{T}$ to a random
vector uniformly distributed on the $\left(n-1\right)$-sphere. For
arbitrary vector $s^{n}\in\mathbb{R}^{n}$, we can easily find an
orthogonal matrix $B$ with the first column $\frac{s^{n}}{\left\Vert s^{n}\right\Vert }$.
Hence $s^{n}$ can be expressed as $s^{n}=\left\Vert s^{n}\right\Vert B\left(1,0,\cdots,0\right)^{T}$.
Then we have $\Psi s^{n}=\left\Vert s^{n}\right\Vert \Psi B\left(1,0,\cdots,0\right)^{T}$.
From Lemma \ref{lem:randmat}, $\Psi B$ has the same distribution
as $\Psi$. Hence $\Psi B\left(1,0,\cdots,0\right)^{T}$ is also uniformly
distributed on the unit $\left(n-1\right)$-sphere, which implies
$\Psi s^{n}$ is uniformly distributed on the $\left(n-1\right)$-sphere
with radius $\left\Vert s^{n}\right\Vert $.
\end{IEEEproof}
Lemma \ref{lem:Random-matrix} implies the resulting vector will be
uniformly distributed on the sphere where the input vector is, if
the transform matrix is randomly and uniformly chosen from the set
of orthogonal matrices. This is a nice property of the random orthogonal
transform, similar to the property of the random permutation operation.
Utilizing the properties, we can establish the following theorem,
the proof of which is given in Appendix \ref{sec:Proof-of-TheoremGauss2}.
\begin{thm}[Orthogonal-Transform based Scheme]
\label{thm:proposedGaussian2} For the Gaussian communication, the
inner bound $\mathcal{R}^{\mathsf{(i)}}$ given in Theorem \ref{thm:proposedGaussian}
can be achieved by the scheme above as well.
\end{thm}
The inner bound $\mathcal{R}^{\mathsf{(i)}}$ can be understood from a\emph{
}geometric point of view. The random orthogonal transform in the proposed
scheme guarantees that given $Z^{n}$, $S^{n}$ has a uniform distribution
on $2^{nR_{\mathsf{K}}}$ small $\left(n-2\right)-$spheres with radius $r_{2}=\sqrt{\frac{n\lambda N_{0}}{P^{\prime}+N_{0}}}$
whose centers are uniformly distributed on the $\left(n-1\right)-$sphere
with center $O$ (the origin) and radius $r_{1}=\sqrt{\frac{n\lambda P^{\prime}}{P^{\prime}+N_{0}}}$.
However, owing to the uniform conditional distribution of the source
given $Z^{n}$ and the lack of secret key, the wiretapper needs at
least $2^{nR_{\mathsf{K}}}(\frac{r_{2}}{\sqrt{nD_{0}}})^{n}$ balls with radius
$\sqrt{nD_{0}}$ to cover these $\left(n-2\right)-$spheres. On the
other hand, under the unconditional case, the source has a uniform
distribution on the $\left(n-1\right)-$sphere with center $O$ and
radius $r_{0}=\sqrt{n\lambda}$. Hence if ignoring $Z^{n}$, the wiretapper
needs at least $(\frac{r_{0}}{\sqrt{nD_{0}}})^{n}$ balls with radius
$\sqrt{nD_{0}}$ to cover the sphere. This results in the inner bound
$\mathcal{R}^{\mathsf{(i)}}$.

\begin{figure}
\centering\includegraphics[width=0.4\textwidth]{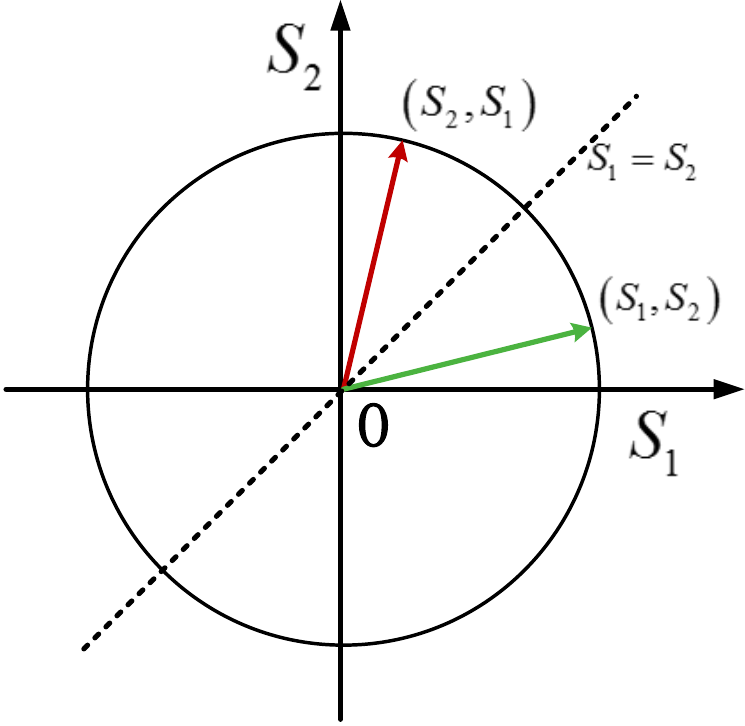} \protect\caption{\label{fig:permu}{\small{}Illustration of permutations of a source
sequence for $n=2$ case. }}
\end{figure}

It seems somewhat counterintuitive that the permutation based scheme
achieves the same performance as the orthogonal-transform based scheme,
as shown by Theorems \ref{thm:proposedGaussian} and \ref{thm:proposedGaussian2};
it is easy to observe that for low-dimension cases, e.g., 2-dimension
case (see Fig. \ref{fig:permu}), permutations cannot always transform
a source sequence into vectors ``uniformly'' (at equal distance)
distributed over a sphere, so why does this property hold (with high
probability) when the dimension goes to infinity? Actually, it indeed
does. This is because as the dimension increases, such ``bad''\footnote{Here a source sequence is said to be ``good'' if its permutations
are ``uniformly'' distributed over a sphere; otherwise it is ``bad''.
Obviously, the permutations of a ``good'' source sequence are also
``good''.} source sequences will occur with vanishing probability. This can
be seen from that\footnote{Here $[S]=\Delta\cdot\textrm{Round}\left(\frac{S}{\Delta}\right)$
and $\mathcal{U}_{\delta}^{n}\left([S]\right)$ is the $\delta$-unified
typical set for $P_{[S]}$; see the proof in \ref{sec:Proof-of-TheoremGauss}. } $\mathbb{P}\left([S]^{n}\in\mathcal{U}_{\delta}^{n}\left([S]\right)\right)\rightarrow1$
as $n\to\infty$ (i.e., besides on the sphere, the source sequence
should also with high probability appear the neighborhoods of the
vectors in $\mathcal{U}_{\delta}^{n}\left([S]\right)$), and moreover,
$\mathcal{U}_{\delta}^{n}\left([S]\right)$ consists of a set of ``good''
source sequences.  Hence the ``good'' source sequences will occur
with high probability as the dimension increases, that is, permutations
will transform an arbitrary source sequence from a high probability
set into vectors ``uniformly'' distributed over a sphere.

\subsection{Comparison with Sign-Change Based Scheme }

\label{subsec:comparison} In previous two subsections, we give an
analysis of the asymptotic performance of permutation based scheme
or orthogonal-transform based scheme. However, is it necessary to
let the blocklength $n$ go to infinity? What if $n$ is set to be
a finite value? In this subsection, we study the simplest finite
blocklength case: $n=1$ (single-letter codes). For this case,
the permutation based scheme is obviously inferior to the asymptotic
case, since for 1 dimension case no permutation exists except for
the source sequence itself. Hence in the following, we mainly consider
the orthogonal-transform based scheme.

For $n=1$, the orthogonal-transform based scheme reduces to a sign-change
based scheme \cite{Kak,Yu-1}. Next we compare the proposed schemes
with this  sign-change based scheme \cite{Kak,Yu-1}. Assume $R_{\mathsf{K}}=1$, and the secret key
is uniformly distributed on $\left\{ 0,1\right\} $.

\emph{Encoding:} Upon observing a source sequence $s$ and a key $k$
, the encoder generates $x$ as follows.
\begin{equation}
x=\alpha\Psi_{k}s,
\end{equation}
where $\alpha=\sqrt{\frac{P^{\prime}}{\lambda}}$ with $0\leq P^{\prime}\leq P$,
and
\begin{equation}
\Psi_{k}=\begin{cases}
-1, & \textrm{if }k=0;\\
1, & \textrm{if }k=1.
\end{cases}\label{eq:tk}
\end{equation}

\emph{Decoding (for Legitimate Receivers):} For legitimate receiver
$\textrm{B}i$, $i=1,2$, upon the received sequence $y_{i}$ and
the key $k$, the decoder reconstructs the source as follows.
\begin{equation}
\widehat{s}_{i}=\beta_{i}\Psi_{k}y_{i},
\end{equation}
where $\beta_{i}=\frac{\sqrt{\lambda P^{\prime}}}{P^{\prime}+N_{i}}$.

It is easy to verify that $(S_{t},K_{t},X_{t},Y_{1,t},Y_{2,t},Z_{t},\widehat{S}_{1,t},\widehat{S}_{2,t})_{t=1}^{\infty}$
are i.i.d. and
\begin{align}
f_{S,Z}\left(s,z\right) & =f_{S}\left(s\right)\cdot\frac{1}{2}\left[f_{V_{0}}\left(z-\alpha s\right)+f_{V_{0}}\left(z+\alpha s\right)\right]\label{eq:-19}\\
 & =f_{Z}\left(z\right)\cdot\frac{1}{2}\left[f_{V_{0}'}\left(s-\beta_{0}z\right)+f_{V_{0}'}\left(s+\beta_{0}z\right)\right],\label{eq:-20}
\end{align}
where $\beta_{0}=\frac{\sqrt{\lambda P^{\prime}}}{P^{\prime}+N_{0}}$,
$f_{V_{0}}$ denotes the probability distribution function (pdf) of
the wiretapper's channel noise $V_{0}$, and $f_{V_{0}'}$ denotes
the pdf of $V_{0}'\sim\mathcal{N}\left(0,\frac{\lambda N_{0}}{P^{\prime}+N_{0}}\right)$.
Given $Z$, $S$ can be regarded as a Gaussian mixture with two components
of equal weight and variance. For such single-letter scheme, in
\cite{Yu-1} we have shown the maximum achievable $R_{\mathsf{L}}$ (or equivalently
the minimum rate needed to code $S$ within distortion $D_{0}$ with
two-sided information $Z$) equals the conditional rate-distortion
function $R_{S|Z}(D_{0})$. The performance of the sign-change based
scheme is given by the following theorem.
\begin{thm}[Sign-Change based Scheme]
\cite{Yu-1}\label{thm:signchange} For the Gaussian communication
with $R_{\mathsf{K}}=1$, the sign-change based scheme above achieves the region
$\mathcal{R}_{\mathsf{sign}}^{\mathsf{(i)}}\subseteq\mathcal{R}$, where
\[
\mathcal{R}_{\mathsf{sign}}^{\mathsf{(i)}}\triangleq\underset{0\leq P^{\prime}\leq P}{\bigcup}\left\{ \begin{array}{l}
\left(R_{\mathsf{K}},R_{\mathsf{L}},P,D_{0},D_{1},D_{2}\right):\\
D_{i}\geq\frac{\lambda N_{i}}{P^{\prime}+N_{i}},i=1,2,\\
R_{\mathsf{L}}\leq R_{S|Z}(D_{0})
\end{array}\right\} ,
\]
with $R_{S|Z}\left(D_{0}\right)$ denoting the conditional rate-distortion
function of $S$ given two-sided information $Z$, defined in \eqref{eq:rd-si-1}.
\end{thm}
Since it is hard (even if possible) to express $R_{S|Z}(D_{0})$ in
closed form, for ease of comparison, we will derive a closed-form
upper bound for $R_{S|Z}(D_{0})$. The result is shown in the following
lemma, and the proof is given in Appendix \ref{sec:Proof-of-Lemma-ub}.
\begin{lem}
\label{lem:upperbound} If $\left(S,Z\right)$ follows the distribution
\eqref{eq:-19} or \eqref{eq:-20}, then
\begin{equation}
R_{S|Z}(D_{0})\leq\min\left\{ R_{S|Z}^{\mathsf{(UB)}}(D_{0}),\frac{1}{2}\log^{+}\left(\frac{\lambda}{D_{0}}\right)\right\} ,\label{eq:-37}
\end{equation}
where
\begin{equation}
R_{S|Z}^{\mathsf{(UB)}}(D_{0})\triangleq\begin{cases}
\frac{\left(\lambda-D_{0}\right)\left(P^{\prime}+N_{0}\right)}{\lambda P^{\prime}}, & \textrm{if }\frac{\lambda N_{0}}{P^{\prime}+N_{0}}<D_{0}\leq\lambda;\\
1+\frac{1}{2}\log\left(\frac{\lambda N_{0}}{D_{0}\left(P^{\prime}+N_{0}\right)}\right), & \textrm{if }0\leq D_{0}\leq\frac{\lambda N_{0}}{P^{\prime}+N_{0}}.
\end{cases}
\end{equation}
\end{lem}
Since $R_{S|Z}(D_{0})$ denotes the minimum rate needed to code $S$
within distortion $D_{0}$ when $Z$ is available at both encoder
and decoder, we can give an interpretation for the upper bound from
the perspective of source coding. First, by ignoring the side information,
we have $R_{S|Z}(D_{0})\leq\frac{1}{2}\log^{+}\left(\frac{\lambda}{D_{0}}\right)$,
where $\frac{1}{2}\log^{+}\left(\frac{\lambda}{D_{0}}\right)$ is
the minimum rate needed to code $S$ without any side information.
Second, if $\frac{\lambda N_{0}}{P^{\prime}+N_{0}}\leq D_{0}\leq\lambda$,
then consider the following timesharing coding strategy.\footnote{Note that the argument here is only available for the inequality \eqref{eq:-37},
and does not apply to the secrecy problem considered in this paper.
For the secrecy problem the wiretapper and henchman cannot benefit
from adopting a timesharing strategy since the constraint \eqref{listsecrecylossy}
or \eqref{mainobj} is to restrict the excess-distortion probability,
instead of the average distortion.} If we  code the secret key $K$ (1 bit per symbol), then using a
linear decoder (similar to those of legitimate users), we can reconstruct
the source within distortion $\frac{\lambda N_{0}}{P^{\prime}+N_{0}}$.
On the other hand, if we do not code anything, then it results in
rate 0 and distortion $\lambda$. By using a timesharing strategy
between these two schemes, we need $\frac{\left(\lambda-D_{0}\right)\left(P^{\prime}+N_{0}\right)}{\lambda P^{\prime}}$
rate to reconstruct the source within distortion $D_{0}$. Finally,
if $0\leq D_{0}\leq\frac{\lambda N_{0}}{P^{\prime}+N_{0}}$, then
we reconstruct the source within distortion $\frac{\lambda N_{0}}{P^{\prime}+N_{0}}$
by using rate 1 to code the secret key, and upon the reconstruction,
we further code the residual error within distortion $D_{0}$ by using
rate $\frac{1}{2}\log\left(\frac{\lambda N_{0}}{D_{0}\left(P^{\prime}+N_{0}\right)}\right)$.

\begin{figure}
\centering \includegraphics[width=0.55\textwidth]{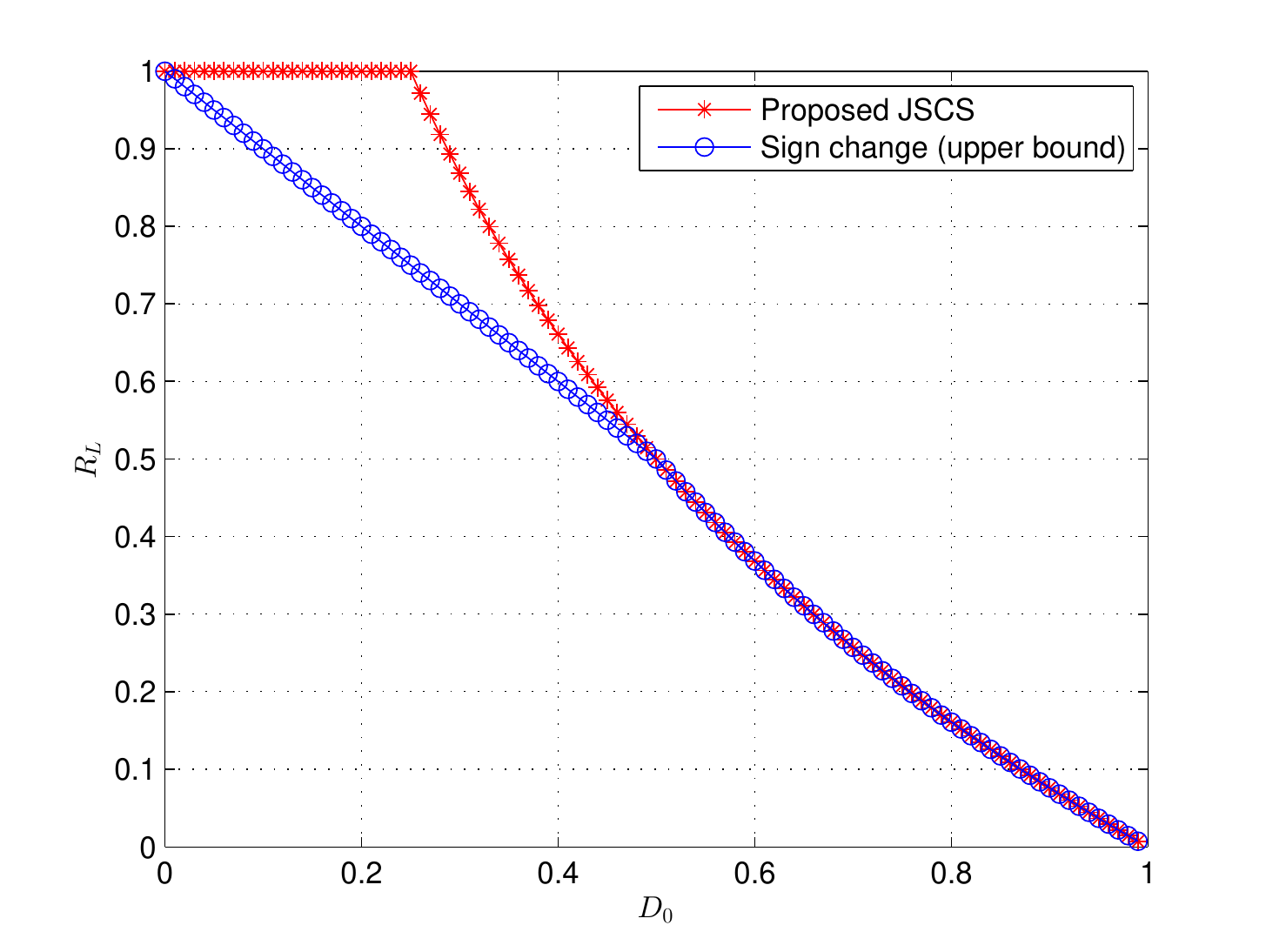} \caption{\label{fig:Gaussianproposedvsothers}Comparison of the achievable
$R_{\mathsf{L}}$ by the proposed (infinite blocklength) schemes and that by
the sign-change based scheme. $\lambda=1,$ $N_{0}=0$ (noiseless
wiretap channel) and\textit{ }$R_{\mathsf{K}}=1$\textit{.}}
\end{figure}

Combining Theorem \ref{thm:signchange} and Lemma \ref{lem:upperbound}
gives us the following result.
\begin{thm}[Outer Bound of $\mathcal{R}_{\mathsf{sign}}^{\mathsf{(i)}}$]
\label{thm:ThmSignUB} For the Gaussian communication with $R_{\mathsf{K}}=1$,
the region achieved by the sign-change based scheme satisfies $\mathcal{R}_{\mathsf{sign}}^{\mathsf{(i)}}\subseteq\mathcal{R}_{\mathsf{sign}}^{\mathsf{(o)}}$,
where
\begin{equation}
\mathcal{R}_{\mathsf{sign}}^{\mathsf{(o)}}\triangleq\underset{0\leq P^{\prime}\leq P}{\bigcup}\left\{ \begin{array}{l}
\left(R_{\mathsf{K}},R_{\mathsf{L}},P,D_{0},D_{1},D_{2}\right):\\
D_{i}\geq\frac{\lambda N_{i}}{P^{\prime}+N_{i}},i=1,2,\\
R_{\mathsf{L}}\leq\min\left\{ R_{S|Z}^{\mathsf{(UB)}}(D_{0}),\frac{1}{2}\log^{+}\left(\frac{\lambda}{D_{0}}\right)\right\}
\end{array}\right\} .\label{eq:-25}
\end{equation}
\end{thm}
\begin{rem}
Observe that only 1 bit/symbol of key can be exploited by the sign-change
based scheme even when $R_{\mathsf{K}}>1$. Hence for that case, its performance
is still that given by Theorem \ref{thm:signchange} and outer bounded
by \ref{eq:-25}.
\end{rem}
From Lemma \ref{lem:upperbound}, it can be observed that when $R_{\mathsf{K}}=1$,
$R_{S|Z}^{\mathsf{(UB)}}(D_{0})=1+\frac{1}{2}\log^{+}\left(\frac{\lambda N_{0}}{D_{0}\left(P^{\prime}+N_{0}\right)}\right)$
for $0\leq D_{0}\leq\frac{\lambda N_{0}}{P^{\prime}+N_{0}}$, and
$R_{S|Z}^{\mathsf{(UB)}}(D_{0})<1=1+\frac{1}{2}\log^{+}\left(\frac{\lambda N_{0}}{D_{0}\left(P^{\prime}+N_{0}\right)}\right)$
for $\frac{\lambda N_{0}}{P^{\prime}+N_{0}}<D_{0}\leq\lambda$. Hence
$\mathcal{R}_{\mathsf{sign}}^{\mathsf{(o)}}\subsetneqq\mathcal{R}^{\mathsf{(i)}}$,
where $\mathcal{R}^{\mathsf{(i)}}$ given in Theorem \ref{thm:proposedGaussian}
denotes the achievable region by the  permutation based scheme or
orthogonal-transform based scheme. This implies for the same $P^{\prime}$,
the sign-change based scheme is strictly inferior to the proposed
schemes under the condition $\frac{\lambda N_{0}}{P^{\prime}+N_{0}}<D_{0}\leq\lambda$.
That is, the single-letter version of orthogonal-transform based
scheme is inferior to the corresponding infinite blocklength version.
To see it clearer, the $R_{\mathsf{L}}$ achieved by the proposed (infinite
blocklength) schemes (given in Theorem \ref{thm:proposedGaussian})
and the upper bound of $R_{\mathsf{L}}$ achieved by the sign-change based
scheme (given in Theorem \ref{thm:ThmSignUB}) are illustrated in
Fig. \ref{fig:Gaussianproposedvsothers}.

\subsection{Outer Bound}

For the Gaussian communication, the following outer bound has been
proven for the system with only one legitimate user \cite{Yu-1}.
\begin{lem}
\cite{Yu-1}\label{thm:singleuser} For the Gaussian communication
with only one legitimate user,
\begin{equation}
\mathcal{R}\subseteq\mathcal{R}^{\mathsf{(o)}}\triangleq\left\{ \begin{array}{l}
\left(R_{\mathsf{K}},R_{\mathsf{L}},P,D_{0},D_{1}\right):\\
D_{1}\geq\frac{\lambda N_{1}}{P+N_{1}},\\
R_{\mathsf{L}}\le\min\left\{ R_{1},\frac{1}{2}\log^{+}\left(\frac{\lambda}{D_{0}}\right)\right\}
\end{array}\right\} ,
\end{equation}
where
\begin{equation}
R_{1}=R_{\mathsf{K}}+\frac{1}{2}\log^{+}\left(\frac{1+\nicefrac{P}{N_{1}}}{1+\nicefrac{P}{N_{0}}}\right)+\frac{1}{2}\log^{+}\left(\frac{D_{1}}{D_{0}}\right).
\end{equation}
\end{lem}
Using this result, we have the following outer bound for the system
with 2 legitimate users (the system considered in this paper).
\begin{thm}[Outer Bound]
\label{thm:broadcast} For the Gaussian communication (with 2 legitimate
users),
\begin{equation}
\mathcal{R}\subseteq\mathcal{R}^{\mathsf{(o)}}\triangleq\left\{ \begin{array}{l}
\left(R_{\mathsf{K}},R_{\mathsf{L}},P,D_{0},D_{1},D_{2}\right):\\
D_{i}\geq\frac{\lambda N_{i}}{P+N_{i}},i=1,2,\\
R_{\mathsf{L}}\le\min\left\{ R_{1},R_{2},\frac{1}{2}\log^{+}\left(\frac{\lambda}{D_{0}}\right)\right\}
\end{array}\right\} ,
\end{equation}
where
\begin{equation}
R_{i}=R_{\mathsf{K}}+\frac{1}{2}\log^{+}\left(\frac{1+\nicefrac{P}{N_{i}}}{1+\nicefrac{P}{N_{0}}}\right)+\frac{1}{2}\log^{+}\left(\frac{D_{i}}{D_{0}}\right),i=1,2.
\end{equation}
\end{thm}
Comparing Theorem \ref{thm:proposedGaussian2} and Corollary \ref{thm:broadcast},
we can identify the optimality of the proposed schemes for the Gaussian
communication. This result is similar to Theorem \ref{thm:optimalitybinary}
for the binary communication.
\begin{thm}[Optimality of the Proposed Schemes]
\label{thm:optimalityGaussian} For the Gaussian communication (with
2 legitimate users), the proposed scheme is optimal if $N_{0}\leq N_{i},D_{0}\geq D_{i}$
or $N_{0}\geq N_{i},D_{0}\leq D_{i}=\frac{\lambda N_{i}}{P+N_{i}}$
holds for $i=1$ or $2$.
\end{thm}
A similar remark to Remark \ref{rem:Theorem--implies} applies to this theorem.

\section{Vector Gaussian Communication}

The proposed schemes are easily extended to vector Gaussian communication
scenarios. Consider an $m$-vector Gaussian source $\boldsymbol{S}\sim\mathcal{N}\left(\mathbf{0},\textrm{diag}\left(\lambda_{1},\lambda_{2},\cdots,\lambda_{m}\right)\right)$\footnote{In this paper, we use bold font to denote vector or matrix, e.g.,
$\left(S_{1},\cdots,S_{m}\right)$ and $\left(s_{1},\cdots,s_{m}\right)$
are denoted by $\boldsymbol{S}$ and $\boldsymbol{s}$, respectively.} transmitted over an $m$-vector Gaussian broadcast channel
\begin{equation}
\boldsymbol{Y}_{i}=\boldsymbol{X}+\boldsymbol{V}_{i},i=1,2,
\end{equation}
where $\boldsymbol{Y}_{i}$ is the channel output vector observed
by the $i$-th legitimate user, and $\boldsymbol{V}_{i}\sim\mathcal{N}(\mathbf{0},\textrm{diag}(N_{i,1},N_{i,2},\cdots,N_{i,m}))$
is an additive Gaussian noise vector. A wiretapper Eve accesses to
another channel output $\boldsymbol{Z}$ through a channel
\begin{equation}
\boldsymbol{Z}=\boldsymbol{X}+\boldsymbol{V}_{0},
\end{equation}
where $\boldsymbol{V}_{0}\sim\mathcal{N}\left(\mathbf{0},\textrm{diag}\left(N_{0,1},N_{0,2},\cdots,N_{0,m}\right)\right)$
is an additive Gaussian noise vector as well. The distortion measures
are set to $d_{\mathsf{B}}\left(\boldsymbol{s},\boldsymbol{\widehat{s}}\right)=d_{\mathsf{E}}\left(\boldsymbol{s},\boldsymbol{\widehat{s}}\right)=\sum_{j=1}^{m}(s_{j}-\widehat{s}_{j})^{2}$,
and the channel cost function is set to $\rho\left(\boldsymbol{x}\right)=\sum_{j=1}^{m}x_{j}^{2}$.

Consider the vectors $\boldsymbol{S},\boldsymbol{X},\boldsymbol{Y}_{i},\boldsymbol{Z},\boldsymbol{\widehat{S}}_{i},\boldsymbol{\widecheck{S}}_{i}$
as super-symbols, then the proposed permutation based scheme can be
applied to the vector Gaussian case directly. The performance of this
scheme can be proven by following similar steps to the proof for the
scalar Gaussian case.

Furthermore, we can apply the proposed orthogonal-transform based
scheme to each subsource-subchannel pair, as shown in the following.

\emph{Codebook (Public Key) Generation:} Generate $m\cdot\mbox{2}^{nR_{\mathsf{K}}}$
random $n\times n$ matrices $Q_{j,k},j\in\left[m\right],k\in\left[\mbox{2}^{nR_{\mathsf{K}}}\right]$
independently whose elements are generated i.i.d. according to $\mathcal{N}\left(0,1\right)$.
Then we apply Gram-Schmidt orthonormalization process on every matrix,
hence all the resulting matrices are orthogonal, and constitute a
subset of orthogonal matrices $\mathcal{C}=\left\{ \Psi_{j,k},j\in\left[m\right],k\in\left[\mbox{2}^{nR_{\mathsf{K}}}\right]\right\} $.
As a public key, the codebook $\mathcal{C}$ is revealed to the sender
and all the receivers (including the wiretapper).

\emph{Encoding:} Upon observing a source sequence $\boldsymbol{s}^{n}=\left(s_{1}^{n},s_{2}^{n},\cdots,s_{m}^{n}\right)$
and a key $k$ , the encoder generates $\boldsymbol{x}^{n}=\left(x_{1}^{n},x_{2}^{n},\cdots,x_{m}^{n}\right)$
as follows.
\begin{equation}
x_{j}^{n}=\alpha_{j}\Psi_{j,k}s_{j}^{n},j\in\left[m\right],
\end{equation}
where $\alpha_{j}=\sqrt{\frac{P_{j}}{\lambda_{j}}}$ with transmitting
power $P_{1},P_{2},\cdots,P_{m}$ such that $0\leq\sum_{j=1}^{m}P_{j}\leq P$.

\emph{Decoding (for Legitimate Users):} For the legitimate user $\textrm{B}i$,
$i=1,2$, upon the received sequence $\boldsymbol{y}_{i}^{n}$ and
the key $k$, the decoder reconstructs the source as follows.
\begin{equation}
\widehat{s}_{i,j}^{n}=\beta_{i,j}\Psi_{j,k}^{T}y_{i,j}^{n},j\in\left[m\right],
\end{equation}
where $\beta_{i,j}=\frac{\sqrt{\lambda_{j}P_{j}}}{P_{j}+N_{i}}$.

The achievable regions by the proposed schemes (permutation based
scheme and orthogonal-transform based scheme) are given in the following
theorem, the proof of which is given in Appendix \ref{sec:Proof-of-TheoremVGauss}.
\begin{thm}[Performance of the Proposed Schemes]
\label{thm:vector} For the vector Gaussian communication, the permutation
based scheme or the orthogonal-transform based scheme above achieves
the same region $\mathcal{R}^{\mathsf{(i)}}\subseteq\mathcal{R}$, where
\[
\mathcal{R}^{\mathsf{(i)}}\triangleq\underset{\begin{array}{c}
P_{1},P_{2},\cdots,P_{m}\geq0,\\
0\leq\sum_{j=1}^{m}P_{j}\leq P
\end{array}}{\bigcup}\left\{ \begin{array}{l}
\left(R_{\mathsf{K}},R_{\mathsf{L}},P,D_{0},D_{1},D_{2}\right):\\
D_{i}\geq\sum_{j=1}^{m}\frac{\lambda_{j}N_{i,j}}{P_{j}+N_{i,j}},i=1,2,\\
R_{\mathsf{L}}\leq\min\left\{ R_{\mathsf{K}}+R_{\boldsymbol{S|Z}}(D_{0}),R_{\boldsymbol{S}}(D_{0})\right\}
\end{array}\right\} ,
\]
with
\begin{align}
\begin{array}{l}
R_{\boldsymbol{S}}(D_{0})\end{array} & =\sum_{j=1}^{m}\frac{1}{2}\log^{+}\left(\frac{\lambda_{j}}{\mu}\right)\label{eq:Rs}\\
R_{\boldsymbol{S|Z}}(D_{0}) & =\sum_{j=1}^{m}\frac{1}{2}\log^{+}\left(\frac{\lambda_{j}N_{0,j}}{\theta\left(P_{j}+N_{0,j}\right)}\right)\label{eq:Rs-z}
\end{align}
and with $\mu$ and $\theta$ such that
\begin{eqnarray}
D_{0} & = & \sum_{j=1}^{m}\min\left\{ \mu,\lambda_{j}\right\} ,\\
D_{0} & = & \sum_{j=1}^{m}\min\left\{ \theta,\frac{\lambda_{j}N_{0,j}}{P_{j}+N_{0,j}}\right\} .
\end{eqnarray}
\end{thm}
\begin{rem}
Actually, in Theorem \ref{thm:vector}, $R_{\boldsymbol{S}}(D_{0})$
denotes the rate-distortion function of the source $\boldsymbol{S}$,
and $R_{\boldsymbol{S|Z}}(D_{0})$ denotes the rate-distortion function
of the source $\boldsymbol{S}$ with the side information $\boldsymbol{Z}$
available at both the encoder and decoder, where $Z_{j}=\sqrt{\frac{P_{j}}{\lambda_{j}}}S_{j}+V{}_{j},j\in\left[m\right]$
with $\boldsymbol{V}\sim\mathcal{N}\left(0,\textrm{diag}\left(N_{0,1},N_{0,2},\cdots,N_{0,m}\right)\right)$
independent of $\boldsymbol{S}$.
\end{rem}

\section{Concluding Remarks}

\label{sec:conclusion}

In this paper, we studied the joint source-channel secrecy problem
for secure source broadcast in the Shannon cipher system, in which
the list secrecy is used to measure the secrecy of communication.
We proposed two secure uncoded schemes: a permutation based scheme
for discrete, scalar Gaussian, and vector Gaussian communications,
and an orthogonal-transform based scheme for the latter two communications.
In these two uncoded schemes, a random permutation or a random orthogonal
transform is cascaded with the traditional uncoded JSCC scheme. The
analysis showed that the proposed schemes outperform  the sign-change
based scheme. Interestingly, by adding the random permutation operation
or the random orthogonal transform into the traditional uncoded scheme,
the proposed uncoded schemes, on one hand, provide a certain level
of secrecy, and on the other hand, do not lose any performance in
terms of the distortions for legitimate users.

Although the proposed
schemes adopt two different random transforms, permutation operation
and orthogonal transform, they are consistent in two aspects: First,
actually the permutation operation  is one kind of  orthogonal transform;
second, for the Gaussian communication, the orthogonal transform can
be also considered as a shift operation that shifts a sequence to
another in the same ``type'', if we treat the Euclidean norm  of
the source sequence as its ``type''\footnote{This kind of type can be called ``weak type'', since the relationship
of it and the weak typicality is similar to that of the traditional
type (empirical distribution) and strong typicality.}.  Furthermore, it is worth noting that different from the common
construction of codebook  in information theory (including \emph{spherical codes} such as
the one used in \cite{Lapidoth}), the codebooks
in the proposed schemes are constructed by generating a sequence of
i.i.d. random permutations or random matrices, instead of a sequence
of i.i.d. random samples. In other words, the codebooks used here
specify a sequence of bijective operations or transforms and hence
they apply to uncoded schemes; while the common codebooks in information
theory only specify a sequence of samples and hence can only be used
in quantization operation (or digital schemes). Furthermore, such
random-permutation or random-matrix based codebook construction can
be also found in \cite{Ahlswede,Kang,Slepian65,Slepian68,Ericson},
where they were used to design digital schemes for communication,
secrecy communication, and antijamming communication problems.  But
different from those works, in our case they were used to design uncoded
schemes, instead of digital schemes.

It is worth noting that the proofs used in this paper follow basic
outline of the proofs in \cite{Schieler}. But different from \cite{Schieler},
 besides the finite alphabet case, we also considered the countably
infinite alphabet and continuous (Gaussian) alphabet cases. Hence
some powerful techniques, including unified typicality, $\mathsf{d}-$tilted
information, geometric analysis, and discretization, are used in our
proofs. Furthermore, the unified typicality used in our proofs is
different from the existing one defined in \cite{Ho}. The unified
typical set defined by us has a good property that the sequences in
it only have (nearly) sub-exponential number of types. This property
coincides with the finite alphabet case, and is of crucial importance
to our proofs. We believe our definition of unified typicality could  be used to further extend the method of types to countably infinite alphabet cases (besides the extension in  \cite{Ho}).
%, to some extent,

\appendices{}

\section{\label{sec:Proof-of-TheoremDM}Proof of Theorem \ref{thm:proposedDM}}

Denote
\begin{align}
 & Z'^{n}\triangleq\Psi_{K}^{-1}\left(Z^{n}\right),\label{eq:-53}\\
 & X'^{n}\triangleq\Psi_{K}^{-1}\left(X^{n}\right),\\
 & Y_{i}^{\prime n}\triangleq\Psi_{K}^{-1}\left(Y_{i}^{n}\right).\label{eq:-52}
\end{align}
Then from the fact that the permutation operation is bijective, we
have that
\begin{align}
 & P_{\mathcal{C}S^{n}KS^{\prime n}X^{n}Y_{i}^{n}Z^{n}X^{\prime n}Y_{i}^{\prime n}Z^{\prime n}\widehat{S}_{i}^{\prime n}\widehat{S}_{i}^{n}}\nonumber \\
 & =P_{K}P_{\mathcal{C}}P_{S^{n}}P_{S^{\prime n}|S^{n}\Psi_{K}}P_{X^{n}|S^{\prime n}}P_{Y_{i}^{n}Z^{n}|X^{n}}P_{\widehat{S}_{i}^{\prime n}|Y_{i}^{n}\Psi_{K}}P_{\widehat{S}_{i}^{n}|\widehat{S}_{i}^{\prime n}}P_{X^{\prime n}|X^{n}\Psi_{K}}P_{Y_{i}^{\prime n}|Y_{i}^{n}\Psi_{K}}P_{Z^{\prime n}|Z^{n}\Psi_{K}}\\
 & =P_{K}P_{\mathcal{C}}P_{S^{n}}P_{S^{\prime n}|S^{n}\Psi_{K}}P_{X^{\prime n}|S^{\prime n}\Psi_{K}}P_{Y_{i}^{\prime n}Z^{\prime n}|X^{\prime n}\Psi_{K}}P_{\widehat{S}_{i}^{\prime n}|Y_{i}^{\prime n}\Psi_{K}}P_{\widehat{S}_{i}^{n}|Y_{i}^{\prime n}\Psi_{K}}P_{X^{n}|X^{\prime n}\Psi_{K}}P_{Y_{i}^{n}|Y_{i}^{\prime n}\Psi_{K}}P_{Z^{n}|Z^{\prime n}\Psi_{K}}\\
 & =P_{K}P_{\mathcal{C}}P_{S^{n}}P_{X^{\prime n}|S^{n}}P_{Y_{i}^{\prime n}Z^{\prime n}|X^{\prime n}}P_{\widehat{S}_{i}^{n}|Y_{i}^{\prime n}}P_{S^{\prime n}|S^{n}\Psi_{K}}P_{\widehat{S}_{i}^{\prime n}|Y_{i}^{\prime n}\Psi_{K}}P_{X^{n}|X^{\prime n}\Psi_{K}}P_{Y_{i}^{n}|Y_{i}^{\prime n}\Psi_{K}}P_{Z^{n}|Z^{\prime n}\Psi_{K}},\label{eq:-5}
\end{align}
and similarly, $P_{\mathcal{C}S^{n}KS^{\prime n}X^{n}Y_{i}^{n}Z^{n}X^{\prime n}Y_{i}^{\prime n}Z^{\prime n}\widehat{S}_{i}^{\prime n}\widehat{S}_{i}^{n}}$
can be also expressed as
\begin{align}
 & P_{\mathcal{C}S^{n}KS^{\prime n}X^{n}Y_{i}^{n}Z^{n}X^{\prime n}Y_{i}^{\prime n}Z^{\prime n}\widehat{S}_{i}^{\prime n}\widehat{S}_{i}^{n}}\nonumber \\
 & =P_{K}P_{\mathcal{C}}P_{S^{\prime n}}P_{X^{n}|S^{\prime n}}P_{Y_{i}^{n}Z^{n}|X^{n}}P_{\widehat{S}_{i}^{\prime n}|Y_{i}^{n}}P_{S^{n}|S^{\prime n}\Psi_{K}}P_{\widehat{S}_{i}^{n}|\widehat{S}_{i}^{\prime n}}P_{X^{\prime n}|X^{n}\Psi_{K}}P_{Y_{i}^{\prime n}|Y_{i}^{n}\Psi_{K}}P_{Z^{\prime n}|Z^{n}\Psi_{K}}.\label{eq:-50}
\end{align}
Hence $\left(\Psi_{K},Z^{n}\right)\rightarrow Z'^{n}\rightarrow S^{n}$
forms a Markov chain. Furthermore, since the permutation operation
does not change the joint distribution of the sequences, we have $P_{S^{n}}P_{X^{\prime n}|S^{n}}P_{Y_{i}^{\prime n}Z^{\prime n}|X^{\prime n}}P_{\widehat{S}_{i}^{n}|Y_{i}^{\prime n}}=P_{S^{\prime n}}P_{X^{n}|S^{\prime n}}P_{Y_{i}^{n}Z^{n}|X^{n}}P_{\widehat{S}_{i}^{\prime n}|Y_{i}^{n}}=\prod P_{S}P_{X|S}P_{Y_{i}Z|X}P_{\widehat{S}_{i}|Y_{i}}$,
where $P_{\widehat{S}_{i}|Y_{i}}\left(\widehat{s}|y\right)\triangleq1\left\{ \widehat{s}=\widehat{s}_{i}\left(y\right)\right\} $
denotes the conditional distribution induced by the decoder $i$,
and $P_{S}P_{X|S}P_{Y_{i}Z|X}P_{\widehat{S}_{i}|Y_{i}}$ is the distribution
given in \eqref{eq:-30}.

Since $\left(S^{n},\widehat{S}_{i}^{n}\right)$ is an i.i.d. sequence,
by the law of large numbers,
\begin{align}
\Pbb\Big[d_{\mathsf{B}}(S^{n},\widehat{S}_{i}^{n})\le\mathbb{E}d_{\mathsf{B}}(S,\widehat{S}_{i})+\epsilon\Big] & \xrightarrow{n\to\infty}1,\label{eq:legitimatedistortion-1-2}
\end{align}
for any $\epsilon>0$. Hence the distortion constraints for legitimate
users are satisfied.

Next we prove the secrecy constraint is also satisfied, i.e., if
\begin{equation}
\limsup_{n\rightarrow\infty}R_{n}<\min\left\{ R_{\mathsf{K}}+R_{S|Z}(D_{0}),R_{S}(D_{0})\right\} ,\label{eq:-3-1}
\end{equation}
then $\mathop{\lim}\limits _{n\to\infty}\mathbb{E}_{\mathcal{C}Z^{n}}\Bigl[\max_{R_{n}\mathsf{Hcodes}}\mathbb{P}\bigl[d_{\mathsf{E}}(S^{n},\widecheck{S}^{n})\le D_{0}\bigr]\Bigr]=0$.
To that end, we need the following lemma.
\begin{lem}
\noindent \cite{Yu-1}\label{lem:probability} For a sequence of random
variables $\left\{ X_{n}\right\} $, and a sequence of events $\left\{ \mathcal{A}_{n}\right\} $,
$\lim_{n\to\infty}\mathbb{P}\left(\mathcal{A}_{n}\right)=0$, if and
only if $\lim_{n\to\infty}\mathbb{P}\left[\mathbb{P}\left(\mathcal{A}_{n}|X_{n}\right)>\tau_{n}\right]=0$
for some sequence $\left\{ \tau_{n}\right\} $ with $\tau_{n}>0$
and $\lim_{n\to\infty}\tau_{n}=0$.
\end{lem}
From Lemma \ref{lem:probability}, to prove the secrecy constraint
we only need to show that if $R_{n}$ satisfies \eqref{eq:-3-1},
then
\begin{align}
 & \mathop{\lim}\limits _{n\to\infty}\mathbb{P}_{\mathcal{C}Z^{n}}\Bigl[\max_{R_{n}\mathsf{Hcodes}}\mathbb{P}\bigl[d_{\mathsf{E}}(S^{n},\widecheck{S}^{n})\le D_{0}\bigr]>\tau_{n}\Bigr]=0,\label{eq:-2-2-2}
\end{align}
for some sequence $\left\{ \tau_{n}\right\} $ with $\tau_{n}>0$
and $\lim_{n\to\infty}\tau_{n}=0$. Next we prove this.

Define event
\begin{align}
 & \mathcal{A}\triangleq\left\{ \left(S^{n},Z'^{n}\right)\in\mathcal{T}_{\delta}^{n}\left(S,Z'\right)\right\} ,
\end{align}
for $\delta>0$. The $\delta$-typical set is defined according to
the notion of strong typicality, see \cite{Gamal}:
\begin{equation}
\Tcal_{\delta}^{n}(S)\triangleq\{s^{n}\in\Scal^{n}:\sum_{s\in\mathcal{S}}\left|T_{s^{n}}\left(s\right)-P_{S}\left(s\right)\right|\leq\delta\},\label{eq:typicality}
\end{equation}
where $T_{s^{n}}$ denotes the type (or empirical distribution) of
$s^{n}$. For simplicity, $\Tcal_{\delta}^{n}(S)$ is also shortly
denoted as $\Tcal_{\delta}^{n}$.

Since $\left(S^{n},Z'^{n}\right)$ is an i.i.d. sequence, from the
fact that the typical set has total probability close to one \cite{Gamal},
we have the following lemma.
\begin{lem}
\label{lem:A} \cite{Gamal} For any $\delta>0$, $\mathbb{P}\left[\mathcal{A}\right]\to1$,
as $n\to\infty$.
\end{lem}
Consider that for each $n$, the optimal $R_{n}$-rate henchman code
that maximizes $\mathbb{P}\bigl[d_{\mathsf{E}}(S^{n},\widecheck{S}^{n})\le D_{0}|\mathcal{C}Z^{n}\bigr]$
is adopted, then we only need to show $\mathop{\lim}\limits _{n\to\infty}\mathbb{P}_{\mathcal{C}Z^{n}}\Bigl[\mathbb{P}\bigl[d_{\mathsf{E}}(S^{n},\widecheck{S}^{n})\le D_{0}|\mathcal{C}Z^{n}\bigr]>\tau_{n}\Bigr]=0$
for these codes. By utilizing Lemmas \ref{lem:probability} and \ref{lem:A},
we have
\begin{align}
 & \mathbb{P}_{\mathcal{C}Z^{n}}\Bigl[\mathbb{P}\bigl[d_{\mathsf{E}}(S^{n},\widecheck{S}^{n})\le D_{0}|\mathcal{C}Z^{n}\bigr]>\tau_{n}\Bigr]\nonumber \\
\leq & \mathbb{P}_{\mathcal{C}Z^{n}}\Bigl[\mathbb{P}\bigl[d_{\mathsf{E}}(S^{n},\widecheck{S}^{n})\le D_{0}|\mathcal{C}Z^{n}\bigr]>\tau_{n},\mathbb{P}\left[\mathcal{A}^{c}|\mathcal{C}Z^{n}\right]\leq\epsilon_{n}\Bigr]+\Pbb\left[\mathbb{P}\left[\mathcal{A}^{c}|\mathcal{C}Z^{n}\right]>\epsilon_{n}\right]\\
\leq & \mathbb{P}_{\mathcal{C}Z^{n}}\Bigl[\mathbb{P}\bigl[d_{\mathsf{E}}(S^{n},\widecheck{S}^{n})\le D_{0},\mathcal{A}|\mathcal{C}Z^{n}\bigr]+\mathbb{P}\left[\mathcal{A}^{c}|\mathcal{C}Z^{n}\right]>\tau_{n},\mathbb{P}\left[\mathcal{A}^{c}|\mathcal{C}Z^{n}\right]\leq\epsilon_{n}\Bigr]+\epsilon_{n}^{\prime}\\
\leq & \mathbb{P}_{\mathcal{C}Z^{n}}\Bigl[\mathbb{P}\bigl[d_{\mathsf{E}}(S^{n},\widecheck{S}^{n})\le D_{0},\mathcal{A}|\mathcal{C}Z^{n}\bigr]>\tau'_{n}\Bigr]+\epsilon_{n}^{\prime},\label{eq:-18-1}
\end{align}
for some $\epsilon_{n}$ and $\epsilon_{n}^{\prime}$ that both vanish
as $n\to\infty$, where $\tau'_{n}=\tau_{n}-\epsilon_{n}$. By choosing
proper $\tau_{n}$, $\tau'_{n}$ can be set to some sequence that
converges to zero sub-exponentially fast (i.e., $\tau'_{n}=2^{-o\left(n\right)}$).
Since $\epsilon_{n}$ vanishes as $n\to\infty$, this guarantees that
$\tau_{n}$ also vanishes as $n\to\infty$.

Owing to the rate constraint, given $(\Ccal,Z^{n})$, the reconstruction
$\widecheck{S}^{n}$ cannot take more than $R_{n}$ values. Denote the
set of possible values as $c(\Ccal,Z^{n})$, then
\begin{align}
\mathbb{P}\bigl[d_{\mathsf{E}}(S^{n},\widecheck{S}^{n})\le D_{0},\mathcal{A}|\mathcal{C}Z^{n}\bigr] & =\mathbb{P}\Big[\min_{\widecheck{s}^{n}\in c(\Ccal,Z^{n})}d_{\mathsf{E}}(S^{n},\widecheck{s}^{n})\leq D_{0},\mathcal{A}|\mathcal{C}Z^{n}\Big].\label{eq:boundMax-1-1}
\end{align}
Now we apply a union bound to the right-hand side of \eqref{eq:boundMax-1-1}
and write
\begin{align}
 & \mathbb{P}\Big[\min_{\widecheck{s}^{n}\in c(\Ccal,Z^{n})}d_{\mathsf{E}}(S^{n},\widecheck{s}^{n})\leq D_{0},\mathcal{A}|\mathcal{C}Z^{n}\Big]\nonumber \\
\leq & \sum_{\widecheck{s}^{n}\in c(\Ccal,Z^{n})}\Pbb\Big[d_{\mathsf{E}}(S^{n},\widecheck{s}^{n})\leq D_{0},\mathcal{A}|\mathcal{C}Z^{n}\Big]\label{eq:-9-1}\\
\leq & 2^{nR_{n}}\max_{\widecheck{s}^{n}\in c(\Ccal,Z^{n})}\Pbb\Big[d_{\mathsf{E}}(S^{n},\widecheck{s}^{n})\leq D_{0},\mathcal{A}|\mathcal{C}Z^{n}\Big]\\
\leq & 2^{nR_{n}}\max_{\widecheck{s}^{n}\in\widecheck{\mathcal{S}}^{n}}\Pbb\Big[d_{\mathsf{E}}(S^{n},\widecheck{s}^{n})\leq D_{0},\mathcal{A}|\mathcal{C}Z^{n}\Big]\\
= & 2^{nR_{n}}\max_{\widecheck{s}^{n}\in\widecheck{\mathcal{S}}^{n}}\sum_{k=1}^{2^{nR_{\mathsf{K}}}}\Pbb\left[K=k|\mathcal{C}Z^{n}\right]\Pbb\left[d_{\mathsf{E}}(S^{n},\widecheck{s}^{n})\leq D_{0},\mathcal{A}|\mathcal{C}Z^{n},K=k\right]\\
= & 2^{n\left(R_{n}-R_{\mathsf{K}}\right)}\max_{\widecheck{s}^{n}\in\widecheck{\mathcal{S}}^{n}}\sum_{k=1}^{2^{nR_{\mathsf{K}}}}\Pbb\left[d_{\mathsf{E}}(S^{n},\widecheck{s}^{n})\leq D_{0},\mathcal{A}|\Psi_{k},Z^{n}\right],\label{eq:-12-1}
\end{align}
where \eqref{eq:-12-1} follows from the Markov chain $\mathcal{C}KZ^{n}\rightarrow\Psi_{K}Z^{n}\rightarrow S^{n}Z^{n}\mathcal{A}$
and $\Pbb\left[K=k|\mathcal{C}=c,Z^{n}=z^{n}\right]=2^{-nR_{\mathsf{K}}}$
(see \eqref{eq:-50}).

Combine \eqref{eq:-18-1}, \eqref{eq:boundMax-1-1}, and \eqref{eq:-12-1},
then we have
\begin{align}
 & \mathbb{P}_{\mathcal{C}Z^{n}}\Bigl[\mathbb{P}\bigl[d_{\mathsf{E}}(S^{n},\widecheck{S}^{n})\le D_{0}|\mathcal{C}Z^{n}\bigr]>\tau_{n}\Bigr]\nonumber \\
 & \leq\mathbb{P}_{\mathcal{C}Z^{n}}\Bigl[\max_{\widecheck{s}^{n}\in\widecheck{\mathcal{S}}^{n}}\sum_{k=1}^{2^{nR_{\mathsf{K}}}}\Pbb\left[d_{\mathsf{E}}(S^{n},\widecheck{s}^{n})\leq D_{0},\mathcal{A}|\Psi_{k},Z^{n}\right]>\tau'_{n}2^{-n\left(R_{n}-R_{\mathsf{K}}\right)}\Bigr]+\epsilon_{n}^{\prime}\label{eq:-26-2}\\
 & \leq\left|\widecheck{\mathcal{S}}^{n}\right|\max_{\widecheck{s}^{n}\in\widecheck{\mathcal{S}}^{n}}\mathbb{P}_{\mathcal{C}Z^{n}}\Big[\sum_{k=1}^{2^{nR_{\mathsf{K}}}}\xi_{k,z^{n}}\left(\widecheck{s}^{n}\right)>\tau'_{n}2^{-n\left(R_{n}-R_{\mathsf{K}}\right)}\Big]+\epsilon_{n}^{\prime},\label{eq:-2}
\end{align}
where
\begin{align}
\xi_{k,z^{n}}\left(\widecheck{s}^{n}\right) & \triangleq\Pbb\left[d_{\mathsf{E}}(S^{n},\widecheck{s}^{n})\leq D_{0},\mathcal{A}|\Psi_{k},Z^{n}\right],
\end{align}
Therefore, if we can show that the probability in \eqref{eq:-2} decays
doubly exponentially fast with $n$, then the proof will be complete.

Consider that given $\widecheck{s}^{n}$ and $z^{n}$, $\xi_{k,z^{n}}\left(\widecheck{s}^{n}\right),k\in\left[2^{nR_{\mathsf{K}}}\right]$
are i.i.d. random variables, with mean

\noindent
\begin{align}
\mathbb{E}_{{\mathcal{C}}}\xi_{k,z^{n}}\left(\widecheck{s}^{n}\right) & =\mathbb{E}_{{\mathcal{C}}}\Pbb\left[d_{\mathsf{E}}(S^{n},\widecheck{s}^{n})\leq D_{0},\mathcal{A}|\Psi_{k},z^{n}\right]\\
 & =\mathbb{E}_{\Psi_{k}}\Pbb\left[d_{\mathsf{E}}(S^{n},\widecheck{s}^{n})\leq D_{0},\mathcal{A}|\Psi_{k},z^{n}\right]
\end{align}
To complete the proof, we need introduce the following lemmas. The
proof of Lemma \ref{lem:typebound} is given in Appendix \ref{sec:Proof-of-Lemma-typebound}.
\begin{lem}
\label{lem:typebound} Assume $S^{n}$ is i.i.d. according to $P_{S}$,
then for any type $t$ of sequences in $\mathcal{S}^{n}$ and any
$\widecheck{s}^{n}$,
\begin{equation}
\Pbb[d_{\mathsf{E}}\left(S^{n},\widecheck{s}^{n}\right)\leq D,S^{n}\in\mathcal{T}_{\delta}^{n}|T_{S^{n}}=t]\leq2^{-n(R_{S}(D)-o(1))},\label{eq:rd-1}
\end{equation}
where $T_{S^{n}}$ denotes the type of $S^{n}$, and $o(1)$ is a
term that vanishes as $\delta\rightarrow0$ and $n\rightarrow\infty$.
\end{lem}
\begin{lem}
\label{lem:typebound-rd-si-2} \cite{Schieler} Fix $P_{S|Z}$ and
$z^{n}$. If $S^{n}$ is distributed according to $\prod_{i=1}^{n}P_{S|Z=z_{i}}$,
then for any $\widecheck{s}^{n}$,
\begin{equation}
\Pbb[d_{\mathsf{E}}\left(S^{n},\widecheck{s}^{n}\right)\leq D,\left(S^{n},z{}^{n}\right)\in\mathcal{T}_{\delta}^{n}|z{}^{n}]\leq2^{-n(R_{S|Z}(D)-o(1))},\label{eq:rdsi}
\end{equation}
where $o(1)$ is a term that vanishes as $\delta\rightarrow0$ and
$n\rightarrow\infty$.
\end{lem}
\begin{lem}
\label{lem:chernoff} \cite{Schieler} If $X^{m}$ is a sequence of
i.i.d. random variables on the interval $[0,a]$ with $\mathbb{E}[X_{i}]=p$,
then
\begin{equation}
\Pbb\Big[\sum_{i=1}^{m}X_{i}>k\Big]\leq\left(\frac{e\!\cdot\!m\!\cdot\!p}{k}\right)^{k/a}.
\end{equation}
\end{lem}
From \eqref{eq:-5}, we have
\begin{align}
\Pbb\left[S^{n}=s^{n}|\Psi_{k},z^{n}\right] & =\prod P_{S|Z}\left(s_{i}|z_{i}^{\prime}\right).\label{eq:-6}
\end{align}

\noindent Hence Lemma \ref{lem:typebound-rd-si-2} implies
\begin{align}
\begin{array}{c}
\xi_{k,z^{n}}\left(\widecheck{s}^{n}\right)\end{array} & =\Pbb\left[d_{\mathsf{E}}(S^{n},\widecheck{s}^{n})\leq D_{0},\mathcal{A}|z^{\prime n}\right]\label{eq:-43-1}\\
 & \leq2^{-n(R_{S|Z}(D_{0})-o(1))}.
\end{align}

\noindent On the other hand,
\begin{align}
\begin{array}{c}
\mathbb{E}_{\mathcal{C}}\xi_{k,z^{n}}\left(\widecheck{s}^{n}\right)\end{array} & \leq\Ebb_{\Psi_{k}}\Pbb[d_{\mathsf{E}}(S^{n},\widecheck{s}^{n})\leq D_{0},S^{n}\in\mathcal{T}_{\delta}^{n}|\Psi_{k},z^{n}]\\
 & =\sum_{s'^{n}}\Pbb[S'^{n}=s'^{n}|z^{n}]\Ebb_{\Psi_{k}}\Pbb[d_{\mathsf{E}}(S^{n},\widecheck{s}^{n})\leq D_{0},S^{n}\in\mathcal{T}_{\delta}^{n}|S'^{n}=s'^{n},\Psi_{k}]\\
 & =\sum_{s'^{n}}\Pbb[S'^{n}=s'^{n}|z^{n}]\Pbb[d_{\mathsf{E}}(S^{n},\widecheck{s}^{n})\leq D_{0},S^{n}\in\mathcal{T}_{\delta}^{n}|S'^{n}=s'^{n}]\\
 & =\sum_{s'^{n}}\Pbb[S'^{n}=s'^{n}|z^{n}]\Pbb[d_{\mathsf{E}}(S^{n},\widecheck{s}^{n})\leq D_{0},S^{n}\in\mathcal{T}_{\delta}^{n}|T_{S^{n}}=T_{s'^{n}}]\\
 & \leq2^{-n(R_{S}(D{}_{0})-o(1))}.\label{eq:-16-1-1}
\end{align}
Using these bounds, we apply Lemma~\ref{lem:chernoff} to the probability
in \eqref{eq:-2} by identifying
\begin{align}
m & =2^{nR_{\mathsf{K}}},\\
a & =2^{-n(R_{S|Z}(D{}_{0})-o(1))},\\
p & \leq2^{-n(R_{S}(D{}_{0})-o(1))},\\
k & =\tau'_{n}2^{-n\left(R_{n}-R_{\mathsf{K}}\right)}.
\end{align}

\noindent Then we have
\begin{equation}
\Pbb\Big[\sum_{k=1}^{2^{nR_{\mathsf{K}}}}\xi_{k,z^{n}}\left(\widecheck{s}^{n}\right)>\tau'_{n}2^{-n\left(R_{n}-R_{\mathsf{K}}\right)}\Big]\leq2^{-n\alpha2^{n\beta}},\label{doubleexp-2-2}
\end{equation}
where
\begin{equation}
\begin{array}{l}
\alpha=R_{S}(D{}_{0})-R_{n}-o(1)\\
\beta=R_{\mathsf{K}}+R_{S|Z}(D{}_{0})-R_{n}-o(1).
\end{array}\label{eq:-27}
\end{equation}
For small enough $\delta$ and large enough $n$, both $\alpha$ and
$\beta$ are positive and bounded away from zero, and \eqref{doubleexp-2-2}
vanishes doubly exponentially fast. Therefore, the expression in \eqref{eq:-2}
vanishes. This completes the proof of Theorem \ref{thm:proposedDM}.

\section{\label{sec:Proof-of-Lemma-typebound}Proof of Lemma \ref{lem:typebound}}

If $\sum_{s\in\mathcal{S}}\left|t\left(s\right)-P_{S}\left(s\right)\right|>\delta$,
then $\Pbb[d_{\mathsf{E}}\left(S^{n},\widecheck{s}^{n}\right)\leq D,S^{n}\in\mathcal{T}_{\delta}^{n}|T_{S^{n}}=t]=0$.
Hence we only need to consider the $t$'s such that $\sum_{s\in\mathcal{S}}\left|t\left(s\right)-P_{S}\left(s\right)\right|\leq\delta$.

Consider
\begin{align}
\Pbb[T_{S^{n}}=t] & =\left|\left\{ s^{\prime n}\in\mathcal{S}^{n}:T_{s^{\prime n}}=t\right\} \right|2^{-n\left(D\left(t||P_{S}\right)+H(t)\right)}\\
 & =2^{-n\left(D\left(t||P_{S}\right)+o(1)\right)}\label{eq:-54}
\end{align}
for any type $t$ of sequences in $\mathcal{S}^{n}$, where $D\left(t||P_{S}\right)$
denotes the relative entropy between $t$ and $P_{S}$, and \eqref{eq:-54}
follows from \eqref{eq:-26}. Moreover, from \cite[Thm. 25]{Sason}
we have
\begin{equation}
D\left(t||P_{S}\right)\leq\log\left(1+\frac{\left(\sum_{s}\left|t\left(s\right)-P_{S}\left(s\right)\right|\right)^{2}}{2P_{S,\min}}\right)\leq\log\left(1+\frac{\delta^{2}}{2P_{S,\min}}\right)\rightarrow0,
\end{equation}
as $\delta\rightarrow0$, where $P_{S,\min}=\min_{s\in\mathcal{S}}P_{S}\left(s\right)$.
Therefore,
\begin{equation}
\Pbb[T_{S^{n}}=t]\geq2^{-no(1)}.\label{eq:-17}
\end{equation}
Utilizing \eqref{eq:-17}, we get
\begin{align}
 & \Pbb[d_{\mathsf{E}}\left(S^{n},\widecheck{s}^{n}\right)\leq D,S^{n}\in\mathcal{T}_{\delta}^{n}|T_{S^{n}}=t]\nonumber \\
 & =\frac{\Pbb[d_{\mathsf{E}}\left(S^{n},\widecheck{s}^{n}\right)\leq D,S^{n}\in\mathcal{T}_{\delta}^{n},T_{S^{n}}=t]}{\Pbb[T_{S^{n}}=t]}\\
 & \leq\frac{\Pbb[d_{\mathsf{E}}\left(S^{n},\widecheck{s}^{n}\right)\leq D,S^{n}\in\mathcal{T}_{\delta}^{n}]}{2^{-no(1)}}.\label{eq:-28}
\end{align}

To complete the proof, we need the following lemma.
\begin{lem}
\label{lem:typebound-1} \cite{Schieler} Assume $S^{n}$ is i.i.d.
according to $P_{S}$, then for any $\widecheck{s}^{n}$,
\begin{equation}
\Pbb[d_{\mathsf{E}}\left(S^{n},\widecheck{s}^{n}\right)\leq D,S^{n}\in\mathcal{T}_{\delta}^{n}]\leq2^{-n(R_{S}(D)-o(1))}.\label{eq:rd-2}
\end{equation}
\end{lem}
\noindent By the lemma above, \eqref{eq:-28} implies that
\begin{align}
 & \Pbb[d_{\mathsf{E}}\left(S^{n},\widecheck{s}^{n}\right)\leq D,S^{n}\in\mathcal{T}_{\delta}^{n}|T_{S^{n}}=t]\leq2^{-n(R_{S}(D)-o(1))}.
\end{align}

\section{\label{sec:Proof-of-TheoremGDM}Proof of Theorem \ref{thm:proposedGDM}}

Define $X'^{n},Y_{i}^{\prime n},Z'^{n}$ same as \eqref{eq:-53}-\eqref{eq:-52},
then the distribution $P_{\mathcal{C}S^{n}KS^{\prime n}X^{n}Y_{i}^{n}Z^{n}X^{\prime n}Y_{i}^{\prime n}Z^{\prime n}\widehat{S}_{i}^{\prime n}\widehat{S}_{i}^{n}}$
also satisfies \eqref{eq:-5} and \eqref{eq:-50}. Similar to the
finite alphabet case, it is easy to show the distortion constraints
for legitimate users are satisfied.

Next following similar steps to the proof for the finite alphabet
case, we prove the secrecy constraint is also satisfied for this case.
Before proving that, we need introduce $\mathsf{d}-$tilted information
and conditional $\mathsf{d}-$tilted information first.

Let $P_{\widecheck{S}^{\star}|S}$ be a distribution that achieves the
rate-distortion function $R_{S}(D)$ (which is not necessarily unique).
Then $\mathsf{d}-$tilted information is defined as follows.
\begin{defn}[$\mathsf{d}-$tilted information \cite{Yu-1}]
\label{defn:id} For $D>D_{\min}\triangleq\inf\left\{ D\colon~R_{S}(D)<\infty\right\} $,
the $\mathsf{d}-$tilted information in $s$ is defined as
\begin{equation}
\jmath_{S}(s,D)=\log\frac{1}{\mathbb{E}\bigl[\exp\bigl(\lambda^{\star}D-\lambda^{\star}d(s,\widecheck{S}^{\star})\bigr)\bigr]},\label{eq:idball}
\end{equation}
where the expectation is with respect to $P_{\widecheck{S}^{\star}}$,
i.e. the unconditional distribution of the reproduction random variable
that achieves $R_{S}(D)$, and
\begin{equation}
\lambda^{\star}=-R_{S}^{\prime}(D).\label{eq:lambdastar}
\end{equation}
\end{defn}
For $\left(S,Z\right)$ that follow the distribution in \eqref{eq:-30},
we define
\begin{equation}
R_{S|Z=z}(\beta)=\min_{P_{\widecheck{S}|S,Z=z}:\mathbb{E}\bigl[d_{\mathsf{E}}(S,\widecheck{S})|Z=z\bigr]\leq\beta}I(S;\widecheck{S}|Z=z).\label{eq:-97}
\end{equation}
Let $P_{\widecheck{S}^{\star}|S,Z=z}$ be a distribution that achieves
$R_{S|Z=z}(\beta)$.   Define $b^{\star}\left(z\right)\triangleq\mathbb{E}_{S,\widecheck{S}^{\star}|Z=z}d(S,\widecheck{S}^{\star})$
with the expectation taken with respect to $P_{S|Z=z}P_{\widecheck{S}^{\star}|S,Z=z}$.
\begin{defn}[Conditional $\mathsf{d}-$tilted information \cite{Yu-1}]
\label{defn:id-1} For $b^{\star}\left(z\right)>\beta_{\min}\left(z\right)\triangleq\inf\left\{ \beta\colon~R_{S|Z=z}(\beta)<\infty\right\} $,
the conditional $\mathsf{d}-$tilted information in $s$ under condition
$Z=z$ is defined as
\begin{equation}
\jmath_{S|Z=z}(s,b^{\star}\left(z\right))=\log\frac{1}{\mathbb{E}_{\widecheck{S}^{\star}|Z=z}\left[\exp\left(\lambda^{\star}\left(z\right)b^{\star}\left(z\right)-\lambda^{\star}\left(z\right)d(s,\widecheck{S}^{\star})\right)\right]},\label{eq:idball-2}
\end{equation}
where the expectation is with respect to $P_{\widecheck{S}^{\star}|Z=z}$,
i.e. the margin distribution of $P_{S|Z=z}P_{\widecheck{S}^{\star}|S,Z=z}$,
and
\begin{equation}
\lambda^{\star}\left(z\right)=-R_{S|Z=z}^{\prime}(b^{\star}\left(z\right)).\label{eq:lambdastar-1}
\end{equation}
\end{defn}
Next we prove the secrecy constraint. To that end, we need re-define

\begin{align}
\mathcal{A}\triangleq & \Bigl\{ S{}^{n}\in\mathcal{U}_{\delta}^{n},\frac{1}{n}\sum_{i=1}^{n}\jmath_{S}(S{}_{i},D_{0})\geq R_{S}(D_{0})-\delta,\nonumber \\
 & \frac{1}{n}\sum_{i=1}^{n}\jmath_{S|Z=Z_{i}^{\prime}}(S{}_{i},b^{\star}(Z_{i}^{\prime}))\geq R_{S|Z}(D_{0})-\delta,\frac{1}{n}\sum_{i=1}^{n}b^{\star}(Z_{i}^{\prime})\geq D_{0}-\delta\Bigr\}
\end{align}
for $\delta>0$. The $\delta$-unified typical set is defined as\footnote{Here the $\delta$-unified typical set is different from the one defined
in \cite{Ho}. Our definition has the benefit that it makes the following
property hold: For each sequence $s^{n}\in\Ucal_{\delta}^{n}(S)$,
$\bigl|\left\{ s^{\prime n}\in\Ucal_{\delta}^{n}(S):T_{s^{\prime n}}=T_{s^{n}}\right\} \bigr|=2^{n\left(H(S)-o(1)\right)}$,
or equivalently, $\Pbb[T_{S^{n}}=T_{s^{n}}]=2^{-no(1)}$, where $o(1)$
is a term that vanishes as $\delta\rightarrow0$ and $n\rightarrow\infty$.
This property coincides with \eqref{eq:-17} for the finite alphabet
case, and it is of crucial importance to our proof here (see \eqref{eq:-17-1}). }
\begin{equation}
\Ucal_{\delta}^{n}(S)\triangleq\Tcal_{\frac{\delta}{\log n}}^{n}(S)\cap\Wcal_{\delta}^{n}(S),\label{eq:typicality-1-1}
\end{equation}
where $\Tcal_{\frac{\delta}{\log n}}^{n}(S)$ defined in \eqref{eq:typicality},
denotes the $\frac{\delta}{\log n}$-strongly typical set, and
\begin{equation}
\Wcal_{\delta}^{n}(S)\triangleq\Bigl\{ s^{n}\in\Scal^{n}:\Bigl|-\frac{1}{n}\log P_{S^{n}}\left(s^{n}\right)-H\left(S\right)\Bigr|\leq\delta\Bigr\},
\end{equation}
denotes the $\delta$-weakly typical set \cite{Cover}. For simplicity,
$\Ucal_{\delta}^{n}(S)$ is also shortly denoted as $\Ucal_{\delta}^{n}$.

Since $\left(S^{n},Z'^{n}\right)$ is an i.i.d. sequence, we have
the following lemma.
\begin{lem}
\label{lem:A2}\cite[Lem. 18]{Yu-1} \cite[Lem. 2]{Ho} Assume $P_{S}$
satisfies $\widetilde{N}_{P_{S}}\left(\frac{\delta'}{\log n}\right)=o\left(\frac{n}{\log^{2}n}\right),\forall0<\delta'\leq1$.
Then for any $\delta>0$, $\mathbb{P}\left[\mathcal{A}\right]\to1$
as $n\to\infty$.
\end{lem}
Then the derivation up to \eqref{eq:-2} still holds, i.e.,
\begin{align}
 & \mathbb{P}_{\mathcal{C}Z^{n}}\Bigl[\mathbb{P}\bigl[d_{\mathsf{E}}(S^{n},\widecheck{S}^{n})\le D_{0}|\mathcal{C}Z^{n}\bigr]>\tau_{n}\Bigr]\nonumber \\
 & \leq\left|\widecheck{\mathcal{S}}^{n}\right|\max_{\widecheck{s}^{n}\in\widecheck{\mathcal{S}}^{n}}\mathbb{P}_{\mathcal{C}Z^{n}}\Big[\sum_{k=1}^{2^{nR_{\mathsf{K}}}}\xi_{k,z^{n}}\left(\widecheck{s}^{n}\right)>\tau'_{n}2^{-n\left(R_{n}-R_{\mathsf{K}}\right)}\Big]+\epsilon_{n}^{\prime},\label{eq:-34}
\end{align}

Therefore, if we can show that the probability in \eqref{eq:-34}
decays doubly exponentially fast with $n$, then the proof will be
complete. To that end, we need introduce the following lemmas. The
proof of Lemma \ref{lem:SPO} is given in Appendix \ref{sec:Proof-of-Lemma-typebound-1}.
\begin{lem}
\label{lem:SPO} Assume $P_{S}$ satisfies $N_{P_{S}}\left(\frac{1}{n}\right)=o\left(\frac{n}{\log n}\right),\Phi_{P_{S}}\left(\frac{1}{n}\right)=o\left(\frac{1}{\log n}\right)$,
and $S^{n}$ is i.i.d. according to $P_{S}$, then for any type $t$
of sequences in $\mathcal{S}^{n}$ and any $\widecheck{s}^{n}\in\widecheck{\mathcal{S}}^{n}$,
\begin{equation}
\Pbb\bigl[d_{\mathsf{E}}(S^{n},\widecheck{s}^{n})\leq D,S^{n}\in\mathcal{U}_{\delta}^{n},\frac{1}{n}\sum_{i=1}^{n}\jmath_{S}(S_{i},D)\geq R_{S}(D)-\delta|T_{S^{n}}=t\bigr]\leq2^{-n(R_{S}(D)-o(1))},\label{eq:rd-1-1-1-3}
\end{equation}
where $o(1)$ is a term that vanishes as $\delta\rightarrow0$ and
$n\rightarrow\infty$.
\end{lem}
\begin{lem}
\label{lem:SPO-si} \cite{Yu-1} Fix $P_{SZ}$ and $z^{n}\in\mathcal{Z}^{n}$.
Assume  given $Z^{n}=z^{n}$, $S^{n}$ is distributed according to
$\prod_{i=1}^{n}P_{S|Z=z_{i}}$, then for any $\widecheck{s}^{n}\in\widecheck{\mathcal{S}}^{n}$,
\begin{align}
 & \Pbb\Bigl[d_{\mathsf{E}}(S^{n},\widecheck{s}^{n})\leq D,\frac{1}{n}\sum_{i=1}^{n}\jmath_{S|Z=z_{i}}(S{}_{i},b^{\star}(z_{i}))\geq R_{S|Z}(D)-\delta,\frac{1}{n}\sum_{i=1}^{n}b^{\star}(z_{i})\geq D-\delta|Z^{n}=z^{n}\Bigr]\nonumber \\
 & \leq2^{-n(R_{S|Z}(D)-o(1))},
\end{align}
where $o(1)$ is a term that vanishes as $\delta\rightarrow0$ and
$n\rightarrow\infty$.
\end{lem}
Apply Lemmas \ref{lem:chernoff}, \ref{lem:SPO} and \ref{lem:SPO-si},
then we have that the probability in \eqref{eq:-34} decays doubly
exponentially fast with $n$. This completes the proof of Theorem
\ref{thm:proposedGDM}.

\section{\label{sec:Proof-of-Lemma-typebound-1}Proof of Lemma \ref{lem:SPO}}

If $\sum_{s\in\mathcal{S}}\left|t\left(s\right)-P_{S}\left(s\right)\right|\leq\frac{\delta}{\log n}$
does not hold, then $\Pbb[d_{\mathsf{E}}\left(S^{n},\widecheck{s}^{n}\right)\leq D,S^{n}\in\mathcal{U}_{\delta}^{n}|T_{S^{n}}=t]=0$.
Hence we only need to consider the $t$'s satisfying $\sum_{s\in\mathcal{S}}\left|t\left(s\right)-P_{S}\left(s\right)\right|\leq\frac{\delta}{\log n}$.

The Lemma 2.6 of \cite{Csiszar} says that for any type $t$ of sequences
in $\mathcal{S}^{n}$,
\begin{equation}
\Pbb[T_{S^{n}}=t]\geq\left(n+1\right)^{-\left|\textrm{supp}\left(t\right)\right|}2^{-nD\left(t||P_{S}\right)},\label{eq:-55}
\end{equation}
where $\textrm{supp}\left(t\right)\triangleq\left\{ s\in\mathcal{S}:t\left(s\right)>0\right\} $
denotes the suppose of $t$.

Now we prove that for any $\delta>0$, $\left|\textrm{supp}\left(t\right)\right|\leq\frac{n}{\log n}\left(\delta+\epsilon_{n}\right)$
holds, where $\epsilon_{n}$ is a term that vanishes as $n\rightarrow\infty$.
To that end, we divide $\mathcal{S}$ into two parts: $\left\{ s:P_{S}\left(s\right)\geq\frac{1}{n}\right\} $
and $\left\{ s:P_{S}\left(s\right)<\frac{1}{n}\right\} $. Then
\begin{align}
\left|\textrm{supp}\left(t\right)\right| & =\left|\left\{ s:t\left(s\right)>0,P_{S}\left(s\right)\geq\frac{1}{n}\right\} \right|+\left|\left\{ s:t\left(s\right)>0,P_{S}\left(s\right)<\frac{1}{n}\right\} \right|\\
 & \leq\left|\left\{ s:P_{S}\left(s\right)\geq\frac{1}{n}\right\} \right|+\left|\left\{ s:t\left(s\right)>0,P_{S}\left(s\right)<\frac{1}{n}\right\} \right|\\
 & =N_{P_{S}}\left(\frac{1}{n}\right)+\sum_{s:P_{S}\left(s\right)<\frac{1}{n}}1\left\{ t\left(s\right)>0\right\} \label{eq:-56}\\
 & \leq N_{P_{S}}\left(\frac{1}{n}\right)+n\sum_{s:P_{S}\left(s\right)<\frac{1}{n}}t\left(s\right),\label{eq:-3}
\end{align}
where \eqref{eq:-56} follows from the definition of $N_{P_{S}}\left(\frac{1}{n}\right)$,
and \eqref{eq:-3} follows from the fact $t\left(s\right)\geq\frac{1}{n}$
for any $s$ such that $t\left(s\right)>0$.

Since $\sum_{s\in\mathcal{S}}\left|t\left(s\right)-P_{S}\left(s\right)\right|\geq\sum_{s:P_{S}\left(s\right)<\frac{1}{n}}\left|t\left(s\right)-P_{S}\left(s\right)\right|\geq\sum_{s:P_{S}\left(s\right)<\frac{1}{n}}\left(t\left(s\right)-P_{S}\left(s\right)\right)$
and $\sum_{s\in\mathcal{S}}\left|t\left(s\right)-P_{S}\left(s\right)\right|\leq\frac{\delta}{\log n}$,
 we have
\begin{align}
\sum_{s:P_{S}\left(s\right)<\frac{1}{n}}t\left(s\right) & \leq\sum_{s:P_{S}\left(s\right)<\frac{1}{n}}P_{S}\left(s\right)+\frac{\delta}{\log n}\\
 & =\Phi_{P_{S}}\left(\frac{1}{n}\right)+\frac{\delta}{\log n}.
\end{align}

Therefore,

\begin{align}
\left|\textrm{supp}\left(t\right)\right| & \leq N_{P_{S}}\left(\frac{1}{n}\right)+n\Phi_{P_{S}}\left(\frac{1}{n}\right)+\frac{\delta n}{\log n}.
\end{align}

Since $N_{P_{S}}\left(\frac{1}{n}\right)=o\left(\frac{n}{\log n}\right),\Phi_{P_{S}}\left(\frac{1}{n}\right)=o\left(\frac{1}{\log n}\right)$,
we have $\left|\textrm{supp}\left(t\right)\right|\leq\frac{n}{\log n}\left(\delta+\epsilon_{n}\right)$.
Therefore, \eqref{eq:-55} implies
\begin{equation}
\Pbb[T_{S^{n}}=t]\geq2^{-n\left(D\left(t||P_{S}\right)+\delta+\epsilon_{n}\right)}.
\end{equation}
Furthermore, $S^{n}\in\mathcal{U}_{\delta}^{n}$ implies $D\left(t||P_{S}\right)\leq2\delta$,
which is obtained by following  part of proof steps of \cite[Thm. 3]{Ho}
(but with$\epsilon$ and $\delta$ replaced with $\delta$ and $\frac{n}{\log n}$,
respectively). Hence it holds that
\begin{equation}
\Pbb[T_{S^{n}}=t]\geq2^{-no(1)},\label{eq:-17-1}
\end{equation}
where $o(1)$ is a term that vanishes as $\delta\rightarrow0$ and
$n\rightarrow\infty$. Utilizing \eqref{eq:-17-1}, we can get
\begin{align}
 & \Pbb[d_{\mathsf{E}}\left(S^{n},\widecheck{s}^{n}\right)\leq D,S^{n}\in\mathcal{U}_{\delta}^{n},\frac{1}{n}\sum_{i=1}^{n}\jmath_{S}(S_{i},D)\geq R_{S}(D)-\delta|T_{S^{n}}=t]\nonumber \\
 & =\frac{\Pbb[d_{\mathsf{E}}\left(S^{n},\widecheck{s}^{n}\right)\leq D,S^{n}\in\mathcal{U}_{\delta}^{n},\frac{1}{n}\sum_{i=1}^{n}\jmath_{S}(S_{i},D)\geq R_{S}(D)-\delta,T_{S^{n}}=t]}{\Pbb[T_{S^{n}}=t]}\\
 & \leq\frac{\Pbb[d_{\mathsf{E}}\left(S^{n},\widecheck{s}^{n}\right)\leq D,\frac{1}{n}\sum_{i=1}^{n}\jmath_{S}(S_{i},D)\geq R_{S}(D)-\delta]}{2^{-no(1)}}.\label{eq:-29}
\end{align}

To complete the proof, we need the following lemma.
\begin{lem}
\label{lem:typebound-1-1} \cite{Yu-1} Assume $\mathcal{S}$ and
$\widecheck{\mathcal{S}}$ are general (not necessarily countable) alphabets,
and $S^{n}$ is i.i.d. drawn from $\mathcal{S}^{n}$ according to
$P_{S}$. Then for any $D>D_{\min}$ ($D_{\min}$ is defined in Definition
\ref{defn:id}) and any $\widecheck{s}^{n}\in\widecheck{\mathcal{S}}^{n}$,
\begin{equation}
\Pbb[d_{\mathsf{E}}\left(S^{n},\widecheck{s}^{n}\right)\leq D,\frac{1}{n}\sum_{i=1}^{n}\jmath_{S}(S_{i},D)\geq R_{S}(D)-\delta]\leq2^{-n(R_{S}(D)-o(1))}.\label{eq:rd-3}
\end{equation}
\end{lem}
\noindent Hence by the lemma above, \eqref{eq:-29} implies that
\begin{align}
 & \Pbb[d_{\mathsf{E}}\left(S^{n},\widecheck{s}^{n}\right)\leq D,S^{n}\in\mathcal{U}_{\delta}^{n},\frac{1}{n}\sum_{i=1}^{n}\jmath_{S}(S_{i},D)\geq R_{S}(D)-\delta|T_{S^{n}}=t]\leq2^{-n(R_{S}(D)-o(1))}.
\end{align}

\section{\label{sec:Proof-of-TheoremGauss}Proof of Theorem \ref{thm:proposedGaussian}}

Similar to the DM cases, it is easy to show the distortion constraints
for legitimate users are satisfied.

Next by following similar steps to the proof for the DM cases, we
prove the secrecy constraint is also satisfied. To that end, we first
need to discretize the source $S$ and the reconstruction $\widecheck{S}$.
Let
\begin{equation}
[S],[\widecheck{S}]\in\mathcal{N}\triangleq\left\{ \cdots,-2\Delta,-\Delta,0,\Delta,2\Delta,\cdots\right\} ,\label{eq:-9}
\end{equation}
be quantized versions of $S$ and $\widecheck{S}$, obtained by mapping
$S$ and $\widecheck{S}$ to the closest quantization point, i.e., $[S]=\Delta\cdot\textrm{Round}\left(\frac{S}{\Delta}\right),[\widecheck{S}]=\Delta\cdot\textrm{Round}\left(\frac{\widecheck{S}}{\Delta}\right)$.
Then we have for any $s^{n}\in\mathbb{R}^{n}$,
\begin{equation}
0\leq\frac{1}{n}\sum_{i=1}^{n}\left(s_{i}-[s_{i}]\right)^{2}\leq\frac{\Delta^{2}}{4}.
\end{equation}
Furthermore, it holds that
\begin{align}
\sqrt{nd(s^{n},\widecheck{s}^{n})} & =||s^{n}-\left[\widecheck{s}\right]^{n}+\left[\widecheck{s}\right]^{n}-\widecheck{s}^{n}||\\
 & \geq||s^{n}-\left[\widecheck{s}\right]^{n}||-||\left[\widecheck{s}\right]^{n}-\widecheck{s}^{n}||\label{eq:-22-1-1}\\
 & \geq||\left[s\right]^{n}-\left[\widecheck{s}\right]^{n}||-||s^{n}-\left[s\right]^{n}||-||\left[\widecheck{s}\right]^{n}-\widecheck{s}^{n}||\\
 & \geq\sqrt{nd(\left[s\right]^{n},\left[\widecheck{s}\right]^{n})}-\Delta
\end{align}
where \eqref{eq:-22-1-1} follows from triangle inequality. Utilizing
this inequality, we have
\begin{align}
\mathbb{P}\left[d(S^{n},\widecheck{s}^{n})\leq D_{0},\mathcal{A}|\Psi_{k},Z^{n}\right] & \leq\mathbb{P}\left[d\left([S]^{n},\left[\widecheck{s}\right]^{n}\right)\le\left(\sqrt{D_{0}}+\Delta\right)^{2},\mathcal{A}|\Psi_{k},Z^{n}\right]\\
 & =\mathbb{P}\left[d\left([S]^{n},\left[\widecheck{s}\right]^{n}\right)\le D'_{0},\mathcal{A}|\Psi_{k},Z^{n}\right],\label{eq:-25-2}
\end{align}
where $D'_{0}\triangleq\left(\sqrt{D_{0}}+\Delta\right)^{2}$.

Reorder the probabilities $P_{\left[S\right]}(\left[s\right]),\left[s\right]\in\mathcal{N}$
in decreasing order, and denote the result as $P_{i},i=1,2,...$.
Then $P_{1}=P_{\left[S\right]}(0),\,P_{2j}=P_{2j+1}=P_{\left[S\right]}(j\Delta),j\geq1$.
Obviously,
\begin{equation}
\Delta f_{S}\left((j+1)\Delta\right)\leq P_{2j}=P_{2j+1}\leq\Delta f_{S}\left(j\Delta\right),
\end{equation}
and hence for Gaussian sources, $P_{2j}=P_{2j+1}=o(e^{-j^{2}})$.
From Remark \ref{rem:The-conditions--}, we have that $P_{\left[S\right]}$
satisfies the conditions \eqref{eq:-46}-\eqref{eq:-47}.

Define event
\begin{align}
\mathcal{A}\triangleq & \Bigl\{ S^{n}\in\mathcal{W}_{\delta}^{n},[S]^{n}\in\mathcal{U}_{\delta}^{n}([S]),\frac{1}{n}\sum_{i=1}^{n}\jmath_{S}([S]_{i},D_{0})\geq R_{[S]}(D_{0})-\delta,\nonumber \\
 & \frac{1}{n}\sum_{i=1}^{n}\jmath_{[S]|Z=Z'_{i}}([S]_{i},b^{\star}(Z'_{i}))\geq R_{[S]|Z}(D_{0})-\delta,\frac{1}{n}\sum_{i=1}^{n}b^{\star}(Z'_{i})\geq D_{0}-\delta\Bigr\}
\end{align}
for $\delta>0$. Observe that the distribution $P_{\mathcal{C}S^{n}KS^{\prime n}X^{n}Y_{i}^{n}Z^{n}X^{\prime n}Y_{i}^{\prime n}Z^{\prime n}\widehat{S}_{i}^{\prime n}\widehat{S}_{i}^{n}}$
also satisfies \eqref{eq:-5} and \eqref{eq:-50}, which implies $\left(S^{n},Z'^{n}\right)$
is an i.i.d. sequence. Hence Lemma \ref{lem:A2} still holds for this
case. Following similar steps to the proof of Theorem \ref{thm:proposedDM},
we can get
\begin{align}
 & \mathbb{P}_{\mathcal{C}Z^{n}}\Bigl[\mathbb{P}\bigl[d_{\mathsf{E}}(S^{n},\widecheck{S}^{n})\le D_{0}|\mathcal{C}Z^{n}\bigr]>\tau_{n}\Bigr]\nonumber \\
 & \leq\mathbb{P}_{\mathcal{C}Z^{n}}\Bigl[\max_{\left[\widecheck{s}\right]^{n}\in\mathcal{N}^{n}}\sum_{k=1}^{2^{nR_{\mathsf{K}}}}\Pbb\left[d([S]^{n},\left[\widecheck{s}\right]^{n})\leq D'_{0},\mathcal{A}|\Psi_{k},Z^{n}\right]>\tau'_{n}2^{-n\left(R_{n}-R_{\mathsf{K}}\right)}\Bigr]+\epsilon_{n}^{\prime}\label{eq:-12}\\
 & =\mathbb{P}_{\mathcal{C}Z^{n}}\Bigl[\max_{\left[\widecheck{s}\right]^{n}\in\mathcal{B}^{n}}\sum_{k=1}^{2^{nR_{\mathsf{K}}}}\Pbb\left[d([S]^{n},\left[\widecheck{s}\right]^{n})\leq D'_{0},\mathcal{A}|\Psi_{k},Z^{n}\right]>\tau'_{n}2^{-n\left(R_{n}-R_{\mathsf{K}}\right)}\Bigr]+\epsilon_{n}^{\prime}\label{eq:-41}\\
 & \leq\left|\mathcal{B}^{n}\right|\max_{\left[\widecheck{s}\right]^{n}\in\mathcal{B}^{n}}\mathbb{P}_{\mathcal{C}Z^{n}}\Big[\sum_{k=1}^{2^{nR_{\mathsf{K}}}}\xi_{k,z^{n}}\left(\left[\widecheck{s}\right]^{n}\right)>\tau'_{n}2^{-n\left(R_{n}-R_{\mathsf{K}}\right)}\Big]+\epsilon_{n}^{\prime},\label{eq:-44}
\end{align}
where
\begin{equation}
\mathcal{B}^{n}\triangleq\left\{ [\widecheck{s}]^{n}\in\mathcal{N}^{n}:\left\Vert [\widecheck{s}]^{n}\right\Vert \leq\sqrt{n\Gamma}\right\} \label{eq:-18}
\end{equation}
with $\sqrt{\Gamma}\triangleq\sqrt{\lambda\left(1+\delta\right)}+\sqrt{\Delta}+\sqrt{D'_{0}}$,
\begin{align}
\xi_{k,z^{n}}\left(\left[\widecheck{s}\right]^{n}\right) & \triangleq\Pbb\left[d\left([S]^{n},\left[\widecheck{s}\right]^{n}\right)\leq D'_{0},\mathcal{A}|\Psi_{k},Z^{n}\right],
\end{align}
and \eqref{eq:-41} follows that $S^{n}$ only appears in the ball
with radius $\sqrt{n\lambda\left(1+\delta\right)}$ which implies
$[S]^{n}$ only appears in the ball with radius $\sqrt{\lambda\left(1+\delta\right)}+\sqrt{\Delta}$
, hence it is sufficient to only consider $\left[\widecheck{s}\right]^{n}\in\mathcal{B}^{n}$
instead of the whole set $\mathcal{N}^{n}$.

Furthermore, observe that
\begin{align}
\left|\mathcal{B}^{n}\right| & \leq\frac{\textrm{Volume of \ensuremath{n-}ball with radius }\sqrt{n\Gamma}+\sqrt{n\Delta^{2}}}{\Delta^{n}}\\
 & =\frac{\pi^{n/2}\left(\sqrt{n\Gamma}+\sqrt{n\Delta^{2}}\right)^{n}}{\Delta^{n}\Gamma\left(\frac{n}{2}+1\right)}\\
 & \leq2^{\frac{n}{2}\log\pi+n\log\left(\sqrt{n\Gamma}+\sqrt{n\Delta^{2}}\right)-n\log\Delta-\frac{n}{2}\log\frac{n}{2e}-\frac{1}{2}\log\pi n+o\left(1\right)}\\
 & =2^{\frac{n}{2}\log\frac{2\pi e\left(\sqrt{\Gamma}+\sqrt{\Delta^{2}}\right)^{2}}{\Delta^{2}}-\frac{1}{2}\log\pi n+o\left(1\right)}.\label{eq:-37-1}
\end{align}

Therefore, if we can show that the probability in \eqref{eq:-44}
decays doubly exponentially fast with $n$, then the proof will be
complete.

Apply Lemmas \ref{lem:chernoff}, \ref{lem:SPO} and \ref{lem:SPO-si},
then we have the probability in \eqref{eq:-44} decays doubly exponentially
fast with $m$. Hence $\mathop{\lim}\limits _{n\to\infty}\mathbb{E}_{\mathcal{C}Z^{n}}\Bigl[\max_{R_{n}\mathsf{Hcodes}}\mathbb{P}\bigl[d_{\mathsf{E}}(S^{n},\widecheck{S}^{n})\le D_{0}\bigr]\Bigr]=0$
if
\begin{equation}
\limsup_{n\rightarrow\infty}R_{n}<\min\left\{ R_{\mathsf{K}}+R_{[S]|Z}(D'_{0}),R_{[S]}(D'_{0})\right\} .\label{eq:-3-1-2}
\end{equation}

To complete the proof, we need to show $R_{[S]|Z}(D'_{0})\geq R_{S|Z}(D{}_{0})$
and $R_{[S]}(D'_{0})\geq R_{S}(D{}_{0})$ as $\Delta\rightarrow0$.
Suppose $P_{[\widecheck{S}]^{*}|[S]Z}$ achieves $R_{[S]|Z}(D'_{0})$,
then $R_{[S]|Z}(D'_{0})=I([S];[\widecheck{S}]^{*}|Z)$ and $\mathbb{E}d([S],[\widecheck{S}]^{*})\le D'_{0}$.
Since for $P_{[S][\widecheck{S}]^{*}|SZ}=P_{[S]|S}P_{[\widecheck{S}]^{*}|[S]Z}$,
\begin{align}
\mathbb{E}d(S,[\widecheck{S}]^{*}) & =\mathbb{E}\left(S-[S]+[S]-[\widecheck{S}]^{*}\right)^{2}\\
 & =\mathbb{E}\left([S]-[\widecheck{S}]^{*}\right)^{2}+\mathbb{E}\left(S-[S]\right)^{2}+2\mathbb{E}\left(S-[S]\right)\left([S]-[\widecheck{S}]^{*}\right)\label{eq:-22-1-1-1-2}\\
 & \leq\mathbb{E}d([S],[\widecheck{S}]^{*})+\Delta\sqrt{\mathbb{E}d([S],[\widecheck{S}]^{*})}\label{eq:-4}\\
 & \le D'_{0}+\Delta\sqrt{D'_{0}},
\end{align}
where \eqref{eq:-4} follows from the Cauchy\textendash Schwarz inequality.
On the other hand, $R_{S|Z}\left(D'_{0}+\Delta\sqrt{D'_{0}}\right)$
is defined as the minimum $I(S;\widecheck{S}|Z)$ such that $\mathbb{E}d(S,\widecheck{S})\le D'_{0}+\Delta\sqrt{D'_{0}}.$
Hence $R_{S|Z}\left(D'_{0}+\Delta\sqrt{D'_{0}}\right)\leq I(S;[\widecheck{S}]^{*}|Z)=I([S];[\widecheck{S}]^{*}|Z)=R_{[S]|Z}(D'_{0})$.
Let $\Delta\rightarrow0$, then we have $R_{S|Z}(D{}_{0})\leq\lim_{\Delta\rightarrow0}R_{[S]|Z}(D'_{0})$.
Similarly, we can prove $R_{S}(D{}_{0})\leq\lim_{\Delta\rightarrow0}R_{[S]}(D'_{0})$.
This completes the proof of Theorem \ref{thm:proposedGaussian}.

\section{\label{sec:Proof-of-TheoremGauss2}Proof of Theorem \ref{thm:proposedGaussian2}}

Denote $X'^{n},Y_{i}^{\prime n},Z'^{n}$ as \eqref{eq:-53}-\eqref{eq:-52},
where $\Psi_{K}\left(\cdot\right)$ denotes the orthogonal transform,
instead of the permutation operation. Then it can be verified that
for Gaussian source-channel pair, the distribution $P_{\mathcal{C}S^{n}KS^{\prime n}X^{n}Y_{i}^{n}Z^{n}X^{\prime n}Y_{i}^{\prime n}Z^{\prime n}\widehat{S}_{i}^{\prime n}\widehat{S}_{i}^{n}}$
also satisfies \eqref{eq:-5} and \eqref{eq:-50}. Hence $\left(\Psi_{K},Z^{n}\right)\rightarrow Z'^{n}\rightarrow S^{n}$
forms a Markov chain.

Similar to the permutation-based scheme, it is easy to show the power
constraint and the distortion constraints for legitimate users are
satisfied. Next by following similar steps to the proof for the permutation-based
scheme, we prove the secrecy constraint is also satisfied. But a slight
difference is that here we use an argument from a geometric point
of view, instead of the one from the view of rate-distortion theory
(or method of types) used for Theorems \ref{thm:proposedDM}, \ref{thm:proposedGDM},
and \ref{thm:proposedGaussian}. Here we do not need to discretize
the source. But we still need to discretize the reconstruction $\widecheck{S}$
as \eqref{eq:-9}, since it will enable us to take the maximizing
operation out of the probability, just as done in  \eqref{eq:-12}-\eqref{eq:-44}.

Define event
\begin{align}
 & \mathcal{A}\triangleq\left\{ \left(S^{n},Z'^{n}\right)\in\Wcal_{\delta}^{n}\left(S,Z\right)\right\} ,
\end{align}
for $\delta>0$. For jointly Gaussian variables $X$ and $Z$, where
$Z=X+U$ and $U$ is independent of $X$, the $\delta$-weakly typical
set and the $\delta$-weakly jointly typical set become

\begin{equation}
\Wcal_{\delta}^{n}(X)\triangleq\Bigl\{ x^{n}\in\Xcal^{n}:\Bigl|\frac{\left\Vert x^{n}\right\Vert ^{2}}{nN_{X}}-1\Bigr|\leq\delta\Bigr\},
\end{equation}
and
\begin{align}
\Wcal_{\delta}^{n}(X,Z)\triangleq\Bigl\{ & \left(x^{n},z^{n}\right)\in\mathbb{R}^{2n}:\Bigl|\frac{\left\Vert x^{n}\right\Vert ^{2}}{nN_{X}}-1\Bigr|\leq\delta,\label{eq:-57}\\
 & \Bigl|\frac{\left\Vert z^{n}\right\Vert ^{2}}{nN_{Z}}-1\Bigr|\leq\delta,\\
 & \Bigl|\frac{\left\Vert x^{n}\right\Vert ^{2}}{nN_{X}}+\frac{\left\Vert z^{n}-x^{n}\right\Vert ^{2}}{nN_{U}}-2\Bigr|\leq\delta\Bigr\},\label{eq:-58}
\end{align}
respectively, where $N_{Z},N_{X}$ and $N_{U}$ denote the variances
of $Z$, $X$ and $U$, respectively.

Since $\left(S^{n},Z'^{n}\right)$ is an i.i.d. sequence, from the
fact that (weakly) typical set has total probability close to one
\cite{Cover}, we have the following lemma.
\begin{lem}
\label{lem:A2-1} \cite{Cover} For any $\delta>0$, $\mathbb{P}\left[\mathcal{A}^{c}\right]\to0$,
as $n\to\infty$.
\end{lem}
Following similar steps to the proof of Theorem \ref{thm:proposedGaussian},
we can get
\begin{align}
 & \mathbb{P}_{\mathcal{C}Z^{n}}\Bigl[\mathbb{P}\bigl[d(S^{n},\widecheck{S}^{n})\le D_{0}|\mathcal{C}Z^{n}\bigr]>\tau_{n}\Bigr]\nonumber \\
 & \leq\left|\mathcal{B}^{n}\right|\max_{\left[\widecheck{s}\right]^{n}\in\mathcal{B}^{n}}\mathbb{P}_{\mathcal{C}Z^{n}}\Big[\sum_{k=1}^{2^{nR_{\mathsf{K}}}}\xi_{k,z^{n}}\left(\left[\widecheck{s}\right]^{n}\right)>\tau'_{n}2^{-n\left(R_{n}-R_{\mathsf{K}}\right)}\Big]+\epsilon_{n}^{\prime},\label{eq:-44-1}
\end{align}
where $\mathcal{B}^{n}$ is given by \eqref{eq:-18}, and
\begin{align}
\xi_{k,z^{n}}\left(\left[\widecheck{s}\right]^{n}\right) & \triangleq\Pbb\left[d\left(S^{n},\left[\widecheck{s}\right]^{n}\right)\leq D'_{0},\mathcal{A}|\Psi_{k},Z^{n}\right].
\end{align}

Since as shown in \eqref{eq:-37-1}, $\left|\mathcal{B}^{n}\right|$ is upper bounded by an exponential function, we only need to show that the probability
in \eqref{eq:-44-1} decays doubly exponentially fast with $n$. To
that end, we need introduce the following lemmas. The proofs of Lemmas
\ref{lem:typebound-rd} and \ref{lem:typebound-rd-si} are given in
Appendixes \ref{sec:Proof-of-Lemma-typebound-rd-si} and \ref{sec:Proof-of-Lemma-typebound-rd},
respectively.
\begin{lem}
\label{lem:typebound-rd-si} Assume $S^{n}=Z^{n}+U^{n}$, where
$Z^{n}\sim\mathcal{N}\left(\mathbf{0},N_{Z}\boldsymbol{I}\right)$
and $U^{n}\sim\mathcal{N}\left(\mathbf{0},N_{U}\boldsymbol{I}\right)$\footnote{Here $\mathbf{0}$ denotes an all-zero vector and $\boldsymbol{I}$ denotes an identity matrix}
are independent, then for any $z^{n},\bar{s}^{n}\in\mathbb{R}{}^{n}$,
\begin{equation}
\Pbb[d\left(S^{n},\bar{s}^{n}\right)\leq D,\left(S^{n},z^{n}\right)\in\mathcal{W}_{\delta}^{n}|z{}^{n}]\leq2^{-n(R_{S|Z}(D)-o(1))},\label{eq:rd-1-1-1-1}
\end{equation}
where $R_{S|Z}(D)=\frac{1}{2}\log^{+}\left(\frac{N_{U}}{D}\right)$,
and $o(1)$ is a term that vanishes as $\delta\rightarrow0$ and $n\rightarrow\infty$.
\end{lem}
\begin{lem}
\label{lem:typebound-rd} Assume $S^{n}\sim\mathcal{N}\left(\mathbf{0},N_{S}\boldsymbol{I}\right)$
and $S'^{n}=\Psi S^{n}$, with $\Psi$ uniformly distributed on orthogonal
matrices set and independent of $S^{n}$, then for any $s'^{n},\bar{s}^{n}\in\mathbb{R}{}^{n}$,
\begin{equation}
\Pbb[d\left(S^{n},\bar{s}^{n}\right)\leq D,S^{n}\in\mathcal{W}_{\delta}^{n}|s'^{n}]\leq2^{-n(R_{S}(D)-o(1))},\label{eq:rd-1-1-1-1-1}
\end{equation}
where $R_{S}(D)=\frac{1}{2}\log^{+}\left(\frac{N_{S}}{D}\right)$,
and $o(1)$ is a term that vanishes as $\delta\rightarrow0$ and $n\rightarrow\infty$.
\end{lem}
Then we have
\begin{align}
\begin{array}{c}
\mathbb{E}_{\mathcal{C}}\xi_{k,z^{n}}\left(\left[\widecheck{s}\right]^{n}\right)\end{array} & \leq\Ebb_{\Psi_{k}}\Pbb[d\left(S^{n},\left[\widecheck{s}\right]^{n}\right)\leq D'_{0},S^{n}\in\mathcal{W}_{\delta}^{n}|\Psi_{k},z^{n}]\\
 & =\Ebb_{\Psi_{k}}\int\Pbb[d\left(S^{n},\left[\widecheck{s}\right]^{n}\right)\leq D'_{0},S^{n}\in\mathcal{W}_{\delta}^{n}|\Psi_{k},z^{n},s'^{n}]f_{S'^{n}|Z^{n}}\left(s'^{n}|\Psi_{k},z^{n}\right)ds'^{n}\\
 & =\int\Pbb_{\Psi_{k}}[d\left(S^{n},\left[\widecheck{s}\right]^{n}\right)\leq D'_{0},S^{n}\in\mathcal{W}_{\delta}^{n}|s'^{n}]f_{S'^{n}|Z^{n}}\left(s'^{n}|z^{n}\right)ds'^{n}\\
 & \leq2^{-n(R_{S}(D'_{0})-o(1))},\label{eq:-16-1}
\end{align}
where \eqref{eq:-16-1} follows from Lemma \ref{lem:typebound-rd}.
Furthermore, Lemma \ref{lem:typebound-rd-si} implies
\begin{align}
\begin{array}{c}
\xi_{k,z^{n}}\left(\left[\widecheck{s}\right]^{n}\right)\end{array} & =\Pbb\left[d\left(S^{n},\left[\widecheck{s}\right]^{n}\right)\leq D'_{0},\mathcal{A}|z'^{n}\right]\label{eq:-43}\\
 & \leq2^{-n(R_{S|Z}(D'_{0})-o(1))},
\end{align}
where \eqref{eq:-16-1} follows from $\Psi_{K}Z^{n}\rightarrow Z'^{n}\rightarrow S^{n}$.

Applying Lemmas \ref{lem:chernoff}, we have that the probability
in \eqref{eq:-44-1} decays doubly exponentially fast with $n$. This
completes the proof of Theorem \ref{thm:proposedGaussian2}.

\section{\label{sec:Proof-of-Lemma-typebound-rd-si}Proof of Lemma \ref{lem:typebound-rd-si}}

Consider that
\begin{align}
 & \Pbb[d\left(S^{n},\bar{s}^{n}\right)\leq D,\left(S^{n},z{}^{n}\right)\in\mathcal{W}_{\delta}^{n}|z{}^{n}]\nonumber \\
 & \leq\Pbb[d\left(S^{n},\bar{s}^{n}\right)\leq D,\frac{\left\Vert z^{n}\right\Vert ^{2}}{nN_{Z}}\in1\pm\delta,\frac{\left\Vert z^{n}\right\Vert ^{2}}{nN_{Z}}+\frac{\left\Vert S^{n}-z{}^{n}\right\Vert ^{2}}{nN_{U}}\in2\pm\delta|z{}^{n}]\\
 & \leq\Pbb[d\left(S^{n},\bar{s}^{n}\right)\leq D,\frac{\left\Vert S^{n}-z{}^{n}\right\Vert ^{2}}{nN_{U}}\in1\pm2\delta|z{}^{n}].\label{eq:}
\end{align}
Denote $R=\frac{1}{n}\left\Vert S^{n}-z{}^{n}\right\Vert ^{2}$, then
\begin{align}
 & \Pbb[d\left(S^{n},\bar{s}^{n}\right)\leq D,\left(S^{n},z{}^{n}\right)\in\mathcal{W}_{\delta}^{n}|z{}^{n}]\nonumber \\
 & \leq\int_{N_{U}\left(1-2\delta\right)}^{N_{U}\left(1+2\delta\right)}f_{R|Z^{n}}\left(r|z^{n}\right)\Pbb[d\left(S^{n},\bar{s}^{n}\right)\leq D|R=r,Z{}^{n}=z{}^{n}]dr.\label{eq:-8}
\end{align}

\begin{figure}
\centering \includegraphics[width=0.45\textwidth]{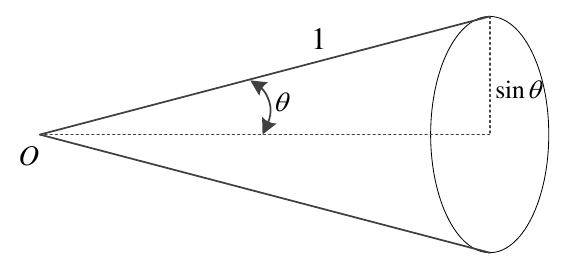} \protect\caption{\label{fig:cap}Cap cut out by the cone on the unit sphere\textit{.}}
\end{figure}

Given $Z^{n}=z{}^{n}$, $S^{n}-z{}^{n}\sim\mathcal{N}\left(\mathbf{0},N_{U}\boldsymbol{I}\right)$,
and on the other hand, Gaussian distribution is isotropic (or invariant
to rotation). Hence under condition of $Z^{n}=z{}^{n}$ and $R=r$,
$S^{n}$ is uniformly distributed over the sphere with center $z^{n}$
and radius $\sqrt{nr}$.
\begin{align}
 & \Pbb[d\left(S^{n},\bar{s}^{n}\right)\leq D|R=r,Z{}^{n}=z{}^{n}]\leq\frac{\Omega\left(\theta\right)}{\Omega\left(\pi\right)},\label{eq:-1}
\end{align}
where
\begin{equation}
\theta=\arcsin\sqrt{\frac{D}{r}},\label{eq:-13}
\end{equation}
and $\Omega\left(\theta\right)$ be solid angle in $n$ space of a
cone with half-angle $\theta$, i.e., the area of a spherical cap
on a unit sphere (see Fig. \ref{fig:cap}). To approximate $\frac{\Omega\left(\theta\right)}{\Omega\left(\pi\right)}$,
we need the following lemma.

\begin{lem}
\label{lem:sphericalcap}\cite{Shannon59} Let $\Omega\left(\theta\right)$
be solid angle in $n$ space of a cone with half-angle $\theta$,
then it holds that
\begin{equation}
\frac{\Omega\left(\theta\right)}{\Omega\left(\pi\right)}=\frac{\sin^{n-1}\theta}{\sqrt{2\pi n}\cos\theta}\left(1+O\left(\frac{1}{n}\right)\right).
\end{equation}
\end{lem}
Lemma \ref{lem:sphericalcap} implies
\begin{align}
\frac{\Omega\left(\theta\right)}{\Omega\left(\pi\right)} & =2^{n\left(\log\sin\theta-\frac{1}{n}\log\left(\sqrt{2\pi n}\sin\theta\cos\theta\right)+\frac{1}{n}\log\left(1+O\left(\frac{1}{n}\right)\right)\right)}\\
 & =2^{n\left(\log\sin\theta+o(1)\right)}.\label{eq:-14}
\end{align}
Combine \eqref{eq:-1}, \eqref{eq:-13} and \eqref{eq:-14}, then
we have for any $r\in N_{U}\left(1\pm2\delta\right)$,
\begin{align}
\Pbb[d\left(S^{n},\bar{s}^{n}\right)\leq D|R=r,Z{}^{n}=z{}^{n}] & \leq2^{n\left(\log\sqrt{\frac{D}{r}}+o(1)\right)}\\
 & \leq2^{n\left(\log\sqrt{\frac{D}{N_{U}\left(1-2\delta\right)}}+o(1)\right)}\label{eq:-7}\\
 & =2^{-n\left(\frac{1}{2}\log\frac{N_{U}}{D}-o(1)\right)},\label{eq:-10}
\end{align}
where $o(1)$ is a term that vanishes as $\delta\rightarrow0$ and
$n\rightarrow\infty$. Combining \eqref{eq:-8} and \eqref{eq:-10}
gives us $\Pbb[d\left(S^{n},\bar{s}^{n}\right)\leq D,\left(S^{n},z{}^{n}\right)\in\mathcal{W}_{\delta}^{n}|z{}^{n}]\leq2^{-n\left(\frac{1}{2}\log\frac{N_{U}}{D}-o(1)\right)}$.
This completes the proof of Lemma \ref{lem:typebound-rd-si}.

\section{\label{sec:Proof-of-Lemma-typebound-rd}Proof of Lemma \ref{lem:typebound-rd}}

Observe that
\begin{align}
 & \Pbb[d\left(S^{n},\bar{s}^{n}\right)\leq D,S^{n}\in\mathcal{W}_{\delta}^{n}|s'^{n}]\nonumber \\
 & =\Pbb[d\left(S^{n},\bar{s}^{n}\right)\leq D,\frac{\left\Vert S^{n}\right\Vert ^{2}}{nN_{S}}\in1\pm\delta|s'^{n}]\\
 & =\Pbb_{\Psi}[d\left(\Psi^{T}s'^{n},\bar{s}^{n}\right)\leq D|s'^{n}]1\left\{ \frac{\left\Vert s'^{n}\right\Vert ^{2}}{nN_{S}}\in1\pm\delta\right\} .\label{eq:-16}
\end{align}

\noindent Since $\Psi$ is uniformly distributed on orthogonal matrices
set (so is $\Psi^{T}$ as stated in Lemma \eqref{lem:randmat}), Lemma
\ref{lem:Random-matrix} implies that for any $s'^{n}$, $\Psi^{T}s'^{n}$
is uniformly distributed over the sphere with center at the origin
$O$ and radius $\left\Vert s'^{n}\right\Vert $. Hence
\begin{align}
 & \Pbb_{\Psi}[d\left(\Psi^{T}s'^{n},\bar{s}^{n}\right)\leq D|s'^{n}]\leq\frac{\Omega\left(\theta\right)}{\Omega\left(\pi\right)},\label{eq:-11}
\end{align}
where as described in the previous section, $\Omega\left(\theta\right)$
denotes solid angle in $n$ space of a cone with half-angle $\theta$,
and
\begin{equation}
\theta=\arcsin\sqrt{\frac{D}{\frac{1}{n}\left\Vert s'^{n}\right\Vert ^{2}}}.\label{eq:-13-2}
\end{equation}

From Lemma \ref{lem:sphericalcap}, we have for any $s'^{n}$ such
that $\frac{\left\Vert s'^{n}\right\Vert ^{2}}{nN_{S}}\in1\pm\delta$,
\begin{align}
\Pbb_{\Psi}[d\left(\Psi^{T}s'^{n},\bar{s}^{n}\right)\leq D|s'^{n}] & \leq2^{n\left(\log\sqrt{\frac{D}{\frac{1}{n}\left\Vert s'^{n}\right\Vert ^{2}}}+o(1)\right)}\\
 & \leq2^{n\left(\log\sqrt{\frac{D}{N_{S}\left(1-\delta\right)}}+o(1)\right)}\\
 & \leq2^{-n\left(\frac{1}{2}\log\frac{N_{S}}{D}+o(1)\right)},\label{eq:-15}
\end{align}
where $o(1)$ is a term that vanishes as $\delta\rightarrow0$ and
$n\rightarrow\infty$. Combining \eqref{eq:-16} and \eqref{eq:-15}
gives us
\begin{align}
 & \Pbb[d\left(S^{n},\bar{s}^{n}\right)\leq D,S^{n}\in\mathcal{W}_{\delta}^{n}|s'^{n}]\leq2^{-n\left(\frac{1}{2}\log\frac{N_{S}}{D}+o(1)\right)}.
\end{align}
This completes the proof of Lemma \ref{lem:typebound-rd}.

\section{\label{sec:Proof-of-Lemma-ub}Proof of Lemma \ref{lem:upperbound}}

By choosing $\widecheck{S}$ to be independent of $Z$, we have $R_{S|Z}(D_{0})=\min\limits _{P_{\widecheck{S}|SZ}:\mathbb{E}d(S,\widecheck{S})\le D_{0}}I(S;\widecheck{S}|Z)\leq\min\limits _{P_{\widecheck{S}|S}:\mathbb{E}d(S,\widecheck{S})\le D_{0}}I(S;\widecheck{S})=R_{S}(D_{0})=\frac{1}{2}\log^{+}\left(\frac{\lambda}{D_{0}}\right)$.
Then we only need to prove $R_{S|Z}(D_{0})\leq R_{S|Z}^{\mathsf{(UB)}}(D_{0})$.

First consider the case of $\frac{\lambda N_{0}}{P^{\prime}+N_{0}}\leq D_{0}\leq\lambda$.
Assume
\begin{equation}
Q=\begin{cases}
1, & \textrm{with probability }p;\\
0, & \textrm{with probability }1-p,
\end{cases}
\end{equation}
independent of $\left(S,Z\right)$, denotes a timesharing random variable,
and also assume
\begin{equation}
\widecheck{S}_{Q}=\begin{cases}
\beta_{0}\Psi_{K}Z, & \textrm{if }Q=1;\\
0, & \textrm{if }Q=0,
\end{cases}
\end{equation}
where $\beta_{0}=\frac{\sqrt{\lambda P^{\prime}}}{P^{\prime}+N_{0}}$
and $\Psi_{K}$ is defined in \eqref{eq:tk}. Then
\begin{align}
\mathbb{E}d(S,\widecheck{S}_{Q}) & =\mathbb{E}_{Q}\mathbb{E}\left[d(S,\widecheck{S}_{Q})|Q\right]\\
 & =p\frac{\lambda N_{0}}{P^{\prime}+N_{0}}+\left(1-p\right)\lambda.
\end{align}
Therefore, to satisfy distortion constraint $\mathbb{E}d(S,\widecheck{S}_{Q})\leq D_{0}$,
it is sufficient to set $p=\frac{\left(\lambda-D_{0}\right)\left(P^{\prime}+N_{0}\right)}{\lambda P^{\prime}}$.
Substituting $\widecheck{S}_{Q}$ into $R_{S|Z}(D_{0})$, we have
\begin{align}
R_{S|Z}(D_{0}) & \leq I(S;\widecheck{S}_{Q}|Z)\\
 & \leq I(S;\widecheck{S}_{Q}Q|Z)\\
 & =I(S;\widecheck{S}_{Q}|QZ)\label{eq:-31}\\
 & =pI(S;\widecheck{S}_{Q}|Z,Q=1)+\left(1-p\right)I(S;\widecheck{S}_{Q}|Z,Q=0)\\
 & =pI(S;\widecheck{S}_{Q}|Z,Q=1)\\
 & \leq pI(S;K|Z,Q=1)\\
 & \leq pH(K)\\
 & =p\label{eq:-32}\\
 & =\frac{\left(\lambda-D_{0}\right)\left(P^{\prime}+N_{0}\right)}{\lambda P^{\prime}},
\end{align}
where \eqref{eq:-31} follows from $Q$ is independent of $\left(S,Z\right)$.

Next consider the case of $0\leq D_{0}\leq\frac{\lambda N_{0}}{P^{\prime}+N_{0}}$.
Observe that
\begin{align}
R_{S|Z}(D_{0}) & =\min\limits _{P_{\widecheck{S}|SZ}:\mathbb{E}d(S,\widecheck{S})\le D_{0}}I(S;\widecheck{S}|Z)\label{eq:-59}\\
 & =\min\limits _{P_{\widecheck{S}|SZK}:\mathbb{E}d(S,\widecheck{S})\le D_{0}}I(S;\widecheck{S}|Z)\label{eq:-51}\\
 & \leq\min\limits _{P_{\widecheck{S}|SZK}:\mathbb{E}d(S,\widecheck{S})\le D_{0}}I(S;\widecheck{S}K|Z)\\
 & =I(S;K|Z)+\min\limits _{P_{\widecheck{S}|SZK}:\mathbb{E}d(S,\widecheck{S})\le D_{0}}I(S;\widecheck{S}|ZK)\\
 & \leq H(K)+\min\limits _{P_{\widecheck{S}|SZK}:\mathbb{E}d(S,\widecheck{S})\le D_{0}}I(S;\widecheck{S}|ZK),\label{eq:-21}
\end{align}
where \eqref{eq:-51} follows since, on one hand, by setting $P_{\widecheck{S}|SZK}$
in \eqref{eq:-51} as $P_{\widecheck{S}|SZ}$ we have \eqref{eq:-59}$\geq$\eqref{eq:-51};
on the other hand, given $P_{SZK}$, both the constraint and the optimization
objective only depend on $P_{\widecheck{S}|SZ}=\sum_{k}P_{\widecheck{S}|SZK}P_{K|SZ}$,
hence it suffices to optimize \eqref{eq:-51} over $P_{\widecheck{S}|SZ}$.

The first term of \eqref{eq:-21} satisfies
\begin{equation}
H(K)=1.\label{eq:-22}
\end{equation}
By \eqref{eq:-19} and \eqref{eq:-20}, we have $S$ and $\Psi_{K}Z$
are jointly Gaussian, i.e.,
\begin{equation}
S=\beta_{0}\Psi_{K}Z+V_{0}'
\end{equation}
where $\beta_{0}=\frac{\sqrt{\lambda P^{\prime}}}{P^{\prime}+N_{0}}$,
$\Psi_{K}$ is defined in \eqref{eq:tk}, and $V_{0}'\sim\mathcal{N}\left(0,\frac{\lambda N_{0}}{P^{\prime}+N_{0}}\right)$
is independent of $\Psi_{K}Z$. Hence we can also write
\begin{equation}
\widecheck{S}^{*}=\beta_{0}\Psi_{K}Z+V_{0}''
\end{equation}
and
\begin{equation}
S=\widecheck{S}^{*}+\Delta V_{0}'',
\end{equation}
where $V_{0}''\sim\mathcal{N}\left(0,\frac{\lambda N_{0}}{P^{\prime}+N_{0}}-D_{0}\right)$
and $\Delta V_{0}''\sim\mathcal{N}\left(0,D_{0}\right)$ are independent
of each other and also independent of $\Psi_{K}Z$. Therefore, we
can bound the second term  in \eqref{eq:-21} as
\begin{align}
\min\limits _{P_{\widecheck{S}|SZK}:\mathbb{E}d(S,\widecheck{S})\le D_{0}}I(S;\widecheck{S}|ZK) & \leq I(S;\widecheck{S}^{*}|ZK)\label{eq:-60}\\
 & =h(S|ZK)-h(S|ZK\widecheck{S}^{*})\\
 & =h(S|K)+h(Z|SK)-h(Z|K)-h(S-\widecheck{S}^{*}|ZK\widecheck{S}^{*})\\
 & \leq\frac{1}{2}\log2\pi e\lambda+\frac{1}{2}\log2\pi eN_{0}-\frac{1}{2}\log2\pi e(P^{\prime}+N_{0})-h(S-\widecheck{S}^{*})\label{eq:-61}\\
 & \leq\frac{1}{2}\log\left(\frac{\lambda N_{0}}{D_{0}\left(P^{\prime}+N_{0}\right)}\right),\label{eq:-23}
\end{align}
where \eqref{eq:-60} follows since $P_{\widecheck{S}^{*}|SZK}$ satisfies
the constraint $\mathbb{E}d(S,\widecheck{S}^{*})\le D_{0}$, and \eqref{eq:-61}
follows since $\Psi_{K}Z\rightarrow\widecheck{S}^{*}\rightarrow S$ forms
a Markov chain and $S-\widecheck{S}^{*}$ is independent of $\widecheck{S}^{*}$.

Combining \eqref{eq:-21}, \eqref{eq:-22} and \eqref{eq:-23} gives
us $R_{S|Z}(D_{0})\leq R_{S|Z}^{\mathsf{(UB)}}(D_{0})$. This completes the
proof of Lemma \ref{lem:upperbound}.

\section{\label{sec:Proof-of-TheoremVGauss}Proof of Theorem \ref{thm:vector}}

Following similar steps to the proof for the scalar Gaussian case,
it is easy to prove $\mathcal{R}^{\mathsf{(i)}}$ is achievable by the permutation
based scheme. However, for the orthogonal-transform based scheme,
the proof for the scalar Gaussian case cannot be applied to the vector
Gaussian case directly, and some details need to be treated specially.
Next we give a proof for this case.

Following similar steps to the proof of Theorem \ref{thm:proposedGaussian2},
it can be shown that for vector Gaussian case, the distortion constraints
and power constraint are satisfied for the tuples given in Theorem
\ref{thm:vector}. Next we prove the secrecy constraint is also satisfied.

Define events
\begin{align}
 & \mathcal{A}_{j}\triangleq\left\{ \left(S_{j}^{n},Z'{}_{j}^{n}\right)\in\mathcal{W}_{\delta}^{n}\left(S_{j},Z_{j}\right)\right\} ,\\
 & \mathcal{A}\triangleq\prod_{j\in[m]}\mathcal{A}_{j},
\end{align}
for $\delta>0$. Similar to Lemma \ref{lem:A2-1}, it can be shown
that for any $\delta>0$, $\Pbb\left[\mathcal{A}_{j}^{c}\right]\to0$,
as $n\to\infty$.

The derivation up to \eqref{eq:-44-1} still holds for vector Gaussian
case. Hence
\begin{align}
 & \mathbb{P}_{\mathcal{C}\boldsymbol{Z}^{n}}\Bigl[\max_{R_{n}\mathsf{Hcodes}}\mathbb{P}\bigl[d(\boldsymbol{S}^{n},\boldsymbol{\widecheck{S}}^{n})\le D_{0}|\mathcal{C}\boldsymbol{Z}^{n}\bigr]>\tau_{n}\Bigr]\nonumber \\
 & \leq\mathbb{P}_{\mathcal{C}\boldsymbol{Z}^{n}}\Bigl[\max_{\boldsymbol{\widecheck{s}}^{n}\in\mathbb{R}^{mn}}\sum_{k=1}^{2^{nR_{\mathsf{K}}}}\Pbb\left[d(\boldsymbol{S}^{n},\left[\boldsymbol{\widecheck{s}}\right]^{n})\leq D'_{0},\mathcal{A}|\Psi_{j,k},j\in[m],\boldsymbol{Z}^{n}\right]>\tau'_{n}2^{-n\left(R_{n}-R_{\mathsf{K}}\right)}\Bigr]\Biggr]+\epsilon_{n}^{\prime},\label{eq:-26-1}
\end{align}
where $D'_{0}\triangleq\left(\sqrt{D_{0}}+m\Delta\right)^{2}$.

Observe that $\Pbb\left[d(\boldsymbol{S}^{n},\left[\boldsymbol{\widecheck{s}}\right]^{n})\leq D_{0},\mathcal{A}|\Psi_{j,k},j\in[m],\boldsymbol{Z}^{n}\right]=\int_{\sum_{j=1}^{m}d(s_{j}^{n},\left[\widecheck{s}_{j}\right]^{n})\leq D'_{0},\mathcal{A}}\prod_{j=1}^{m}f\left(s_{j}^{n}|\Psi_{j,k},z_{j}^{n}\right)ds_{j}^{n}$.
One may expect to exchange $\int$ with $\prod$, in order to write
the expression as $\prod_{j=1}^{m}\Pbb\left[d(S_{j}^{n},\left[\widecheck{s}_{j}\right]^{n})\leq d_{j},\mathcal{A}_{j}|\Psi_{j,k},Z_{j}^{n}\right]$
for some $d_{j},j\in[m]$ such that $\sum_{j=1}^{m}d_{j}\leq D'_{0}$.
However, obviously this is not feasible. To address this problem,
we need to discretize the source, and then eliminate the $\int$ operation
since after discretization it becomes a $\sum$ operation with the
number of summands polynomial in $n$.

Discretize $S$ by $\left[S\right]=\Delta\cdot\textrm{Round}\left(\frac{S}{\Delta}\right)$.
Then we have
\begin{align}
 & \mathbb{P}\left[d(\boldsymbol{S}^{n},\left[\boldsymbol{\widecheck{s}}\right]^{n})\leq D'_{0},\mathcal{A}|\Psi_{j,k},j\in[m],\boldsymbol{Z}^{n}\right]\nonumber \\
 & \leq\mathbb{P}\left[d\left(\left[\boldsymbol{S}\right]^{n},\left[\boldsymbol{\widecheck{s}}\right]^{n}\right)\le D''_{0},\mathcal{A}|\Psi_{j,k},j\in[m],\boldsymbol{Z}^{n}\right],\label{eq:-25-1}
\end{align}
where $D''_{0}\triangleq\left(\sqrt{D'_{0}}+m\Delta\right)^{2}$.
In addition, observe that
\begin{align}
 & \mathbb{P}\left[d\left(\left[\boldsymbol{S}\right]^{n},\left[\boldsymbol{\widecheck{s}}\right]^{n}\right)\le D'_{0},\mathcal{A}|\Psi_{j,k},j\in[m],\boldsymbol{Z}^{n}\right]\nonumber \\
 & =\mathbb{P}\Bigl[\sum_{j=1}^{m}d(\left[S_{j}\right]^{n},\left[\widecheck{s}_{j}\right]^{n})\leq D'_{0},\mathcal{A}|\Psi_{j,k},j\in[m],\boldsymbol{Z}^{n}\Bigr]\\
 & =\sum_{\boldsymbol{d}\in\mathcal{D}^{mn}}\prod_{j=1}^{m}\Pbb\left[d(\left[S_{j}\right]^{n},\left[\widecheck{s}_{j}\right]^{n})=d_{j},\mathcal{A}_{j}|\Psi_{j,k},Z_{j}^{n}\right]\label{eq:-38}\\
 & \leq\sum_{\boldsymbol{d}\in\mathcal{D}^{mn}}\prod_{j=1}^{m}\Pbb\left[d(S_{j}^{n},\left[\widecheck{s}_{j}\right]^{n})\leq d_{j}^{\prime},\mathcal{A}_{j}|\Psi_{j,k},Z_{j}^{n}\right],\label{eq:-36}
\end{align}
where $d_{j}^{\prime}\triangleq\left(\sqrt{d_{j}}+\Delta\right)^{2}$,
and
\begin{align*}
\mathcal{D}^{mn} & \triangleq\left\{ d(\left[\boldsymbol{s}\right]^{n},\left[\boldsymbol{\widecheck{s}}\right]^{n}):\left[\boldsymbol{s}\right]^{n},\left[\boldsymbol{\widecheck{s}}\right]^{n}\in\mathcal{B}^{mn},d\left(\left[\boldsymbol{s}\right]^{n},\left[\boldsymbol{\widecheck{s}}\right]^{n}\right)\le D'_{0}\right\} \\
\mathcal{B}^{mn} & \triangleq\left\{ \left[\boldsymbol{\widecheck{s}}\right]^{n}\in\mathcal{N}^{mn}:\left\Vert \left[\widecheck{s}_{j}\right]^{n}\right\Vert \leq\sqrt{n\Gamma_{j}},1\leq j\leq m\right\}
\end{align*}
with $\sqrt{\Gamma_{j}}\triangleq\sqrt{\lambda_{j}\left(1+\delta\right)}+\Delta+\sqrt{mD'_{0}}$.
\eqref{eq:-38} follows from that $\mathcal{A}_{j}$ implies $\left\Vert \left[S_{j}\right]^{n}\right\Vert \leq\sqrt{n\lambda_{j}\left(1+\delta\right)}+\Delta$,
hence it is sufficient to only consider the case of $\left\Vert \left[\widecheck{s}_{j}\right]^{n}\right\Vert \leq\sqrt{n\Gamma_{j}}$.
\eqref{eq:-36} is obtained by using triangle inequality again.

Combining \eqref{eq:-26-1}, \eqref{eq:-25-1} and \eqref{eq:-36}
gives us
\begin{align}
 & \mathbb{P}_{\mathcal{C}\boldsymbol{Z}^{n}}\Bigl[\max_{R_{n}\mathsf{Hcodes}}\mathbb{P}\bigl[d(\boldsymbol{S}^{n},\boldsymbol{\widecheck{S}}^{n})\le D_{0}|\mathcal{C}\boldsymbol{Z}^{n}\bigr]>\tau_{n}\Bigr]\nonumber \\
 & \leq\left|\mathcal{B}^{mn}\right|\max_{\left[\boldsymbol{\widecheck{s}}\right]^{n}\in\mathcal{B}^{mn}}\mathbb{P}_{\mathcal{C}\boldsymbol{Z}^{n}}\Big[\sum_{k=1}^{2^{nR_{\mathsf{K}}}}\xi_{k,\boldsymbol{z}^{n}}\left(\left[\boldsymbol{\widecheck{s}}\right]^{n}\right)>\tau'_{n}2^{-n\left(R_{n}-R_{\mathsf{K}}\right)}\Big]+\epsilon_{n}^{\prime},\label{eq:-4-2}
\end{align}
where
\begin{align}
\xi_{k,\boldsymbol{z}^{n}}\left(\left[\boldsymbol{\widecheck{s}}\right]^{n}\right) & \triangleq\sum_{\boldsymbol{d}\in\mathcal{D}^{mn}}\prod_{j=1}^{m}\Pbb\left[d(S_{j}^{n},\left[\widecheck{s}_{j}\right]^{n})\leq d_{j}^{\prime},\mathcal{A}_{j}|\Psi_{j,k},Z_{j}^{n}\right].
\end{align}

\noindent Given $\left[\boldsymbol{\widecheck{s}}\right]^{n}$ and $\boldsymbol{z}^{n}$,
$\xi_{k,\boldsymbol{z}^{n}}\left(\left[\boldsymbol{\widecheck{s}}\right]^{n}\right),k\in\left[2^{nR_{\mathsf{K}}}\right]$
are i.i.d. with mean
\begin{align}
\mathbb{E}_{{\mathcal{C}}}\xi_{k,\boldsymbol{z}^{n}}\left(\left[\boldsymbol{\widecheck{s}}\right]^{n}\right) & =\mathbb{E}_{{\mathcal{C}}}\sum_{\boldsymbol{d}\in\mathcal{D}^{mn}}\prod_{j=1}^{m}\Pbb\left[d(S_{j}^{n},\left[\widecheck{s}_{j}\right]^{n})\leq d_{j}^{\prime},\mathcal{A}_{j}|\Psi_{j,k},Z_{j}^{n}\right]\\
 & =\sum_{\boldsymbol{d}\in\mathcal{D}^{mn}}\prod_{j=1}^{m}\mathbb{E}_{\Psi_{j,k}}\Pbb\left[d(S_{j}^{n},\left[\widecheck{s}_{j}\right]^{n})\leq d_{j}^{\prime},\mathcal{A}_{j}|\Psi_{j,k},Z_{j}^{n}\right].
\end{align}

To bound \eqref{eq:-4-2}, we need to bound $\left|\mathcal{B}^{mn}\right|$
and $\left|\mathcal{D}^{mn}\right|$ first. Similar to \eqref{eq:-37-1},
it can be shown
\begin{align}
\left|\mathcal{B}^{mn}\right| & \leq2^{\frac{n}{2}\sum_{j=1}^{m}\log\frac{2\pi e\left(\sqrt{\Gamma_{j}}+\sqrt{\Delta^{2}}\right)^{2}}{\Delta^{2}}+O\left(\log n\right)}.\label{eq:-39}
\end{align}
In addition, for $\left[s_{j}\right]^{n},\left[\widecheck{s}_{j}\right]^{n}$
such that $\left\Vert \left[s_{j}\right]^{n}\right\Vert \leq\sqrt{n\Gamma_{j}}$,
$\left\Vert \left[\widecheck{s}_{j}\right]^{n}\right\Vert \leq\sqrt{n\Gamma_{j}}$,
using triangle inequality we have
\begin{align}
d(\left[s_{j}\right]^{n},\left[\widecheck{s}_{j}\right]^{n}) & =\left\Vert \left[s_{j}\right]^{n}-\left[\widecheck{s}_{j}\right]^{n}\right\Vert ^{2}\leq4n\Gamma_{j}.
\end{align}
Combine it with
\begin{align}
d(\left[s_{j}\right]^{n},\left[\widecheck{s}_{j}\right]^{n}) & =\frac{1}{n}\sum_{i=1}^{n}\left(\left[s_{i,j}\right]-\left[\widecheck{s}_{i,j}\right]\right)^{2}=\frac{\Delta^{2}}{n}\sum_{i=1}^{n}\left(l_{i,j}-\widecheck{l}_{i,j}\right)^{2},
\end{align}
where $l_{i,j}\triangleq\textrm{Round}\left(\frac{s_{i,j}}{\Delta}\right)$
and $\widecheck{l}_{i,j}\triangleq\textrm{Round}\left(\frac{\widecheck{s}_{i,j}}{\Delta}\right)$
are both integers, then we have
\begin{align}
\sum_{i=1}^{n}\left(l_{i,j}-\widecheck{l}_{i,j}\right)^{2} & \leq\frac{4n^{2}\Gamma_{j}}{\Delta^{2}}.
\end{align}
In addition, $\sum_{i=1}^{n}\left(l_{i,j}-\widecheck{l}_{i,j}\right)^{2}\in\mathbb{N}\cup\left\{ 0\right\} $,
hence

\begin{equation}
\left|\mathcal{D}^{mn}\right|\leq\prod_{j=1}^{m}\left(\frac{4n^{2}\Gamma_{j}}{\Delta^{2}}+1\right).
\end{equation}
That is, $\left|\mathcal{D}^{mn}\right|$ is bounded by a polynomial
term of $n$.

If we can show that the probability in \eqref{eq:-4-2} decays doubly
exponentially fast with $n$, then the proof will be complete. To
that end, by using Lemma \ref{lem:typebound-rd-si} we have
\begin{align}
\xi_{k,\boldsymbol{z}^{n}}\left(\left[\boldsymbol{\widecheck{s}}\right]^{n}\right) & \leq\left|\mathcal{D}^{mn}\right|2^{-n(\sum_{j=1}^{m}R_{S_{j}|Z_{j}}(d_{j}^{\prime})-o(1))}\label{eq:-40}\\
 & =2^{-n(\sum_{j=1}^{m}R_{S_{j}|Z_{j}}(d_{j}^{\prime})-o(1))}\label{eq:-45}\\
 & \leq2^{-nm(R_{\boldsymbol{S|Z}}(D_{0}+\epsilon_{\Delta})-o(1))},
\end{align}
where $R_{\boldsymbol{S|Z}}(D)$ given in \eqref{eq:Rs-z} denotes
the conditional rate-distortion function for source $\boldsymbol{S}$
with side information $\boldsymbol{Z}$ at both encoder and decoder,
$o(1)$ is a term that vanishes as $\delta\rightarrow0$ and $n\rightarrow\infty$,
and $\epsilon_{\Delta}$ is a term that vanishes as $\Delta\rightarrow0$.
\eqref{eq:-45} follows from that $\left|\mathcal{D}^{mn}\right|$
grows only polynomially fast with $n$.

From Lemma~\ref{lem:typebound-rd}, we have for any $\boldsymbol{z}^{n}$,
\begin{equation}
\mathbb{E}_{{\mathcal{C}}}\xi_{k,\boldsymbol{z}^{n}}\left(\left[\boldsymbol{\widecheck{s}}\right]^{n}\right)\leq2^{-nm(R_{\boldsymbol{S}}(D_{0}+\epsilon_{\Delta})-o(1))},\label{eq:rd-1-1-1-2}
\end{equation}
where $R_{\boldsymbol{S}}(D)$ given in \eqref{eq:Rs} denotes the
point-to-point rate-distortion function for $\boldsymbol{S}$.

Using these bounds and applying Lemmas \ref{lem:chernoff}, we have
that the probability in \eqref{eq:-4-2} decays doubly exponentially
fast with $n$. This completes the proof of Theorem \ref{thm:vector}.

\end{document}